\documentclass[11pt,notitlepage]{article}
\usepackage[margin=1in]{geometry}
\usepackage{epsfig}
\usepackage{caption}
\usepackage{booktabs}
\usepackage{natbib}
\usepackage{subfigure}
\usepackage{placeins}
\usepackage{amsmath}
\usepackage[usenames,dvipsnames,svgnames,table]{xcolor}
\usepackage{amssymb}
\usepackage{setspace}
\usepackage{graphicx} 
\usepackage{times}
\usepackage{amsthm}
\usepackage{hyperref}
\usepackage{float}
\usepackage{adjustbox}
\hypersetup{bookmarks=true, unicode=false, pdftoolbar=true, pdfmenubar=true, pdffitwindow=false, pdfstartview={FitH}, pdfcreator={}, pdfproducer={}, pdfkeywords={} {} {}, pdfnewwindow=true, colorlinks=true, linkcolor=blue, citecolor=Green, filecolor=magenta, urlcolor=cyan,}
\usepackage{authblk}
\usepackage{multirow}
\usepackage{parskip}
\usepackage{multibib}
\newcites{S}{References}

\graphicspath{{main_figs/}{si_figs/}{si_survey_figs/}}

\newcommand{\ttitle}[0]{Consensus and fragmentation in academic publication preferences}

\newcommand{\BiologyTopUnique}{92}
\newcommand{\BiologyMidUnique}{183}
\newcommand{\BiologyLowUnique}{252}
\newcommand{\BiologyTopNameA}{Nature}
\newcommand{\BiologyTopPercentA}{22.3}
\newcommand{\BiologyTopNameB}{Science}
\newcommand{\BiologyTopPercentB}{22.1}
\newcommand{\BiologyTopNameC}{Cell}
\newcommand{\BiologyTopPercentC}{7.9}
\newcommand{\BiologyTopNameD}{PNAS}
\newcommand{\BiologyTopPercentD}{4.2}
\newcommand{\BiologyTopNameE}{Nature Neuroscience}
\newcommand{\BiologyTopPercentE}{2.4}

\newcommand{\BusinessTopUnique}{50}
\newcommand{\BusinessMidUnique}{125}
\newcommand{\BusinessLowUnique}{159}
\newcommand{\BusinessTopNameA}{The Accounting Review}
\newcommand{\BusinessTopPercentA}{14.3}
\newcommand{\BusinessTopNameB}{Academy of Management Journal}
\newcommand{\BusinessTopPercentB}{11.7}
\newcommand{\BusinessTopNameC}{Journal of Finance}
\newcommand{\BusinessTopPercentC}{8.8}
\newcommand{\BusinessTopNameD}{Management Science}
\newcommand{\BusinessTopPercentD}{7.5}
\newcommand{\BusinessTopNameE}{Administrative Science Quarterly}
\newcommand{\BusinessTopPercentE}{6.8}

\newcommand{\ChemistryTopUnique}{40}
\newcommand{\ChemistryMidUnique}{84}
\newcommand{\ChemistryLowUnique}{134}
\newcommand{\ChemistryTopNameA}{Journal of the American Chemical Society}
\newcommand{\ChemistryTopPercentA}{22.3}
\newcommand{\ChemistryTopNameB}{Science}
\newcommand{\ChemistryTopPercentB}{20.9}
\newcommand{\ChemistryTopNameC}{Nature}
\newcommand{\ChemistryTopPercentC}{18.0}
\newcommand{\ChemistryTopNameD}{Nature Chemistry}
\newcommand{\ChemistryTopPercentD}{3.9}
\newcommand{\ChemistryTopNameE}{Analytical Chemistry}
\newcommand{\ChemistryTopPercentE}{2.9}

\newcommand{\ComputerTopUnique}{97}
\newcommand{\ComputerMidUnique}{152}
\newcommand{\ComputerLowUnique}{155}
\newcommand{\ComputerTopNameA}{Symposium on Principles of Programming Languages (POPL)}
\newcommand{\ComputerTopPercentA}{3.6}
\newcommand{\ComputerTopNameB}{IEEE Transactions on Software Engineering}
\newcommand{\ComputerTopPercentB}{3.0}
\newcommand{\ComputerTopNameC}{Annual Meeting of the Association for Computational Linguistics (ACL)}
\newcommand{\ComputerTopPercentC}{3.0}
\newcommand{\ComputerTopNameD}{USENIX Security Symposium (UsenixSec)}
\newcommand{\ComputerTopPercentD}{3.0}
\newcommand{\ComputerTopNameE}{International Conference on Software Engineering (ICSE)}
\newcommand{\ComputerTopPercentE}{3.0}

\newcommand{\EconomicsTopUnique}{35}
\newcommand{\EconomicsMidUnique}{76}
\newcommand{\EconomicsLowUnique}{113}
\newcommand{\EconomicsTopNameA}{The American Economic Review}
\newcommand{\EconomicsTopPercentA}{41.9}
\newcommand{\EconomicsTopNameB}{Econometrica}
\newcommand{\EconomicsTopPercentB}{12.9}
\newcommand{\EconomicsTopNameC}{Quarterly Journal of Economics}
\newcommand{\EconomicsTopPercentC}{11.3}
\newcommand{\EconomicsTopNameD}{Journal of Finance}
\newcommand{\EconomicsTopPercentD}{5.9}
\newcommand{\EconomicsTopNameE}{Journal of Political Economy}
\newcommand{\EconomicsTopPercentE}{4.8}

\newcommand{\EngineeringTopUnique}{107}
\newcommand{\EngineeringMidUnique}{154}
\newcommand{\EngineeringLowUnique}{180}
\newcommand{\EngineeringTopNameA}{Science}
\newcommand{\EngineeringTopPercentA}{9.3}
\newcommand{\EngineeringTopNameB}{Nature}
\newcommand{\EngineeringTopPercentB}{9.3}
\newcommand{\EngineeringTopNameC}{Journal of Fluid Mechanics}
\newcommand{\EngineeringTopPercentC}{5.6}
\newcommand{\EngineeringTopNameD}{Environmental Science \& Technology}
\newcommand{\EngineeringTopPercentD}{4.7}
\newcommand{\EngineeringTopNameE}{Journal of Structural Engineering-asce}
\newcommand{\EngineeringTopPercentE}{3.7}

\newcommand{\HistoryTopUnique}{58}
\newcommand{\HistoryMidUnique}{111}
\newcommand{\HistoryLowUnique}{136}
\newcommand{\HistoryTopNameA}{The American Historical Review}
\newcommand{\HistoryTopPercentA}{34.7}
\newcommand{\HistoryTopNameB}{The Journal of American History}
\newcommand{\HistoryTopPercentB}{12.1}
\newcommand{\HistoryTopNameC}{Past \& Present}
\newcommand{\HistoryTopPercentC}{4.6}
\newcommand{\HistoryTopNameD}{Isis}
\newcommand{\HistoryTopPercentD}{3.5}
\newcommand{\HistoryTopNameE}{William and Mary Quarterly}
\newcommand{\HistoryTopPercentE}{2.3}

\newcommand{\MathematicsTopUnique}{66}
\newcommand{\MathematicsMidUnique}{103}
\newcommand{\MathematicsLowUnique}{125}
\newcommand{\MathematicsTopNameA}{Annals of Mathematics}
\newcommand{\MathematicsTopPercentA}{26.9}
\newcommand{\MathematicsTopNameB}{Journal of the American Statistical Association}
\newcommand{\MathematicsTopPercentB}{7.5}
\newcommand{\MathematicsTopNameC}{Inventiones Mathematicae}
\newcommand{\MathematicsTopPercentC}{6.5}
\newcommand{\MathematicsTopNameD}{Journal of the American Mathematical Society}
\newcommand{\MathematicsTopPercentD}{4.3}
\newcommand{\MathematicsTopNameE}{Journal for Research in Mathematics Education}
\newcommand{\MathematicsTopPercentE}{3.2}

\newcommand{\MedicineTopUnique}{52}
\newcommand{\MedicineMidUnique}{93}
\newcommand{\MedicineLowUnique}{88}
\newcommand{\MedicineTopNameA}{JAMA}
\newcommand{\MedicineTopPercentA}{11.4}
\newcommand{\MedicineTopNameB}{Nature}
\newcommand{\MedicineTopPercentB}{10.5}
\newcommand{\MedicineTopNameC}{The New England Journal of Medicine}
\newcommand{\MedicineTopPercentC}{9.5}
\newcommand{\MedicineTopNameD}{Science}
\newcommand{\MedicineTopPercentD}{5.7}
\newcommand{\MedicineTopNameE}{American Journal of Epidemiology}
\newcommand{\MedicineTopPercentE}{3.8}

\newcommand{\PhilosophyTopUnique}{39}
\newcommand{\PhilosophyMidUnique}{60}
\newcommand{\PhilosophyLowUnique}{77}
\newcommand{\PhilosophyTopNameA}{The Philosophical Review}
\newcommand{\PhilosophyTopPercentA}{25.7}
\newcommand{\PhilosophyTopNameB}{Ethics}
\newcommand{\PhilosophyTopPercentB}{11.9}
\newcommand{\PhilosophyTopNameC}{Mind}
\newcommand{\PhilosophyTopPercentC}{8.9}
\newcommand{\PhilosophyTopNameD}{Philosophy of Science}
\newcommand{\PhilosophyTopPercentD}{5.0}
\newcommand{\PhilosophyTopNameE}{Journal of the History of Philosophy}
\newcommand{\PhilosophyTopPercentE}{4.0}

\newcommand{\PhysicsTopUnique}{41}
\newcommand{\PhysicsMidUnique}{79}
\newcommand{\PhysicsLowUnique}{121}
\newcommand{\PhysicsTopNameA}{Physical Review Letters}
\newcommand{\PhysicsTopPercentA}{31.5}
\newcommand{\PhysicsTopNameB}{Nature}
\newcommand{\PhysicsTopPercentB}{19.7}
\newcommand{\PhysicsTopNameC}{The Astrophysical Journal}
\newcommand{\PhysicsTopPercentC}{14.2}
\newcommand{\PhysicsTopNameD}{Science}
\newcommand{\PhysicsTopPercentD}{9.1}
\newcommand{\PhysicsTopNameE}{The Astrophysical Journal Letters}
\newcommand{\PhysicsTopPercentE}{2.8}

\newcommand{\PsychologyTopUnique}{124}
\newcommand{\PsychologyMidUnique}{211}
\newcommand{\PsychologyLowUnique}{240}
\newcommand{\PsychologyTopNameA}{Journal of Personality and Social Psychology}
\newcommand{\PsychologyTopPercentA}{8.6}
\newcommand{\PsychologyTopNameB}{Science}
\newcommand{\PsychologyTopPercentB}{6.4}
\newcommand{\PsychologyTopNameC}{Psychological Science}
\newcommand{\PsychologyTopPercentC}{5.0}
\newcommand{\PsychologyTopNameD}{Nature}
\newcommand{\PsychologyTopPercentD}{4.4}
\newcommand{\PsychologyTopNameE}{Nature Neuroscience}
\newcommand{\PsychologyTopPercentE}{4.2}

\newcommand{\SociologyTopUnique}{71}
\newcommand{\SociologyMidUnique}{137}
\newcommand{\SociologyLowUnique}{164}
\newcommand{\SociologyTopNameA}{American Sociological Review}
\newcommand{\SociologyTopPercentA}{37.5}
\newcommand{\SociologyTopNameB}{American Journal of Sociology}
\newcommand{\SociologyTopPercentB}{11.2}
\newcommand{\SociologyTopNameC}{Demography}
\newcommand{\SociologyTopPercentC}{4.9}
\newcommand{\SociologyTopNameD}{American Anthropologist}
\newcommand{\SociologyTopPercentD}{3.4}
\newcommand{\SociologyTopNameE}{Science}
\newcommand{\SociologyTopPercentE}{2.6}
\newcommand{\BiologyPercentJIF}{88.63}
\newcommand{\BusinessPercentJIF}{84.6}
\newcommand{\ChemistryPercentJIF}{88.08}
\newcommand{\ComputerSciencePercentJIF}{57.47}
\newcommand{\EconomicsPercentJIF}{85.75}
\newcommand{\EngineeringPercentJIF}{80.16}
\newcommand{\HistoryPercentJIF}{80.55}
\newcommand{\MathematicsPercentJIF}{85.95}
\newcommand{\MedicinePercentJIF}{89.53}
\newcommand{\PhilosophyPercentJIF}{78.43}
\newcommand{\PhysicsPercentJIF}{78.33}
\newcommand{\PsychologyPercentJIF}{85.07}
\newcommand{\SociologyPercentJIF}{84.85}
\newcommand{\TotalParticipants}{3,620}

\newcommand{\TotalVenues}{8,044}
\newcommand{\TotalVenueComparisons}{163,002}
\newcommand{\ParticipantsInRelevantFields}{3,510}
\newcommand{\ParticipantsWithAspirations}{2,194}
\newcommand{\PercentWithoutUpset}{55.7}
\newcommand{\PercentOfCompsUpset}{3.3}
\newcommand{\ObsAvgUpsetRank}{0.51}
\newcommand{\SimAvgUpsetRank}{0.6}
\newcommand{\AcademiaMenPercentPorC}{68.49}
\newcommand{\AcademiaWomenPercentPorC}{24.99}
\newcommand{\AcademiaUnknownPercentPorC}{6.52}
\newcommand{\AcademiaAssistantProfessorPercentPorC}{4.56}
\newcommand{\AcademiaAssociateProfessorPercentPorC}{28.77}
\newcommand{\AcademiaFullProfessorPercentPorC}{62.05}
\newcommand{\AcademiaCountPorC}{3,510}
\newcommand{\BiologyMenPercentPorC}{68.8}
\newcommand{\BiologyWomenPercentPorC}{24.61}
\newcommand{\BiologyUnknownPercentPorC}{6.59}
\newcommand{\BiologyAssistantProfessorPercentPorC}{4.65}
\newcommand{\BiologyAssociateProfessorPercentPorC}{29.26}
\newcommand{\BiologyFullProfessorPercentPorC}{60.47}
\newcommand{\BiologyCountPorC}{516}
\newcommand{\BusinessMenPercentPorC}{71.99}
\newcommand{\BusinessWomenPercentPorC}{19.61}
\newcommand{\BusinessUnknownPercentPorC}{8.4}
\newcommand{\BusinessAssistantProfessorPercentPorC}{11.48}
\newcommand{\BusinessAssociateProfessorPercentPorC}{32.21}
\newcommand{\BusinessFullProfessorPercentPorC}{54.06}
\newcommand{\BusinessCountPorC}{357}
\newcommand{\ChemistryMenPercentPorC}{80.43}
\newcommand{\ChemistryWomenPercentPorC}{13.19}
\newcommand{\ChemistryUnknownPercentPorC}{6.38}
\newcommand{\ChemistryAssistantProfessorPercentPorC}{1.28}
\newcommand{\ChemistryAssociateProfessorPercentPorC}{22.98}
\newcommand{\ChemistryFullProfessorPercentPorC}{70.21}
\newcommand{\ChemistryCountPorC}{235}
\newcommand{\ComputerScienceMenPercentPorC}{72.15}
\newcommand{\ComputerScienceWomenPercentPorC}{17.81}
\newcommand{\ComputerScienceUnknownPercentPorC}{10.05}
\newcommand{\ComputerScienceAssistantProfessorPercentPorC}{3.2}
\newcommand{\ComputerScienceAssociateProfessorPercentPorC}{27.4}
\newcommand{\ComputerScienceFullProfessorPercentPorC}{63.93}
\newcommand{\ComputerScienceCountPorC}{219}
\newcommand{\EconomicsMenPercentPorC}{80.36}
\newcommand{\EconomicsWomenPercentPorC}{12.95}
\newcommand{\EconomicsUnknownPercentPorC}{6.7}
\newcommand{\EconomicsAssistantProfessorPercentPorC}{12.95}
\newcommand{\EconomicsAssociateProfessorPercentPorC}{21.88}
\newcommand{\EconomicsFullProfessorPercentPorC}{60.71}
\newcommand{\EconomicsCountPorC}{224}
\newcommand{\EngineeringMenPercentPorC}{80.0}
\newcommand{\EngineeringWomenPercentPorC}{16.08}
\newcommand{\EngineeringUnknownPercentPorC}{3.92}
\newcommand{\EngineeringAssistantProfessorPercentPorC}{3.53}
\newcommand{\EngineeringAssociateProfessorPercentPorC}{30.59}
\newcommand{\EngineeringFullProfessorPercentPorC}{62.35}
\newcommand{\EngineeringCountPorC}{255}
\newcommand{\HistoryMenPercentPorC}{59.52}
\newcommand{\HistoryWomenPercentPorC}{33.81}
\newcommand{\HistoryUnknownPercentPorC}{6.67}
\newcommand{\HistoryAssistantProfessorPercentPorC}{1.9}
\newcommand{\HistoryAssociateProfessorPercentPorC}{31.9}
\newcommand{\HistoryFullProfessorPercentPorC}{60.0}
\newcommand{\HistoryCountPorC}{210}
\newcommand{\MathematicsMenPercentPorC}{83.72}
\newcommand{\MathematicsWomenPercentPorC}{11.16}
\newcommand{\MathematicsUnknownPercentPorC}{5.12}
\newcommand{\MathematicsAssistantProfessorPercentPorC}{2.33}
\newcommand{\MathematicsAssociateProfessorPercentPorC}{28.84}
\newcommand{\MathematicsFullProfessorPercentPorC}{63.26}
\newcommand{\MathematicsCountPorC}{215}
\newcommand{\MedicineMenPercentPorC}{48.36}
\newcommand{\MedicineWomenPercentPorC}{46.72}
\newcommand{\MedicineUnknownPercentPorC}{4.92}
\newcommand{\MedicineAssistantProfessorPercentPorC}{3.28}
\newcommand{\MedicineAssociateProfessorPercentPorC}{31.97}
\newcommand{\MedicineFullProfessorPercentPorC}{59.02}
\newcommand{\MedicineCountPorC}{122}
\newcommand{\PhilosophyMenPercentPorC}{68.7}
\newcommand{\PhilosophyWomenPercentPorC}{26.09}
\newcommand{\PhilosophyUnknownPercentPorC}{5.22}
\newcommand{\PhilosophyAssistantProfessorPercentPorC}{2.61}
\newcommand{\PhilosophyAssociateProfessorPercentPorC}{34.78}
\newcommand{\PhilosophyFullProfessorPercentPorC}{58.26}
\newcommand{\PhilosophyCountPorC}{115}
\newcommand{\PhysicsMenPercentPorC}{79.73}
\newcommand{\PhysicsWomenPercentPorC}{14.29}
\newcommand{\PhysicsUnknownPercentPorC}{5.98}
\newcommand{\PhysicsAssistantProfessorPercentPorC}{3.32}
\newcommand{\PhysicsAssociateProfessorPercentPorC}{22.26}
\newcommand{\PhysicsFullProfessorPercentPorC}{70.43}
\newcommand{\PhysicsCountPorC}{301}
\newcommand{\PsychologyMenPercentPorC}{53.74}
\newcommand{\PsychologyWomenPercentPorC}{39.25}
\newcommand{\PsychologyUnknownPercentPorC}{7.01}
\newcommand{\PsychologyAssistantProfessorPercentPorC}{3.74}
\newcommand{\PsychologyAssociateProfessorPercentPorC}{28.97}
\newcommand{\PsychologyFullProfessorPercentPorC}{63.08}
\newcommand{\PsychologyCountPorC}{428}
\newcommand{\SociologyMenPercentPorC}{47.28}
\newcommand{\SociologyWomenPercentPorC}{46.96}
\newcommand{\SociologyUnknownPercentPorC}{5.75}
\newcommand{\SociologyAssistantProfessorPercentPorC}{1.6}
\newcommand{\SociologyAssociateProfessorPercentPorC}{33.23}
\newcommand{\SociologyFullProfessorPercentPorC}{60.7}
\newcommand{\SociologyCountPorC}{313}

\newcommand{\NSurveyed}{71,390}
\newcommand{\NSurveyedStarted}{4,448}
\newcommand{\NSurveyedCompleted}{3,462}
\newcommand{\PercSurveyedStart}{6.23}
\newcommand{\PercSurveyedCompleted}{4.85}
\newcommand{\NStudied}{3,510}
\newcommand{\NStudiedNotSurveyed}{140}
\newcommand{\PercStudiedNotSurveyed}{3.99}

\newcommand{\AnthropologyEmailedCount}{1951}
\newcommand{\AnthropologyResponseCount}{118}
\newcommand{\AnthropologyResponseRate}{6.05}
\newcommand{\BiochemistryEmailedCount}{2715}
\newcommand{\BiochemistryResponseCount}{146}
\newcommand{\BiochemistryResponseRate}{5.38}
\newcommand{\BiologyEmailedCount}{4385}
\newcommand{\BiologyResponseCount}{324}
\newcommand{\BiologyResponseRate}{7.39}
\newcommand{\BusinessEmailedCount}{8783}
\newcommand{\BusinessResponseCount}{418}
\newcommand{\BusinessResponseRate}{4.76}
\newcommand{\ChemistryEmailedCount}{3277}
\newcommand{\ChemistryResponseCount}{175}
\newcommand{\ChemistryResponseRate}{5.34}
\newcommand{\CivilEngineeringEmailedCount}{2519}
\newcommand{\CivilEngineeringResponseCount}{109}
\newcommand{\CivilEngineeringResponseRate}{4.33}
\newcommand{\ComputerScienceEmailedCount}{5579}
\newcommand{\ComputerScienceResponseCount}{187}
\newcommand{\ComputerScienceResponseRate}{3.35}
\newcommand{\EconomicsEmailedCount}{3451}
\newcommand{\EconomicsResponseCount}{162}
\newcommand{\EconomicsResponseRate}{4.69}
\newcommand{\HistoryEmailedCount}{4264}
\newcommand{\HistoryResponseCount}{175}
\newcommand{\HistoryResponseRate}{4.1}
\newcommand{\MathematicsEmailedCount}{4691}
\newcommand{\MathematicsResponseCount}{189}
\newcommand{\MathematicsResponseRate}{4.03}
\newcommand{\MechanicalEngineeringEmailedCount}{3576}
\newcommand{\MechanicalEngineeringResponseCount}{115}
\newcommand{\MechanicalEngineeringResponseRate}{3.22}
\newcommand{\PhilosophyEmailedCount}{1929}
\newcommand{\PhilosophyResponseCount}{96}
\newcommand{\PhilosophyResponseRate}{4.98}
\newcommand{\PhysicsAstronomyEmailedCount}{5805}
\newcommand{\PhysicsAstronomyResponseCount}{306}
\newcommand{\PhysicsAstronomyResponseRate}{5.27}
\newcommand{\PsychologyEmailedCount}{6376}
\newcommand{\PsychologyResponseCount}{352}
\newcommand{\PsychologyResponseRate}{5.52}
\newcommand{\SociologyEmailedCount}{2418}
\newcommand{\SociologyResponseCount}{221}
\newcommand{\SociologyResponseRate}{9.14}
\newcommand{\AcademiaMeanPrestigePorC}{24.85}
\newcommand{\AcademiaSdPrestigePorC}{20.55}
\newcommand{\BiologyMeanPrestigePorC}{27.51}
\newcommand{\BiologySdPrestigePorC}{20.46}
\newcommand{\BusinessMeanPrestigePorC}{24.99}
\newcommand{\BusinessSdPrestigePorC}{19.28}
\newcommand{\ChemistryMeanPrestigePorC}{26.53}
\newcommand{\ChemistrySdPrestigePorC}{20.51}
\newcommand{\ComputerScienceMeanPrestigePorC}{21.72}
\newcommand{\ComputerScienceSdPrestigePorC}{18.46}
\newcommand{\EconomicsMeanPrestigePorC}{24.66}
\newcommand{\EconomicsSdPrestigePorC}{20.82}
\newcommand{\EngineeringMeanPrestigePorC}{24.85}
\newcommand{\EngineeringSdPrestigePorC}{21.24}
\newcommand{\HistoryMeanPrestigePorC}{21.66}
\newcommand{\HistorySdPrestigePorC}{18.16}
\newcommand{\MathematicsMeanPrestigePorC}{24.48}
\newcommand{\MathematicsSdPrestigePorC}{20.5}
\newcommand{\MedicineMeanPrestigePorC}{22.7}
\newcommand{\MedicineSdPrestigePorC}{20.62}
\newcommand{\PhilosophyMeanPrestigePorC}{20.48}
\newcommand{\PhilosophySdPrestigePorC}{18.5}
\newcommand{\PhysicsMeanPrestigePorC}{22.53}
\newcommand{\PhysicsSdPrestigePorC}{19.28}
\newcommand{\PsychologyMeanPrestigePorC}{29.33}
\newcommand{\PsychologySdPrestigePorC}{23.67}
\newcommand{\SociologyMeanPrestigePorC}{20.08}
\newcommand{\SociologySdPrestigePorC}{17.38}
\newcommand{\AcademiaAssistantProfessorPercentAARC}{27.33}
\newcommand{\AcademiaAssociateProfessorPercentAARC}{27.63}
\newcommand{\AcademiaCountAARC}{361,287}
\newcommand{\AcademiaFullProfessorPercentAARC}{45.04}
\newcommand{\AcademiaMeanPrestigeAARC}{33.19}
\newcommand{\AcademiaMenPercentAARC}{58.5}
\newcommand{\AcademiaSdPrestigeAARC}{26.45}
\newcommand{\AcademiaUnknownPercentAARC}{5.25}
\newcommand{\AcademiaWomenPercentAARC}{36.25}
\newcommand{\BiologyAssistantProfessorPercentAARC}{29.15}
\newcommand{\BiologyAssociateProfessorPercentAARC}{24.4}
\newcommand{\BiologyCountAARC}{59,486}
\newcommand{\BiologyFullProfessorPercentAARC}{46.45}
\newcommand{\BiologyMeanPrestigeAARC}{29.02}
\newcommand{\BiologyMenPercentAARC}{62.89}
\newcommand{\BiologySdPrestigeAARC}{23.68}
\newcommand{\BiologyUnknownPercentAARC}{5.72}
\newcommand{\BiologyWomenPercentAARC}{31.39}
\newcommand{\BusinessAssistantProfessorPercentAARC}{29.13}
\newcommand{\BusinessAssociateProfessorPercentAARC}{28.64}
\newcommand{\BusinessCountAARC}{24,776}
\newcommand{\BusinessFullProfessorPercentAARC}{42.23}
\newcommand{\BusinessMeanPrestigeAARC}{36.21}
\newcommand{\BusinessMenPercentAARC}{65.49}
\newcommand{\BusinessSdPrestigeAARC}{27.71}
\newcommand{\BusinessUnknownPercentAARC}{6.99}
\newcommand{\BusinessWomenPercentAARC}{27.53}
\newcommand{\ChemistryAssistantProfessorPercentAARC}{23.51}
\newcommand{\ChemistryAssociateProfessorPercentAARC}{22.05}
\newcommand{\ChemistryCountAARC}{13,912}
\newcommand{\ChemistryFullProfessorPercentAARC}{54.44}
\newcommand{\ChemistryMeanPrestigeAARC}{33.23}
\newcommand{\ChemistryMenPercentAARC}{72.35}
\newcommand{\ChemistrySdPrestigeAARC}{26.7}
\newcommand{\ChemistryUnknownPercentAARC}{6.1}
\newcommand{\ChemistryWomenPercentAARC}{21.55}
\newcommand{\ComputerScienceAssistantProfessorPercentAARC}{28.32}
\newcommand{\ComputerScienceAssociateProfessorPercentAARC}{26.08}
\newcommand{\ComputerScienceCountAARC}{22,045}
\newcommand{\ComputerScienceFullProfessorPercentAARC}{45.6}
\newcommand{\ComputerScienceMeanPrestigeAARC}{33.2}
\newcommand{\ComputerScienceMenPercentAARC}{71.73}
\newcommand{\ComputerScienceSdPrestigeAARC}{27.37}
\newcommand{\ComputerScienceUnknownPercentAARC}{8.99}
\newcommand{\ComputerScienceWomenPercentAARC}{19.28}
\newcommand{\EconomicsAssistantProfessorPercentAARC}{26.86}
\newcommand{\EconomicsAssociateProfessorPercentAARC}{23.7}
\newcommand{\EconomicsCountAARC}{9,503}
\newcommand{\EconomicsFullProfessorPercentAARC}{49.44}
\newcommand{\EconomicsMeanPrestigeAARC}{29.65}
\newcommand{\EconomicsMenPercentAARC}{73.36}
\newcommand{\EconomicsSdPrestigeAARC}{25.84}
\newcommand{\EconomicsUnknownPercentAARC}{5.19}
\newcommand{\EconomicsWomenPercentAARC}{21.46}
\newcommand{\EngineeringAssistantProfessorPercentAARC}{25.85}
\newcommand{\EngineeringAssociateProfessorPercentAARC}{24.22}
\newcommand{\EngineeringCountAARC}{38,024}
\newcommand{\EngineeringFullProfessorPercentAARC}{49.93}
\newcommand{\EngineeringMeanPrestigeAARC}{30.58}
\newcommand{\EngineeringMenPercentAARC}{75.71}
\newcommand{\EngineeringSdPrestigeAARC}{25.79}
\newcommand{\EngineeringUnknownPercentAARC}{7.53}
\newcommand{\EngineeringWomenPercentAARC}{16.76}
\newcommand{\HistoryAssistantProfessorPercentAARC}{16.23}
\newcommand{\HistoryAssociateProfessorPercentAARC}{34.08}
\newcommand{\HistoryCountAARC}{9,271}
\newcommand{\HistoryFullProfessorPercentAARC}{49.69}
\newcommand{\HistoryMeanPrestigeAARC}{31.79}
\newcommand{\HistoryMenPercentAARC}{58.19}
\newcommand{\HistorySdPrestigeAARC}{27.13}
\newcommand{\HistoryUnknownPercentAARC}{2.39}
\newcommand{\HistoryWomenPercentAARC}{39.41}
\newcommand{\MathematicsAssistantProfessorPercentAARC}{23.83}
\newcommand{\MathematicsAssociateProfessorPercentAARC}{23.26}
\newcommand{\MathematicsCountAARC}{14,596}
\newcommand{\MathematicsFullProfessorPercentAARC}{52.91}
\newcommand{\MathematicsMeanPrestigeAARC}{33.59}
\newcommand{\MathematicsMenPercentAARC}{72.02}
\newcommand{\MathematicsSdPrestigeAARC}{28.07}
\newcommand{\MathematicsUnknownPercentAARC}{8.18}
\newcommand{\MathematicsWomenPercentAARC}{19.8}
\newcommand{\MedicineAssistantProfessorPercentAARC}{31.03}
\newcommand{\MedicineAssociateProfessorPercentAARC}{24.86}
\newcommand{\MedicineCountAARC}{32,127}
\newcommand{\MedicineFullProfessorPercentAARC}{44.12}
\newcommand{\MedicineMeanPrestigeAARC}{26.89}
\newcommand{\MedicineMenPercentAARC}{58.59}
\newcommand{\MedicineSdPrestigeAARC}{21.72}
\newcommand{\MedicineUnknownPercentAARC}{5.95}
\newcommand{\MedicineWomenPercentAARC}{35.45}
\newcommand{\PhilosophyAssistantProfessorPercentAARC}{20.34}
\newcommand{\PhilosophyAssociateProfessorPercentAARC}{28.93}
\newcommand{\PhilosophyCountAARC}{4,918}
\newcommand{\PhilosophyFullProfessorPercentAARC}{50.73}
\newcommand{\PhilosophyMeanPrestigeAARC}{34.63}
\newcommand{\PhilosophyMenPercentAARC}{66.51}
\newcommand{\PhilosophySdPrestigeAARC}{27.61}
\newcommand{\PhilosophyUnknownPercentAARC}{2.75}
\newcommand{\PhilosophyWomenPercentAARC}{30.74}
\newcommand{\PhysicsAssistantProfessorPercentAARC}{20.14}
\newcommand{\PhysicsAssociateProfessorPercentAARC}{19.56}
\newcommand{\PhysicsCountAARC}{13,758}
\newcommand{\PhysicsFullProfessorPercentAARC}{60.3}
\newcommand{\PhysicsMeanPrestigeAARC}{28.41}
\newcommand{\PhysicsMenPercentAARC}{77.71}
\newcommand{\PhysicsSdPrestigeAARC}{26.2}
\newcommand{\PhysicsUnknownPercentAARC}{5.09}
\newcommand{\PhysicsWomenPercentAARC}{17.2}
\newcommand{\PsychologyAssistantProfessorPercentAARC}{25.45}
\newcommand{\PsychologyAssociateProfessorPercentAARC}{27.08}
\newcommand{\PsychologyCountAARC}{15,399}
\newcommand{\PsychologyFullProfessorPercentAARC}{47.47}
\newcommand{\PsychologyMeanPrestigeAARC}{38.2}
\newcommand{\PsychologyMenPercentAARC}{48.34}
\newcommand{\PsychologySdPrestigeAARC}{27.51}
\newcommand{\PsychologyUnknownPercentAARC}{3.41}
\newcommand{\PsychologyWomenPercentAARC}{48.25}
\newcommand{\SociologyAssistantProfessorPercentAARC}{24.98}
\newcommand{\SociologyAssociateProfessorPercentAARC}{29.89}
\newcommand{\SociologyCountAARC}{8,115}
\newcommand{\SociologyFullProfessorPercentAARC}{45.12}
\newcommand{\SociologyMeanPrestigeAARC}{39.01}
\newcommand{\SociologyMenPercentAARC}{49.28}
\newcommand{\SociologySdPrestigeAARC}{28.47}
\newcommand{\SociologyUnknownPercentAARC}{3.5}
\newcommand{\SociologyWomenPercentAARC}{47.22}

\newcommand{\ComputerScienceLOOPredAccuracy}{66.5}
\newcommand{\EconomicsLOOPredAccuracy}{79.2}

\newcommand{\ComputerScienceTopFiveAgreement}{16.4}
\newcommand{\EconomicsTopFiveAgreement}{71.3}

\newcommand{\TopFiveLOOPredCorrR}{0.91}

\newcommand{\MinPercentOfCompsUpsetByField}{2.2}
\newcommand{\MaxPercentOfCompsUpsetByField}{4.1}

\newcommand{\FigFourMenPrefMean}{0.83}
\newcommand{\FigFourWomenPrefMean}{0.78}
\newcommand{\FigFourMenAspMean}{0.73}
\newcommand{\FigFourWomenAspMean}{0.7}

\newcommand{\FigFourPrefMostPrestigiousMean}{0.82}
\newcommand{\FigFourPrefLeastPrestigiousMean}{0.77}
\newcommand{\FigFourPrestigePrefPValue}{0.028}
\newcommand{\FigFourAspMostPrestigiousMean}{0.77}
\newcommand{\FigFourAspLeastPrestigiousMean}{0.62}

\newcommand{\OverallPrefMean}{0.81}
\newcommand{\OverallAspMean}{0.72}
\newcommand{\FigFiveOverallDiffAtPrestigeTen}{1.19}
\newcommand{\FigFiveOverallDiffPPAtPrestigeTen}{23.8}
\newcommand{\FigFiveBiologyIndivPredAtTen}{3.6}
\newcommand{\FigFiveBiologyFieldPredAtTen}{2.0}
\newcommand{\FigFiveMathematicsIndivPredAtTen}{1.9}
\newcommand{\FigFiveMathematicsFieldPredAtTen}{0.36}

\begin{document}
\author[1]{Ian Van Buskirk}
\author[2]{Marilena Hohmann}
\author[3,4]{Ekaterina Landgren}
\author[5]{Johan Ugander}
\author[1,6,7]{Aaron Clauset}
\author[1,6,7]{Daniel B. Larremore\thanks{daniel.larremore@colorado.edu}}
\affil[1]{Department of Computer Science, University of Colorado Boulder, Boulder, CO, USA}
\affil[2]{Copenhagen Center for Social Data Science, University of Copenhagen, Copenhagen, Denmark}
\affil[3]{Cooperative Institute for Research in Environmental Sciences, University of Colorado Boulder, Boulder, CO, USA}
\affil[4]{Department of Environmental Social Sciences, Stanford Doerr School of Sustainability, Stanford University, Stanford, CA, USA}
\affil[5]{Department of Statistics \& Data Science, Yale University, New Haven, CT, USA}
\affil[6]{BioFrontiers Institute, University of Colorado Boulder, Boulder, CO, USA}
\affil[7]{Santa Fe Institute, Santa Fe, NM, USA}
\date{}
\title{\ttitle}
\maketitle
\begin{abstract}
Academic publishing requires solving a collective coordination problem: among thousands of possible publication venues, which deserve a community's attention? A clear consensus helps scholars allocate attention, match submissions to appropriate outlets, and evaluate scholars for hiring and promotion. Yet preferences are not centrally coordinated---they emerge within each field over time. Here we ask whether all fields have arrived at similar solutions to this coordination problem, and whether preferences vary systematically with individual characteristics. Using an adaptive survey of 3,510 US tenure-track faculty yielding 163,002 pairwise comparisons across 8,044 venues, we show that fields occupy a wide spectrum of coordination. Economics, Chemistry, and Physics exhibit strong consensus, with respondents agreeing on elite venues and accurately predicting one another's choices. Computer Science and Engineering show fragmented preferences distributed across hundreds of outlets with minimal overlap. Within fields, preferences correlate with institutional prestige---faculty at elite institutions prefer higher-ranked venues---and with gender, as men prefer higher-ranked venues than women even after accounting for prestige and career stage. Scholars realize their personal preferences more successfully than their respective fields' consensus preferences, indicating that heterogeneity, not just selective hierarchy, shapes publishing outcomes. Journal Impact Factors explain only 64\% of preference choices, systematically undervaluing what fields actually prefer. These results quantify how publication preferences vary across the structural diversity of academic fields.
\end{abstract}
\captionsetup{font=small}

\section*{Introduction}

Across the academic enterprise, millions of papers are published each year, and authors continuously face questions of where, among thousands of possible venues, they would prefer for their work to appear. Across surveys and disciplines, the two factors most consistently cited as driving these preferences are a venue's reputation or prominence and its topical scope~\cite{rowlands2005scholarly, tenopir2016motivates, rowley2022factors}, though researchers also weigh review quality, open access options, speed of review, and publication fees~\cite{rowley2022factors, wijewickrema2017journal, xu2023factors}.

With sufficient consensus on these dimensions, a community's aggregated preferences reflect a collective solution to an attention allocation problem. By transforming an intractable individual decision (which of thousands of papers deserve my attention?) into a manageable collective one (which venues does our community hold in high regard?), a clear consensus enables communities to estimate the scope and potential importance of articles from venue alone, to match submissions to appropriate outlets, and---in some fields and at some institutions---to make hiring and promotion decisions based on the venues on a scholar's curriculum vitae~\cite{niles2020we, heckman2020publishing, bryce2020journal}. On the other hand, deviations from any putative consensus represent fundamental disagreements about attention allocation, with the potential to undermine these coordinating benefits.

A rich literature has studied publication preferences, typically one field at a time, using methods including citation analysis~\cite{gross1927college, bradshaw2016rank}, expert surveys~\cite{serenko2011comparing, herrmann2011going},
and analysis of manuscript submission flows~\cite{calcagno2012flows, garfinkel2024academic}. This work has produced venue rankings across numerous fields---including economics~\cite{heckman2020publishing}, marketing~\cite{steward2010comprehensive}, statistics~\cite{theoharakis2003statisticians}, medicine~\cite{saha2003impact}, and AI~\cite{serenko2011comparing}---and has shown that preferences correlate with respondents' own publication records~\cite{herrmann2011going, bryce2020journal}, that consensus varies across subfields within a discipline~\cite{theoharakis2003statisticians, barrick2019ranking}, and that citation-based metrics imperfectly capture stated preferences~\cite{serenko2011comparing, bradshaw2016rank, saha2003impact}. Yet because publication preferences are not centrally coordinated---they emerge and evolve independently in each field over time---these field-specific studies leave three important questions unanswered. First, have all fields arrived at similar levels of consensus in their publication preferences? Second, do preferences vary systematically with individual characteristics such as institutional prestige, career stage, or gender? And third, to what extent are scholars effective in realizing their preferences by publishing in preferred venues?

To answer these questions, we surveyed \ParticipantsInRelevantFields~US tenure-track faculty using a three-stage adaptive survey (Materials and Methods). In the first stage, respondents named a top-tier venue that they would most like to publish in, which we refer to as {\it aspirations}. In the second stage, respondents assembled a set of venues they like to, hope to, or want to publish in---a  {\it consideration set}. In the third stage, we confronted each respondent with a sequence of pairwise comparisons between venues from their consideration set to learn their publication {\it preferences}. Combined with professional and demographic information (field, institution, gender, academic title, and institutional prestige~\cite{wapman2022quantifying}), these three data types---aspirations, consideration sets, and preferences---enable a field-by-field analysis of the structure of publication preferences across 13 academic fields.

From these data, we define several key quantities. First, we named each field's most common aspiration its {\it flagship venue}. Second, to measure the strength of consensus across consideration sets, we identified the three most commonly selected venues in each field and then measured the fraction of respondents who selected those venues. Third, we used each respondent's pairwise venue comparisons to compute their {\it individual rankings}, and then computed a set of {\it field-level rankings} from the union of comparison data for respondents from each field. In total, our dataset comprises \TotalVenueComparisons~pairwise comparisons among \TotalVenues~venues. This study was approved by the University of Colorado IRB, and deidentified data and open-source code are available (Materials and Methods).

Respondents to our survey do not constitute a representative sample of academia, yet we have reason to believe that their responses can nevertheless address our research questions. Our sample is restricted to tenure-track faculty at US PhD-granting institutions, and senior faculty, men, and those at more prestigious institutions responded at higher rates (SI Appendix). However, our analyses show no statistical differences in responses by academic title, and our sample contains sufficient representation across institutional prestige and gender to examine differences by these covariates.

\section*{Results}

\subsection*{A spectrum of concentration and consensus}
\label{sec:fields}

\begin{figure*}[t]
    \centering
    \includegraphics[width=1.0\linewidth, trim=0 10 0 0pt, clip]{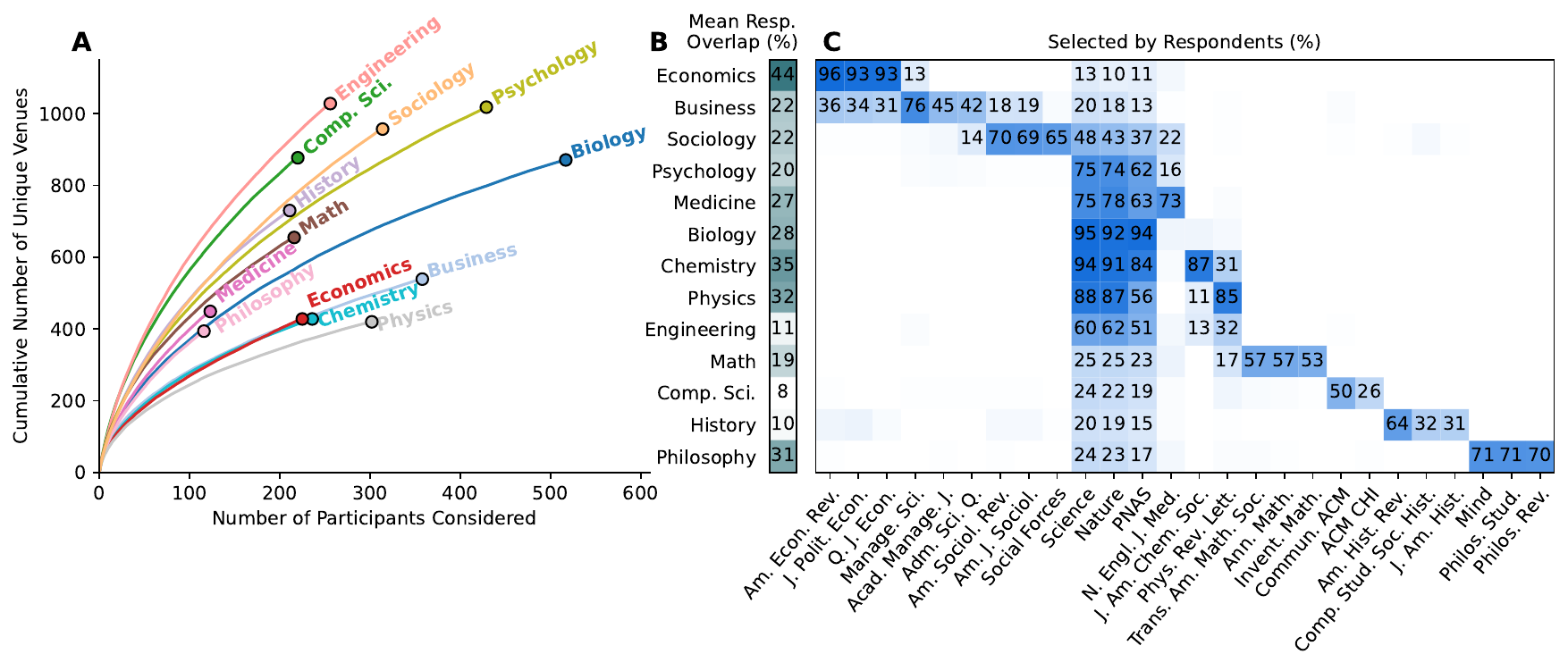}
    \caption{\textbf{Fields vary widely in the diversity of their consideration sets and consensus around common venues.}
    (A) Venue accumulation curves show the growing number of unique venues among the combined consideration sets of an increasing number of respondents, sampled uniformly at random without replacement for each field and averaged over $100$ realizations. Circles and labels indicate the total number of respondents and unique venues selected for each field. 
    (B) Within-field overlap quantifies the average percentage of respondents in each field that have selected venues chosen by others in their field. 
    (C) For each of the three most commonly selected venues in each field, an annotated heatmap shows the percentage of respondents per field who selected that venue. Annotations are suppressed for values under 10\%. 
    }
    \label{fig:overlap}
\end{figure*}

Our data show unambiguously that academic fields vary widely in their levels of consensus and preference coordination. At the most basic level, respondents' consideration sets---the unordered sets of venues in which each respondent would be interested in publishing---reveal a spectrum from more to less consolidated fields. For instance, the union of $100$ consideration sets, randomly sampled from Physics respondents, contains just $247$ unique venues, similar to Economics ($266$), Chemistry ($275$), and Business ($282$), yet equal-sized samples from Computer Science and Engineering contain $567$ and $598$ venues, respectively. These differences are preserved across sampling efforts, reflected in venue accumulation curves (Fig.~\ref{fig:overlap}A) which count the average number of unique venues contained in the union of an increasing number of respondents' consideration sets (akin to species accumulation curves used in empirical ecology~\cite{colwell1994estimating}). Moreover, variation in consideration set diversity is not explained by variation in the number of venues respondents selected (Pearson $\rho=-0.28$; $p=0.36$), despite moderate differences in the average size of consideration sets (range: $13.0$ to $17.7$ venues).

Alternative measures of consensus computed from consideration sets reinforce both the wide variability between fields and their ordering. For instance, the average respondents in Economics, Chemistry, and Physics share 44\%, 35\% and 32\% of their overall consideration sets with others in their respective fields, compared with just 8\% and 11\% for respondents in Computer Science and Engineering, respectively (Fig.~\ref{fig:overlap}B). Or, under a measure of common agreement, focusing on the three most commonly selected venues in each field, 93\textendash96\% of respondents in Economics selected the most popular venues, compared with 24\textendash50\% for Computer Science and 31\textendash64\% for History (Fig.~\ref{fig:overlap}C). Importantly, these measures of consensus are highly mutually correlated (venue accumulation vs mean overlap, Spearman $r=-0.89$; $p<10^{-4}$; mean overlap vs top-3, Spearman $r=0.85$; $p<10^{-4}$), revealing that regardless of how one reasonably analyzes consideration sets, fields can be arranged on a spectrum of venue preference coordination. On one side are high-consensus and concentrated fields (Economics, Physics, and Chemistry) and on the other are low-consensus and deconcentrated fields (Computer Science and Engineering). 

The tabulation of the most commonly selected venues in each field reveals two further points about preferences across fields. First, {\it Science}, {\it Nature}, and {\it the Proceedings of the National Academies of Science } were selected by over 50\% of respondents in only six fields (Psychology, Medicine, Biology, Chemistry, Physics, and Engineering), while less than 25\% chose any of those venues in Economics, Business, Mathematics, Computer Science, History, and Philosophy, meaning that the appeal of these three general science journals is restricted to a subset of the academy (Fig.~\ref{fig:overlap}C). Second, outside of general science journals, the most popular venues in one field are rarely considered relevant in others---a clear indication that venue topic matters---yet exceptions do exist: respondents in Business indicated a preference for popular Economics and Sociology venues, and respondents in multiple fields overlapped in their preferences for {\it Journal of the American Chemical Society} and {\it Physical Review Letters} (Fig.~\ref{fig:overlap}C).


\subsection*{Preferences from Pairwise Comparisons} 

Respondents' pairwise comparisons reveal their preferences over the venues in their consideration sets.
On average, respondents provided $46$ such comparisons each,
from which we computed scalar rankings using the SpringRank algorithm~\cite{de2018physical} at both the individual level (using each respondent's responses in isolation) and the field level (by aggregating responses for all respondents in each field; Materials and Methods). 

\begin{figure}[t]
    \centering
    \includegraphics[width=0.55\linewidth]{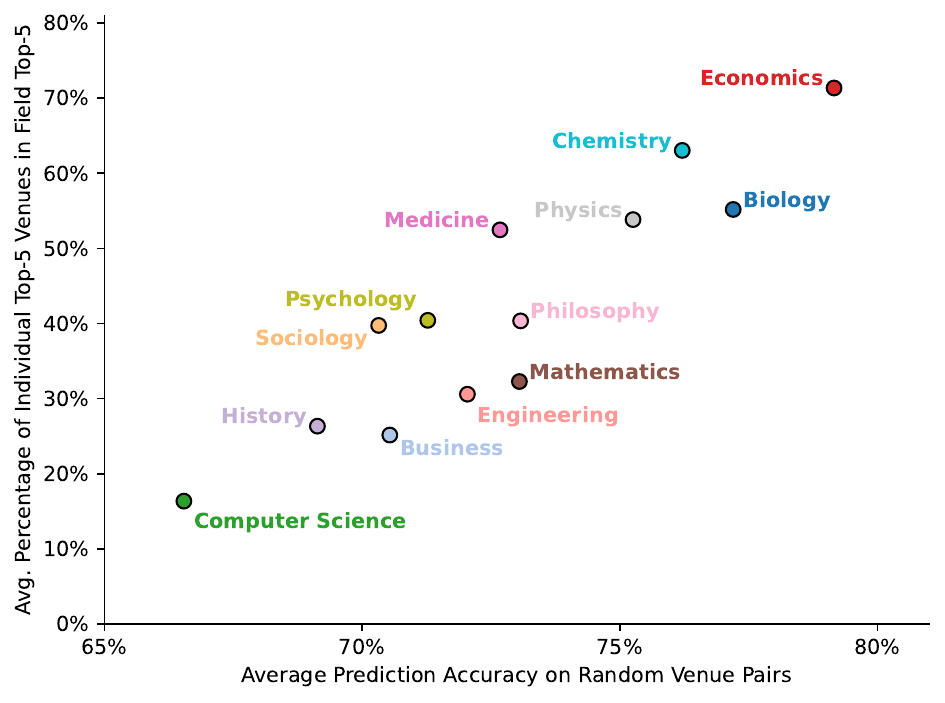}
    \caption{\textbf{Fields vary widely in their consensus around top-5 venues and the general alignment of their preferences.} Whether one uses a field-level ranking constructed from the pairwise comparisons of all respondents in a field to predict an individual's pairwise comparisons (horizontal axis) or assesses the degree to which individuals in a field agree on a top-5 (vertical axis), some fields (e.g.~Economics) demonstrate a much more overlapping and organized set of preferences than others (e.g.~Computer science).
    }
    \label{fig:pref_agreement}
\end{figure}

Do field-level rankings represent an actual consensus or simply the aggregation of mutually conflicting responses? We first answered this question via a prediction task, evaluating how well each respondent's pairwise choices could be predicted by a field-level ranking which excluded their own responses---a form of leave-one-out accuracy. All field-level rankings outperformed the baseline 50\% accuracy of random guessing, but prediction accuracy varied considerably by field, ranging from \ComputerScienceLOOPredAccuracy\% in Computer Science to \EconomicsLOOPredAccuracy\% in Economics (Fig.~\ref{fig:pref_agreement}). 

This simple prediction approach to measuring the strength of preference consensus is unable to resolve the situation in which respondents agree on the top $k$ venues, but disagree about the ordering thereof. Indeed, in some fields, a small number of venues are culturally canonized as members of a named elite group---a {\it top 5} journal in contemporary Economics~\cite{heckman2020publishing} or a {\it big 3} in 1970s Sociology~\cite{gaston1979big}---or are explicitly codified into institutional rule books for tenure and promotion evaluations~\cite{steward2010comprehensive, bryce2020journal}, but are otherwise unordered. Consequently, as a second measure of preference consensus, we directly computed the agreement between individuals' top-5 venues and those of their respective field.

Top-5 agreement provides a similar portrait of preference consensus across fields. Not only do Computer Science and Economics again anchor two ends of the spectrum with \ComputerScienceTopFiveAgreement\% and \EconomicsTopFiveAgreement\% agreement, respectively, but this measure is also highly correlated with prediction accuracy
(Pearson $\rho = \TopFiveLOOPredCorrR$; $p<10^{-4}$; Fig.~\ref{fig:pref_agreement}). In all measures of consensus thus far, Economics, Chemistry, Biology, and Physics consistently emerge as high-consensus fields.

At the individual level, respondents are almost perfectly self-consistent: only \PercentOfCompsUpset\% of comparisons disagreed with inferred rankings (range: \MinPercentOfCompsUpsetByField\%\textendash\MaxPercentOfCompsUpsetByField\% across fields), and \PercentWithoutUpset\% of respondents were entirely self-consistent (SI Appendix). This implies that the wide variation in the strength of preference consensus across fields is due to meaningfully different preferences between generally self-consistent individuals, and not variation in self-consistency by field. Those self-inconsistencies that do occur tend to affect lower-ranked venues, even after taking into account the structure of the adaptive survey design (SI Appendix), implying that preferences for lower-ranked venues are less clearly resolved, and thus more likely to produce inconsistent pairwise choices.

\begin{figure*}
    \centering
    \includegraphics[width=1.0\linewidth]{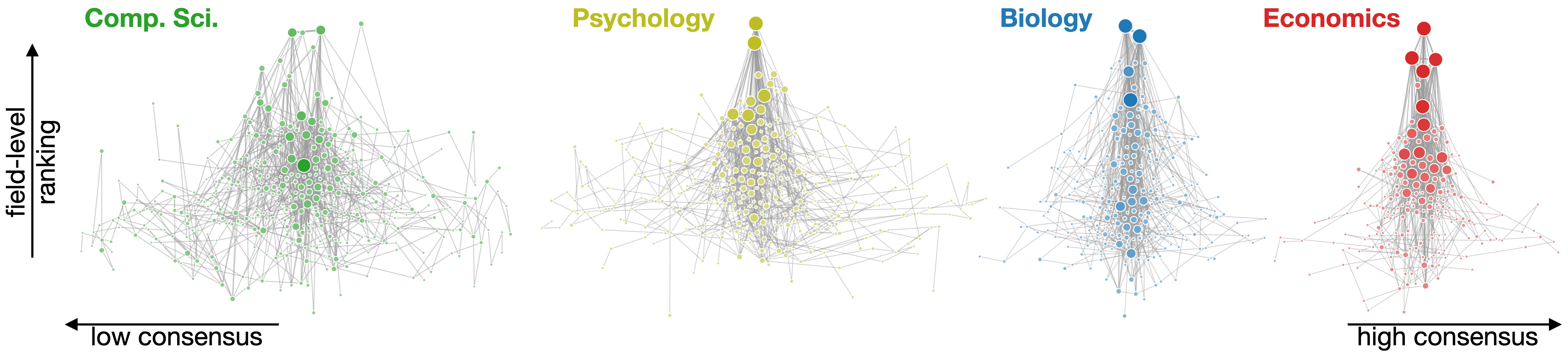}
    \caption{\textbf{Preference network embeddings for four fields arranged from low to high consensus.} Network embeddings illustrate variation in consideration set and preference similarity by field, derived from 50 randomly chosen respondents in each field. Venues (circles) are sized by the frequencies with which they appear in consideration sets, and are vertically positioned by field-level preference scores. Links (lines) are drawn to connect each respondent's first-ranked venue to their second-ranked venue to their third-ranked venue, and so on, allowing nodes to find horizontal positions via a force-directed graph layout. See SI Appendix, Fig.~\ref{sfig:network-sketches} for networks for all fields.}
    \label{fig:networks}
\end{figure*}

Our results paint a relatively clear portrait of consensus and preferences across fields. Whether measured by the predictability of respondents' choices, the consistency of top-5 rankings, the popularity of common venues, or ecological venue accumulation curves, academic fields populate a spectrum from high-consensus fields such as Economics, Chemistry, or Biology, to low-consensus fields such as Engineering, History, or Computer Science. In high-consensus fields, the most popular venues are also the most preferred and there is broad agreement about which venues those are, while in low-consensus fields the most popular venues are less likely to be considered and less likely to be top-ranked. Network embeddings provide an intuitive depiction of these properties across fields (Figs.~\ref{fig:networks} and \ref{sfig:network-sketches}), illustrating that fields solve preference coordination and consensus problems to considerably different degrees.

\subsection*{Prestige, Gender, and Academic Rank}
\label{sec:prestige-gender}

\begin{figure*}[t]
    \centering
    \includegraphics[width=\linewidth]{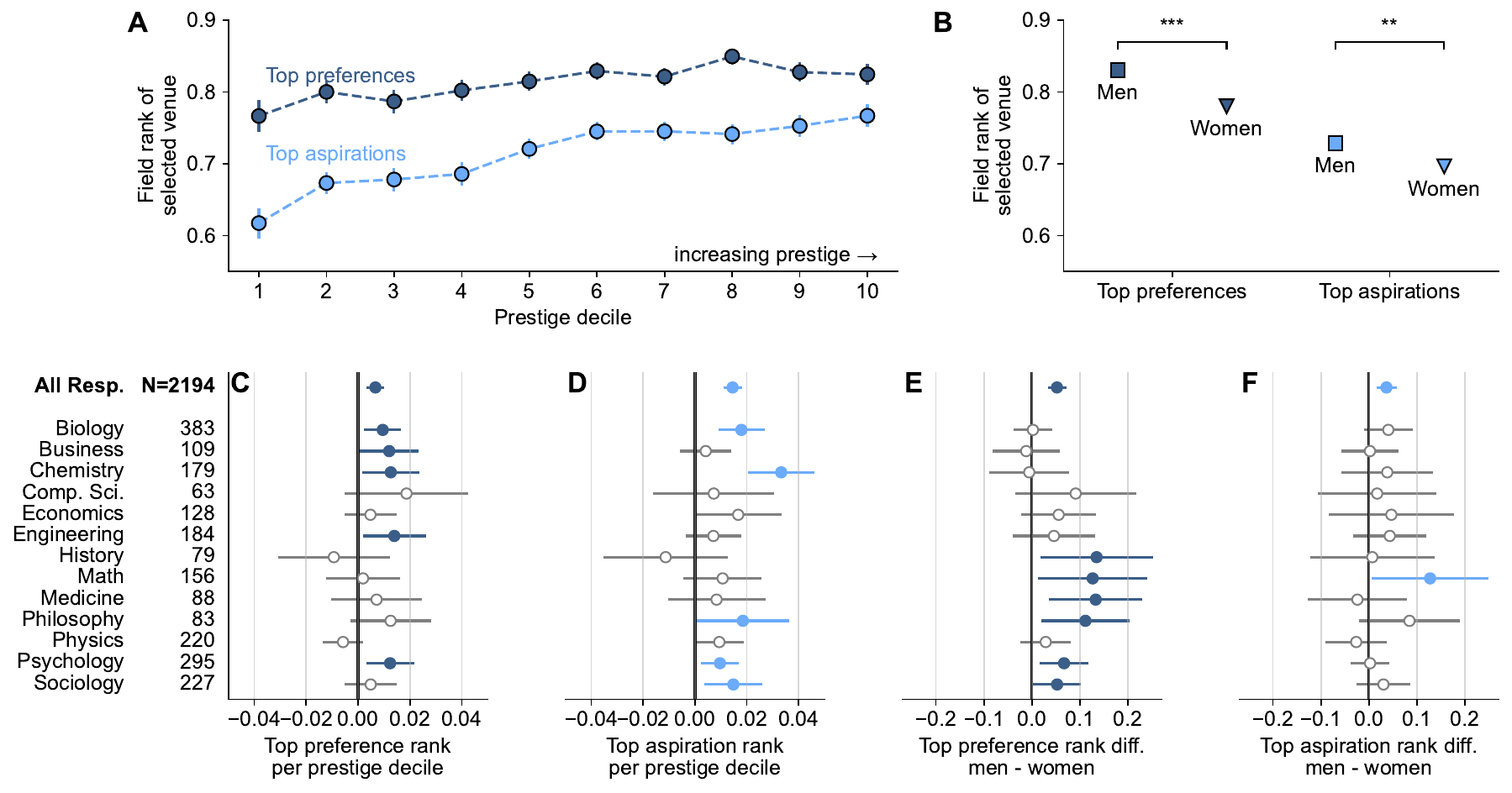}
    \caption{\textbf{Variation in top preferences and top aspirations by prestige and gender.} (A,B) Mean 0-to-1 normalized field-level rank of respondents top preferences and top aspirations, binned by institutional prestige decile and by gender.  (C-F) Multiple linear regression coefficients showing the association between (C,D) a one-decile increase in institutional prestige (E,F) men vs women, for top preferences and top aspirations, controlling for career stage, with fields and sample sizes indicated. Error bars represent $95\%$ confidence intervals, with color indicating statistical significant $\alpha=0.05$, and grey indicating n.s. Asterisks indicate statistical significance via two-sided $z$-test, $^{***}$ $p=1.0\times 10^{-6}$, $^{**}$ $p=2.2\times 10^{-3}$.
    }
    \label{fig:researcher_characteristics}
\end{figure*}

Publication preferences are acquired, learned, or formed over the course of an education and research career, raising the possibility that some of the factors known to shape academic careers may also correlate with preferences. Among the factors measured by our survey are career stage, institutional prestige, and gender, which have been shown to correlate with other quantities of interest in the science of science, such as research productivity and novelty~\cite{way2017misleading, tripodi2025tenure}, funded labor and faculty placement~\cite{zhang2022labor, wapman2022quantifying}, citations~\cite{lariviere2013bibliometrics}, and the impacts of parenthood~\cite{morgan2021unequal}. While many approaches to measuring associations between preferences and these covariates are possible, we focused our analysis on the field-level rank of each respondent's top preferred venue---a measure of how highly others rank that venue---normalized to range from $1$ (the field's and individual's top preferences coincide) to $0$ (the field's least preferred venue is the individual's top preference). 

Preferences correlate with institutional prestige. On average, top-preferred venues had a field-level rank of \FigFourPrefMostPrestigiousMean~for respondents at the most prestigious institutions, compared to \FigFourPrefLeastPrestigiousMean~for those at the least prestigious ones (Fig.~\ref{fig:researcher_characteristics}A, two-sided $z$-test $p=\FigFourPrestigePrefPValue$), indicating that faculty at more prestigious institutions prefer more highly rated venues. A significant association between preferences and prestige persists even after including career stage and gender in a linear regression model (Fig.~\ref{fig:researcher_characteristics}C; see Materials and Methods). Applying this regression model to respondents one field at a time, prestige and top-preference remained significantly correlated in five fields, were not significant in the remaining eight (Fig.~\ref{fig:researcher_characteristics}C), and in no field was the correlation reversed.

Compared to men, women tended to prefer venues that are ranked lower in their field. Men's top preferred venues had an average field-level rank of \FigFourMenPrefMean, compared to \FigFourWomenPrefMean~for women (Fig.~\ref{fig:researcher_characteristics}B, two-sided $z$-test $p=1.0\times 10^{-6}$), and this pattern, too, persisted in regression models combining all respondents, despite the additional inclusion of prestige and career stage. Differences by gender cannot be fully explained by differences in gender composition across fields (SI Appendix, Fig.~\ref{fig:regression_robustness}). Again applying regression models field by field, men's preferences for higher ranked venues  were significant in six individual fields, not significantly different in the remaining seven (Fig.~\ref{fig:researcher_characteristics}E), and in no field was this association reversed. 

Despite the observation that preferences and  aspirations are likely learned or acquired over the course of a career, we found no evidence of associations between preferences or aspirations with career stage (Fig.~\ref{sfig:full_prestige_gender_regression}). This null result, combined with clear associations with both institutional prestige and gender, points to a process in which preferences and aspirations are learned prior to hiring as faculty, and differential acculturation by gender and institutional prestige.

Preferences and aspirations are related but distinct. The rank of aspirations (i.e., the venues that respondents named as most wanting to publish in), like the rank of top preferences, increases with prestige (e.g., \FigFourAspMostPrestigiousMean~vs \FigFourAspLeastPrestigiousMean~for faculty in the highest vs lowest decile of institutional prestige; two-sided $z$-test $p<10^{-6}$; Fig.~\ref{fig:researcher_characteristics}A), and is higher for men than for women (\FigFourMenAspMean~vs \FigFourWomenAspMean, two-sided $z$-test $p=0.002$; Fig.~\ref{fig:researcher_characteristics}B). These associations persisted in regression models of all respondents, but cannot be established with statistical certainty when analyzing most individual fields (Figs.~\ref{fig:researcher_characteristics}D, F). 

Our results show a clear aspiration-preference gap, however, which persists for all values of prestige and for both men and women, such that the respondents' top preferred venues are ranked markedly higher than their top aspirations (Figs.~\ref{fig:researcher_characteristics}A,B). This gap is wider at the bottom of the prestige hierarchy compared to the top, indicating that aspirations vary more sharply with institutional prestige than preferences do. 
Gaps between preferences and aspirations have been documented in other domains~\cite{bruch2018aspirational, hoxby2012missing}, but here the existence of a gap is not guaranteed, because in principle, respondents could have revealed their top aspiration to also be their most preferred venue, but our results show clearly that this is not the case, with top preferences ranking $\OverallPrefMean$ at the field level and aspirations just $\OverallAspMean$, on average over all respondents (two-sided $z$-test, $p<10^{-6}$). 


\subsection*{Realizing Preferences}

\begin{figure*}[t]
    \centering
    \includegraphics[width=\linewidth]{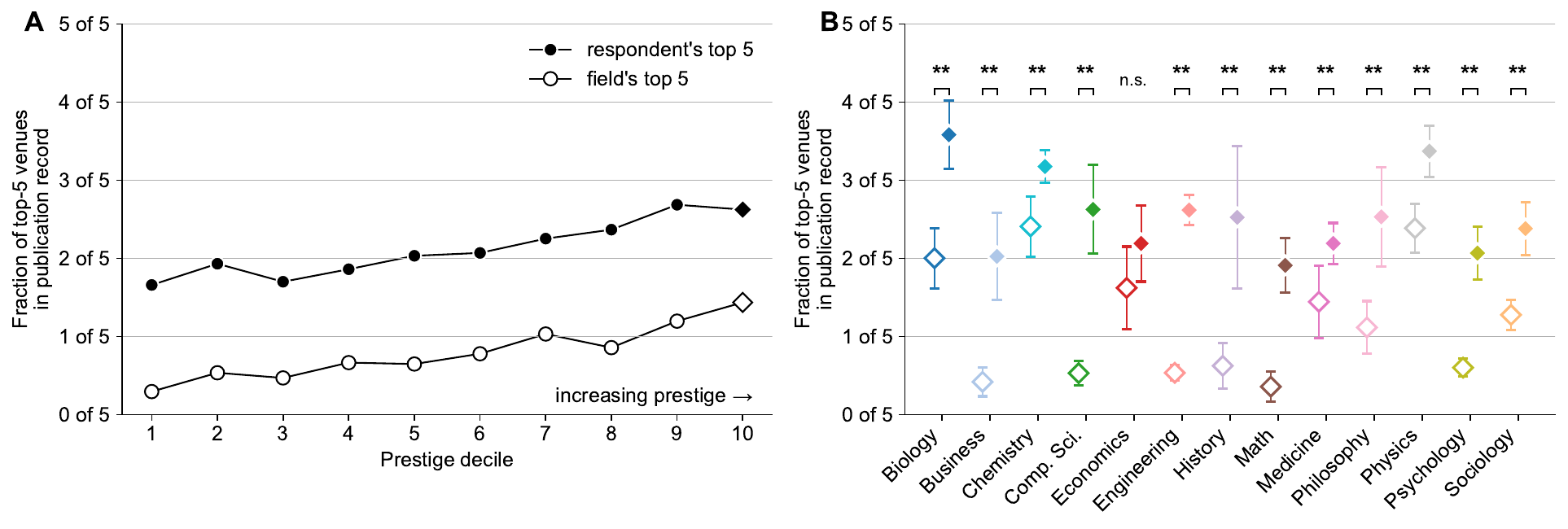}
    \caption{\textbf{Preference realization varies by institutional prestige and reference group}. (A) The average fraction of top-5 venues found in respondent's publication records---a measure of preference realization---is stratified by prestige decile, and separated by which reference group defines the top-5. Solid markers use each individual respondent's top-5, and open markers use field-level top-5. Diamond markers indicate values at the highest prestige decile, to indicate they are comparable to panel B. (B) Post-regression estimates of preference realization in the top-prestige decile for individual top-5 (solid marker) and field-level (open marker) top-5. Brackets and asterisks indicate a significant difference in means (Benjamini-Hochberg corrected for multiple comparisons; $\alpha=0.01$); n.s., not significant. See SI Appendix, Table~\ref{stab:top5_regression_results} for complete regression table.} 
    \label{fig:histogram_publication_history}
\end{figure*}

Preferences may or may not be realized due to the competitive and selective nature of academic publishing. To understand the gap between preference and achievement, we examined two distinct measures, computed from each respondent's Google Scholar profile. First, we computed the fraction of each respondent's ranking's top-5 list they have ever published in.  Second, we repeated this calculation using their respective fields' consensus top-five lists. We then modeled these preference realization rates, or ``top-5 tick rates,''  as a function of institutional prestige, which has been shown to correlate with elite publication rates in other work~\cite{wellmon2017publication}, and field.

Respondents were more successful in realizing their own preferences than those established by field consensus, both overall and in all fields except Economics (where the difference was not significant), with an average difference of \FigFiveOverallDiffAtPrestigeTen~more ticked venues (\FigFiveOverallDiffPPAtPrestigeTen pp; $p<0.01$, Benjamini-Hochberg corrected; Fig.~\ref{fig:histogram_publication_history}). For instance, Biology faculty in the highest prestige decile had ticked an average of \FigFiveBiologyIndivPredAtTen~of their personal top five, compared with \FigFiveBiologyFieldPredAtTen~of the field's consensus, while for faculty in Mathematics, the gap was \FigFiveMathematicsIndivPredAtTen~vs. \FigFiveMathematicsFieldPredAtTen~(Fig.~\ref{fig:histogram_publication_history}B, SI Appendix, Table.~\ref{stab:top5_regression_results}). 

Systematic differences between personal and field-level preference realization might derive from three potential mechanisms: (i) strategic adaptation, in which respondents adjust preferences toward more achievable goals, (ii) learning, in which respondents come to value venues through experience publishing there, or (iii) preference heterogeneity (as distinct from adaptation or learning), in which respondents simply hold and realize different preferences than the consensus. If preference adaptation were highly prevalent, we would expect to see the personal-field gap shrink or disappear at elite institutions where researchers can more readily access consensus-preferred venues. Instead, the gap persists uniformly across the institutional prestige spectrum, despite the fact that preference realization rates do indeed increase with institutional prestige, both empirically (Fig.~\ref{fig:histogram_publication_history}A) and via regression, both overall and in 9 of 13 individual fields (SI Appendix, Table~\ref{stab:top5_regression_results}). 

Similarly, if differences were explained purely by idiosyncratic fit between researchers and venues (pure preference heterogeneity), one would not expect such a strong relationship between institutional prestige and top-5 tick rates. Instead, the prestige gradient suggests genuine stratification in access to selective venues.

Together, these patterns indicate that both hierarchy and heterogeneity structure venue preference realization, i.e., academic publishing. More prestigious institutions genuinely correlate with better placement outcomes, whether measured against an individual's or a field's preferences, while simultaneously, researchers at all levels exhibit preferences that diverge from---and are better satisfied than---field consensus would predict.

\subsection*{Journal Impact Factor}

Citation counts and citation-derived Journal Impact Factors (JIFs) have been used to rank publication venues for nearly 100 years~\cite{gross1927college} for purposes ranging from prioritization of library subscriptions~\cite{gross1927college} to faculty reappointment, tenure, and promotion evaluations~\cite{niles2020we}, despite numerous technical and consequential critiques of these practices (see \citep{lariviere2019journal} for a comprehensive review). Given this widely employed journal metric, we explored the extent to which JIFs explain actual preferences --- that is, whether preferences captured in our survey simply reflect differences in venue citedness.

Among comparisons for which both venues' JIFs exist (77\% of venues, 69\% of comparisons), respondents chose the higher JIF venue in just 64\% of comparisons, compared to 71\% agreement with survey-derived field-consensus rankings over the same comparisons. Among individual fields, JIF was most predictive in Biology (71\% accuracy) and least predictive in History (54\% accuracy), was available for the largest fraction of comparisons in Business (81\%) and smallest in Computer Science (19\%), and consistently predicted fewer comparisons than data-derived rankings (range: 4-13pp fewer; SI Fig.~\ref{fig:jif_comparison_accuracy}). Thus, when available for venues at all, JIF imperfectly correlates with consensus preferences and is an inconsistent predictor of individual choices~\cite{serenko2011comparing, extejt1990behavioral, bradshaw2016rank, bryce2020journal}.

Because JIF relies solely on field-independent citation counts, it systematically misaligns with field-specific publication preferences in predictable ways. When general-interest venues like {\it Nature}, {\it Science}, and {\it PNAS} are central to a field---as in Psychology or Chemistry (Fig.~\ref{fig:overlap})---JIF undervalues them. Conversely, when these venues are peripheral to a field, as in Mathematics or Economics, JIF overvalues them (Supp.\ Table.~\ref{stab:venue_rankings_multidisc}).

The same pattern holds for field-specific flagships which rank an average of 5.9 positions higher in field-level consensus rankings than their JIF would predict (Supp.\ Table~\ref{stab:venue_rankings_flagships}). Together, these patterns reveal a general limitation of impact factors: beyond incomplete coverage and predictive accuracy, impact factors undervalue what each field prefers most, while overvaluing what fields consider peripheral.

\section*{Discussion}

Using a three-stage adaptive survey of 3,510 US tenure-track faculty across 13 academic fields, we demonstrate that publication preferences vary dramatically in their degree of coordination and consensus. Fields occupy a spectrum from high consensus (Economics, Chemistry, Physics) to low consensus (Computer Science, Engineering), with this variation evident across multiple measures: venue accumulation curves, consideration set overlap, top-5 agreement, and pairwise prediction accuracy. Within fields, preferences correlate with institutional prestige—faculty at more prestigious institutions prefer higher-ranked venues—and with gender, as men prefer higher-ranked venues than women even after controlling for prestige and career stage. We also document a persistent aspiration–preference gap: scholars' top preferences rank higher than their stated aspirations, with this gap widening at less prestigious institutions. Scholars more successfully publish in their personal top-5 venues than in field-consensus venues, indicating that preference heterogeneity—not just hierarchical selectivity—structures academic publishing. Finally, Journal Impact Factor poorly predicts actual preferences (64\% accuracy) and systematically undervalues field-specific flagship venues.

Our findings provide direct empirical support for critiques of Journal Impact Factor as a tool for research assessment. 
This pattern is consistent with prior work finding imperfect correlations between JIF and stated preferences: $\rho=0.51$–$0.62$ for AI journals \cite{serenko2011comparing}, $\rho=0.68$–$0.84$ for ecology \cite{bradshaw2016rank}, and a 39\% ``perception gap'' between UK business faculty rankings and official journal guides \cite{bryce2020journal}. 
Our results suggest that evaluators who rely on JIF in lieu of field-specific knowledge may systematically misjudge scholars, particularly when flagship journals in a given field have modest impact factors relative to general-interest alternatives.

This work has clear and cautionary implications for academic evaluation. The stakes are high: in economics, faculty with three publications in ``top 5'' journals experience a 310\% increase in tenure rates compared to equally productive colleagues without such publications \cite{heckman2020publishing}, and 89\% of UK business schools formally incorporate journal ranking lists into promotion and tenure decisions \cite{bryce2020journal}. When asked what drives journal selection, faculty often answer with a version of ``what will help in my tenure case'' \cite{niles2020we}. For hiring committees, our results suggest that venue-based judgments are most reliable in high-consensus fields where evaluators can reasonably assume shared preferences. In low-consensus fields like Computer Science or Engineering, where preferences fragment across hundreds of venues with minimal overlap, evaluators from outside the scholar's subfield may inadvertently impose their own preferences rather than applying community standards. For funding agencies conducting multidisciplinary review, our findings underscore that ``what we all know'' about venue prestige varies substantially by field. Panelists from high-consensus fields may assume that their colleagues share similar clarity about elite venues, when in fact such consensus may not exist. 

Our finding that preferences correlate with institutional prestige aligns with prior work documenting steep hierarchies in faculty hiring \cite{clauset2015systematic, wapman2022quantifying} and the diffusion of scientific ideas \cite{morgan2018prestige}. Faculty at more prestigious institutions not only produce more faculty and publish more papers, but also—as we show here—prefer higher-ranked venues. Whether this reflects socialization (elite institutions acculturate students and faculty into preferring elite venues) or sorting (individuals with such preferences disproportionately obtain positions at elite institutions) remains an open question. Similar ambiguity surrounds our gender findings. Men prefer higher-ranked venues than women, even after controlling for prestige and career stage. Possible mechanisms include differential risk aversion in targeting venues, unequal access to mentorship and professional networks that transmit preference norms, or systematic differences in research topics that correlate with venue preferences. Our data cannot distinguish among these mechanisms, but the persistence of the gender gap across fields and after controls suggests it is not simply a compositional artifact.

Despite clear associations with prestige and gender, we find no evidence that preferences or aspirations vary with faculty rank, a null result consistent with prior surveys finding no significant preference differences by tenure status \cite{extejt1990behavioral}, suggesting that  preference formation occurs during training. Graduate students and postdocs, immersed in lab cultures and exposed to advisor preferences, may acquire stable publication preferences before entering the faculty job market. If so, the prestige and gender associations we observe likely originate during training rather than through on-the-job learning. Any theory of preference formation must simultaneously explain why prestige and gender associations exist but post-training career stage associations do not. A clear extension of this work would survey graduate students and postdoctoral researchers, ideally tracking the same cohort across career transitions, to identify when and how preferences crystallize. Such longitudinal data could also reveal whether preferences adapt to realized publication success or remain stable despite it.

This work is subject to a number of important limitations. First, our sampling frame captures only US tenure-track faculty with appointments in 2020, and responses within this frame were not uniform: senior faculty, men, and those at more prestigious institutions responded at higher rates. While we find no statistical differences in preferences by career stage, our sample may not generalize to faculty in other countries, non-tenure-track researchers, or those in industry or government positions. Future surveys could broaden this frame to include international faculty, graduate students, and researchers outside academia.

Second, self-reported preferences are not revealed preferences. Respondents told us what they prefer, but we do not observe their actual submission behavior. Stated and revealed journal preferences may diverge due to strategic considerations, risk aversion, coauthor compromises, or post-hoc rationalization \cite{calcagno2012flows, rowley2022factors}, and methods for inferring revealed preferences do exist: Calcagno et al.\ \cite{calcagno2012flows} analyzed manuscript flows among 923 biology journals using resubmission patterns, and Garfinkel et al.\ \cite{garfinkel2024academic} used submission histories with rank-order logit models to infer preference orderings across business school disciplines. 
Such data, if available at scale across fields, would complement our survey-based approach and allow direct comparison of stated and revealed preference hierarchies.

Third, our study does not attempt to decompose the determinants of preference. We measure what researchers prefer (and the near-perfect self consistencey of those preferences), but not why. Across other surveys, the two most consistently cited factors are venue reputation (status) and topical scope \cite{rowlands2005scholarly, tenopir2016motivates}, though researchers also weigh review quality, speed, open access options, and publication fees \cite{rowley2022factors, wijewickrema2017journal, xu2023factors}. Moreover, ``quality'' itself is not unidimensional: assessments depend on whether one emphasizes review process rigor, methodological standards, or impact \cite{herrmann2011going, polonsky2006multi}. Our pairwise comparison design captures aggregate preferences but cannot disentangle these components. Future work could directly survey the weight researchers place on different factors, or use conjoint experiments to estimate marginal utilities for specific journal attributes.

Our results raise several stimulating questions about the nature and consequences of preference consensus. First, we suggest that shared preferences function as a form of invisible  infrastructure
by coordinating behavior---reducing uncertainty for authors, reviewers, and evaluators---without explicit negotiation. Underpinning this infrastructure, venue status operates as a shared signal, emerging because it solves a coordination problem:
conditioned on topical fit, high-status venues attract higher-quality submissions, which reinforces their status and attracts further high-quality work, while authors seek the status conferred by publishing in these venues
and thus target the highest-status journal for which their work is appropriate.
High-consensus fields like Economics have effectively built such infrastructure: the widely recognized ``top 5'' journals serve as focal points that reduce uncertainty for authors, reviewers, hiring committees, and promotion boards alike. Once institutionally codified—embedded in tenure guidelines and departmental norms—these focal points become self-reinforcing, driving citations upward and further entrenching their prominence~\cite{drivas2020matthew}. Low-consensus fields lack this infrastructure, requiring actors to rely on local or subfield-specific signals, which may increase coordination costs and also preserve flexibility.

On the other hand, high consensus is not unambiguously beneficial. While coordination infrastructure reduces transaction costs, it may also entrench gatekeeping by concentrating evaluative authority in a small number of venues. Narrow consensus could exclude methodological diversity, topical heterogeneity, or innovative work that does not fit established templates. Conversely, low consensus may reflect epistemic pluralism, distributed innovation, or the natural fragmentation of rapidly evolving fields. 

\section*{Materials and Methods}

\subsection*{Survey administration}
The survey was conducted in two waves with initial invitations sent in January and February of 2024 and follow-up emails sent in September and October of 2024. We contacted faculty at U.S. PhD-granting institutions, drawing on the Academic Analytics Research Center (AARC) census of tenured and tenure-track assistant, associate, and full professors. 
Further details on the sample, its representativeness, and response rates can be found in SI Section~\ref{sec:SI-survey}. This study has been IRB approved (University of Colorado Boulder IRB protocol no.~23-0454).

\subsection*{Respondent Characteristics}
\label{sec:matmeth-respondents}

Respondents selected a field from 19 options, equivalent to the root-level concepts used by OpenAlex circa 2022; analyses are limited to fields with at least 100 respondents (\ParticipantsInRelevantFields~total).
This reduced our initial sample of \TotalParticipants~respondents to \ParticipantsInRelevantFields~respondents and removes the following fields: Art, Environmental Science, Geography, Geology, Materials Science, and Political Science.

Institutional prestige was computed from faculty hiring flows~\cite{wapman2022quantifying}, based on AARC data through 2023. We assign prestige ranks based on respondents' main institutional affiliation at the time of data collection, using self-reported affiliations when available and 2023 employment data from the AARC otherwise. Gender is based on survey responses where available, supplemented by name-based classification via the \texttt{nomquamgender}~\cite{buskirk2023opensource}. Analyses by gender include only respondents labeled as men or women (SI Appendix). Academic titles reflect respondents' self-reports in the survey (Supplementary Materials~\ref{sec:SI-survey}).

\subsection*{Rescaled and Normalized Ranks}

We use respondents' pairwise comparisons to generate preference rankings with the SpringRank algorithm~\cite{de2018physical}, described in detail in the SI Appendix. Individual-level rankings are computed using only each respondent's pairwise comparisons (regularization parameter $\alpha = 0$), while field-level consensus rankings incorporate all comparisons made by respondents within the same field ($\alpha = 20$). For aggregate rankings, setting $\alpha > 0$ ensures that venues with very few comparisons do not disproportionately influence the results by occasionally rising to the top. The results are not sensitive to the specific value of $\alpha$. We rescale so that a one-rank difference corresponds to a 75\% win probability, then normalize to $[0,1]$.

\subsection*{Publication histories}
We obtain respondents' publication histories from Google Scholar. We matched respondents to Google Scholar profiles by name and affiliation, retrieving publication lists for $2,436$ respondents ($2,275$ with prestige data).

\section*{Data, Materials, and Software Availability}
To the extent permitted by our IRB protocol, which protects respondent privacy, data and code have been made available at \href{https://github.com/LarremoreLab/academic-publication-preferences}{github.com/LarremoreLab/academic-publication-preferences}.

\section*{Acknowledgements}
The authors are grateful to all survey respondents who generously contributed their time. This work was supported by Air Force Office of Scientific Research Award FA9550-19-1-0329 (IVB, AC, DBL), the Cooperative Institute for Research in Environmental Sciences Visiting Fellows Program funded by NOAA Cooperative Agreement NA22OAR4320151 (EL), the Stanford Doerr School of Sustainability Dean’s Sustainability Leaders Postdoctoral Fellowship Program (EL), and NSF Alan T. Waterman Award SMA-2226343 (DBL).


\bibliographystyle{unsrt}
\bibliography{prefs}

\begin{thebibliography}{10}

\bibitem{wapman2022quantifying}
K~Hunter Wapman, Sam Zhang, Aaron Clauset, and Daniel~B Larremore.
\newblock Quantifying hierarchy and dynamics in us faculty hiring and
  retention.
\newblock {\em Nature}, 610(7930):120--127, 2022.

\bibitem{morgan2022socioeconomic}
Allison~C Morgan, Nicholas LaBerge, Daniel~B Larremore, Mirta Galesic, Jennie~E
  Brand, and Aaron Clauset.
\newblock Socioeconomic roots of academic faculty.
\newblock {\em Nature Human Behaviour}, 6(12):1625--1633, 2022.

\bibitem{spoon2023gender}
Katie Spoon, Nicholas LaBerge, K~Hunter Wapman, Sam Zhang, Allison~C Morgan,
  Mirta Galesic, Bailey~K Fosdick, Daniel~B Larremore, and Aaron Clauset.
\newblock Gender and retention patterns among us faculty.
\newblock {\em Science Advances}, 9(42):eadi2205, 2023.

\bibitem{priem2022openalex}
Jason Priem, Heather Piwowar, and Richard Orr.
\newblock Openalex: A fully-open index of scholarly works, authors, venues,
  institutions, and concepts.
\newblock {\em arXiv preprint arXiv:2205.01833}, 2022.

\bibitem{collegescorecard2023}
{U.S. Department of Education}.
\newblock College scorecard data.
\newblock \url{https://collegescorecard.ed.gov/data}, 2023.
\newblock Accessed September 26, 2023.

\bibitem{clarivate2024}
Clarivate.
\newblock Journal citation reports, 2024.
\newblock Accessed: 2024-10-07.

\bibitem{vanbuskirk2024wiserank}
Ian Van~Buskirk.
\newblock wiserank, August 2024.
\newblock \url{https://github.com/LarremoreLab/wiserank}.

\bibitem{peng2021neural}
Hao Peng, Qing Ke, Ceren Budak, Daniel~M Romero, and Yong-Yeol Ahn.
\newblock Neural embeddings of scholarly periodicals reveal complex
  disciplinary organizations.
\newblock {\em Science Advances}, 7(17):eabb9004, 2021.

\bibitem{van2008visualizing}
Laurens Van~der Maaten and Geoffrey Hinton.
\newblock Visualizing data using t-sne.
\newblock {\em Journal of machine learning research}, 9(11), 2008.

\bibitem{wattenberg2016how}
Martin Wattenberg, Fernanda Viégas, and Ian Johnson.
\newblock How to use t-sne effectively.
\newblock {\em Distill}, 2016.

\bibitem{de2018physical}
Caterina De~Bacco, Daniel~B Larremore, and Cristopher Moore.
\newblock A physical model for efficient ranking in networks.
\newblock {\em Science advances}, 4(7):eaar8260, 2018.

\end{thebibliography}

\clearpage

\begin{center}
\Large \textbf{Supporting Information for\\
\ttitle}
\end{center}

\FloatBarrier

\renewcommand{\thefigure}{S\arabic{figure}}
\setcounter{figure}{0}

\renewcommand{\thetable}{S\arabic{table}}
\setcounter{table}{0}

\renewcommand{\theequation}{S\arabic{equation}}
\setcounter{equation}{0}

\renewcommand{\thepage}{S\arabic{page}} 
\setcounter{page}{1}

\footnotesize

\tableofcontents

\clearpage

\small
\captionsetup{font=small}

\section{Supplementary Figures}

\FloatBarrier

\begin{figure}[H]
    \centering
    \includegraphics[width=1.0\linewidth]{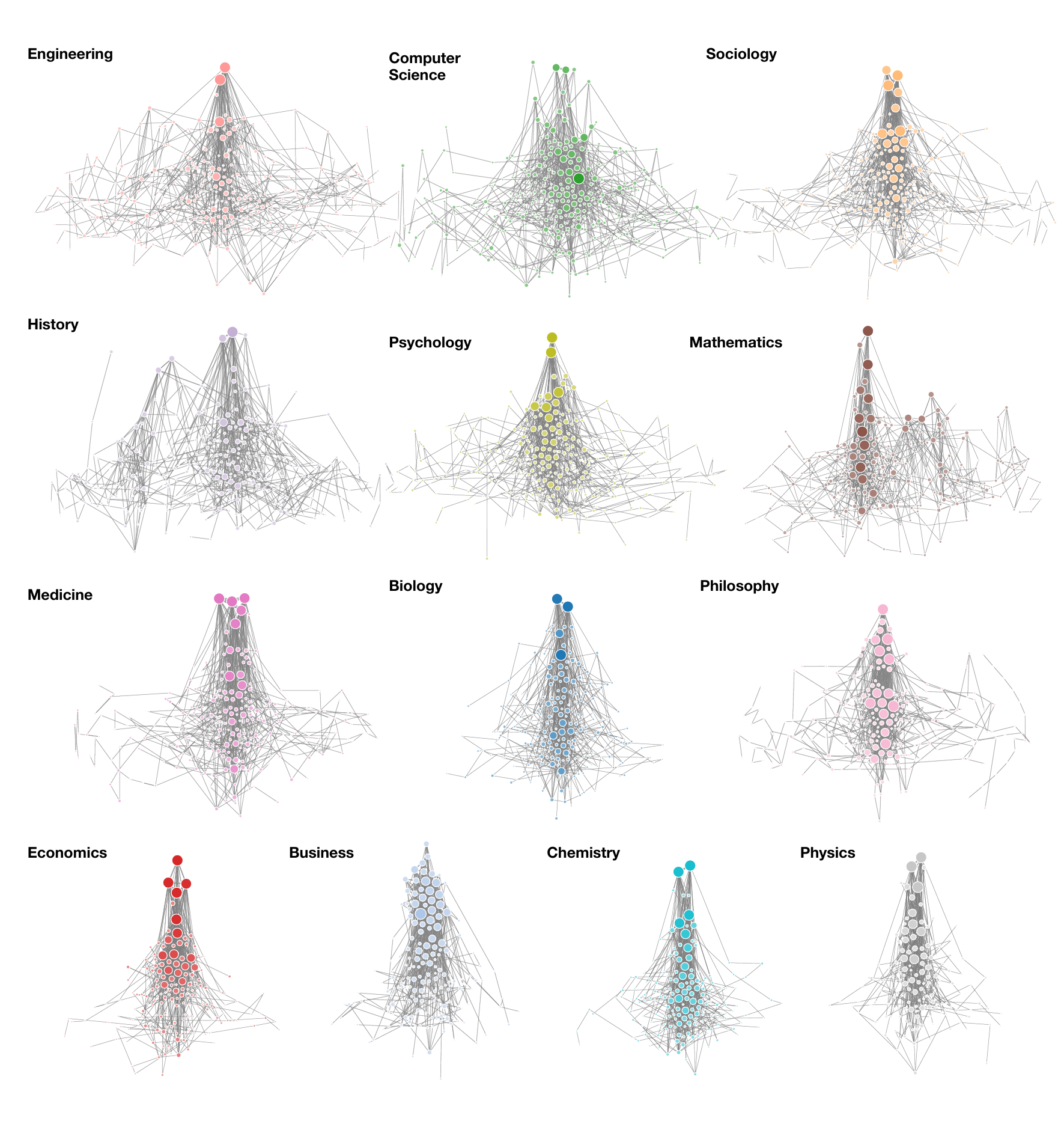}
    \caption{\textbf{Preference network sketches.} Each panel represents the consideration sets and preferences of 50 randomly chosen respondents from each field. Venues (circles) are sized by consideration set popularity and vertically positioned by field-level consensus ranks, while links (lines) connect each respondent's first-ranked venue to their second-ranked, third-ranked, and so on. Fields are arranged from top-left to bottom-right in order of decreasing consideration set diversity. Fig.~3 in the main text shows four selected fields; all 13 fields are shown here.}
    \label{sfig:network-sketches}
    \addcontentsline{toc}{subsubsection}{Figure \ref{sfig:network-sketches}: Preference network sketches}
\end{figure}

\FloatBarrier

\clearpage
\section{Supplementary Tables}

\begin{table}[htbp]
\footnotesize
\centering
\begin{tabular}{lllcc}
\toprule
\textbf{Field} & \textbf{N} & \textbf{Reference group} & \textbf{Slope} & \textbf{Prestige = 10} \\
 & &  & \textbf{(95\% CI)} & \textbf{(95\% CI)} \\
\midrule
\rowcolor{gray!20}
Academia & 2275 & Respondent's Top 5 & 0.022 (0.017, 0.028) & 0.525 (0.481, 0.568) \\
 & & Field's Top 5 & 0.022 (0.017, 0.027) & 0.287 (0.243, 0.331) \\
\addlinespace
\rowcolor{gray!20}
Biology & 341 & Respondent's Top 5 & 0.043 (0.027, 0.060) & 0.716 (0.629, 0.804) \\
 & & Field's Top 5 & 0.038 (0.023, 0.052) & 0.400 (0.323, 0.477) \\
\addlinespace
\rowcolor{gray!20}
Business & 234 & Respondent's Top 5 & 0.006 (-0.015, 0.026) & 0.405 (0.293, 0.516) \\
 & & Field's Top 5 & 0.004 (-0.002, 0.011) & 0.084 (0.047, 0.121) \\
\addlinespace
\rowcolor{gray!20}
Chemistry & 146 & Respondent's Top 5 & 0.031 (0.023, 0.039) & 0.635 (0.593, 0.677) \\
 & & Field's Top 5 & 0.034 (0.020, 0.049) & 0.482 (0.405, 0.559) \\
\addlinespace
\rowcolor{gray!20}
Computer science & 152 & Respondent's Top 5 & 0.030 (0.009, 0.051) & 0.525 (0.411, 0.639) \\
 & & Field's Top 5 & 0.015 (0.009, 0.020) & 0.106 (0.075, 0.137) \\
\addlinespace
\rowcolor{gray!20}
Economics & 153 & Respondent's Top 5 & 0.036 (0.018, 0.055) & 0.438 (0.340, 0.536) \\
 & & Field's Top 5 & 0.030 (0.011, 0.050) & 0.324 (0.219, 0.430) \\
\addlinespace
\rowcolor{gray!20}
Engineering & 191 & Respondent's Top 5 & 0.015 (0.008, 0.022) & 0.523 (0.485, 0.562) \\
 & & Field's Top 5 & 0.009 (0.006, 0.013) & 0.107 (0.088, 0.126) \\
\addlinespace
\rowcolor{gray!20}
History & 105 & Respondent's Top 5 & 0.002 (-0.037, 0.040) & 0.505 (0.322, 0.687) \\
 & & Field's Top 5 & 0.004 (-0.008, 0.016) & 0.125 (0.067, 0.182) \\
\addlinespace
\rowcolor{gray!20}
Mathematics & 126 & Respondent's Top 5 & 0.011 (-0.002, 0.024) & 0.382 (0.312, 0.452) \\
 & & Field's Top 5 & 0.007 (0.000, 0.015) & 0.071 (0.033, 0.110) \\
\addlinespace
\rowcolor{gray!20}
Medicine & 68 & Respondent's Top 5 & 0.028 (0.018, 0.037) & 0.438 (0.386, 0.490) \\
 & & Field's Top 5 & 0.037 (0.019, 0.054) & 0.289 (0.196, 0.381) \\
\addlinespace
\rowcolor{gray!20}
Philosophy & 62 & Respondent's Top 5 & 0.019 (-0.005, 0.043) & 0.506 (0.379, 0.634) \\
 & & Field's Top 5 & 0.023 (0.010, 0.036) & 0.223 (0.156, 0.290) \\
\addlinespace
\rowcolor{gray!20}
Physics & 173 & Respondent's Top 5 & 0.035 (0.023, 0.047) & 0.674 (0.609, 0.740) \\
 & & Field's Top 5 & 0.032 (0.020, 0.044) & 0.477 (0.415, 0.539) \\
\addlinespace
\rowcolor{gray!20}
Psychology & 305 & Respondent's Top 5 & 0.014 (0.002, 0.027) & 0.413 (0.345, 0.482) \\
 & & Field's Top 5 & 0.011 (0.007, 0.015) & 0.120 (0.097, 0.143) \\
\addlinespace
\rowcolor{gray!20}
Sociology & 219 & Respondent's Top 5 & 0.015 (0.003, 0.028) & 0.476 (0.409, 0.544) \\
 & & Field's Top 5 & 0.023 (0.016, 0.030) & 0.255 (0.217, 0.293) \\
\bottomrule
\end{tabular}
\caption{\textbf{Linear regression results predicting the fraction of top-5 venues in publication records as a function of institutional prestige decile.} \normalfont{For each field, we report regression slopes and predicted values at prestige decile 10 (highest prestige), along with 95\% confidence intervals. The ``Respondent's Top 5'' rows show results using each respondent's self-identified top 5 venues in their field, while the ``Field's Top 5'' rows use the field-consensus top 5 venues. Slopes indicate the change in fraction of top-5 publications per one-decile increase in institutional prestige. All fractions range from 0 (0 of 5 publications) to 1 (5 of 5 publications). $N$ indicates the number of respondents in each field or Academia. These results correspond to Fig.~5 in the main text.}}
\label{stab:top5_regression_results}
\addcontentsline{toc}{subsubsection}{Table \ref{stab:top5_regression_results}: Linear regression results predicting the fraction of top-5 venues in publication records as a function of institutional prestige decile}
\end{table}


\begin{table}[htbp]
\centering
\small

\begin{tabular}{@{}lllrrr@{}}
\toprule
Venue & Field & Rank (Pref.) & Rank (JIF) & Diff. \\
\midrule
Nature & Biology & 1 & 1 & 0 \\
Nature & Chemistry & 2 & 2 & 0 \\
Nature & Physics & 1 & 1 & 0 \\
Nature & Medicine & 3 & 5 & $+2$ \\
Nature & Psychology & 2 & 5 & $+3$ \\
Nature & Engineering & 2 & 1 & $-1$ \\
Nature & Sociology & 4 & 5 & $+1$ \\
Nature & Computer science & 1 & 1 & 0 \\
\rowcolor{gray!25}
Nature & Mathematics & 9 & 1 & $-8$ \\
\rowcolor{gray!25}
Nature & Philosophy & 4 & 1 & $-3$ \\
\rowcolor{gray!25}
Nature & History & 3 & 1 & $-2$ \\
\rowcolor{gray!25}
Nature & Business & 2 & 1 & $-1$ \\
\rowcolor{gray!25}
Nature & Economics & 10 & 1 & $-9$ \\
\midrule
Science & Biology & 2 & 3 & $+1$ \\
Science & Chemistry & 1 & 4 & $+3$ \\
Science & Physics & 2 & 3 & $+1$ \\
Science & Medicine & 2 & 7 & $+5$ \\
Science & Psychology & 1 & 7 & $+6$ \\
Science & Engineering & 1 & 3 & $+2$ \\
Science & Sociology & 2 & 6 & $+4$ \\
Science & Computer science & 2 & 2 & 0 \\
\rowcolor{gray!25}
Science & Mathematics & 8 & 2 & $-6$ \\
\rowcolor{gray!25}
Science & Philosophy & 2 & 2 & 0 \\
\rowcolor{gray!25}
Science & History & 4 & 2 & $-2$ \\
\rowcolor{gray!25}
Science & Business & 1 & 2 & $+1$ \\
\rowcolor{gray!25}
Science & Economics & 5 & 2 & $-3$ \\
\midrule
PNAS & Biology & 5 & 10 & $+5$ \\
PNAS & Chemistry & 4 & 12 & $+8$ \\
PNAS & Physics & 9 & 9 & 0 \\
PNAS & Medicine & 8 & 14 & $+6$ \\
PNAS & Psychology & 7 & 13 & $+6$ \\
PNAS & Engineering & 3 & 8 & $+5$ \\
PNAS & Sociology & 5 & 8 & $+3$ \\
PNAS & Computer science & 3 & 5 & $+2$ \\
\rowcolor{gray!25}
PNAS & Mathematics & 14 & 3 & $-11$ \\
\rowcolor{gray!25}
PNAS & Philosophy & 11 & 4 & $-7$ \\
\rowcolor{gray!25}
PNAS & History & 7 & 3 & $-4$ \\
\rowcolor{gray!25}
PNAS & Business & 12 & 10 & $-1$ \\
\rowcolor{gray!25}
PNAS & Economics & 25 & 8 & $-17$ \\
\bottomrule
\end{tabular}

\caption{\textbf{Field-Level Preference Rankings vs. Journal Impact Factor Rankings for Multidisciplinary Venues.} {\normalfont Gray-shaded rows indicate fields where the venue is not commonly chosen ($\leq25\%$ of respondents in that field). Difference = Rank (JIF) $-$ Rank (Pref.); positive values indicate that JIF undervalues the venue relative to the field's preferences, while negative values indicate that JIF overvalues the venue. Only venues chosen by at least 10\% of a field's respondents are included in the underlying rankings.}}
\label{stab:venue_rankings_multidisc}
\addcontentsline{toc}{subsubsection}{Table \ref{stab:venue_rankings_multidisc}: Field-Level Preference Rankings vs. Journal Impact Factor Rankings for Multidisciplinary Venues}
\end{table}


\begin{table}[htbp]
\centering
\small

\begin{tabular}{@{}lllrrr@{}}
\toprule
Venue & Field & Rank (Pref.) & Rank (JIF) & Diff. \\
\midrule
The American Economic Review & Economics & 1 & 5 & $+4$ \\
\rowcolor{gray!25}
The American Economic Review & Business & 5 & 7 & $+2$ \\
\midrule
The Accounting Review & Business & 3 & 33 & $+30$ \\
\midrule
American Sociological Review & Sociology & 1 & 12 & $+11$ \\
\rowcolor{gray!25}
American Sociological Review & Business & 28 & 16 & $-12$ \\
\midrule
Journal of Personality and Social Psychology & Psychology & 13 & 14 & $+1$ \\
\rowcolor{gray!25}
Journal of Personality and Social Psychology & Sociology & 25 & 13 & $-12$ \\
\rowcolor{gray!25}
Journal of Personality and Social Psychology & Business & 30 & 21 & $-9$ \\
\midrule
Journal of the American Chemical Society & Chemistry & 3 & 11 & $+8$ \\
\rowcolor{gray!25}
Journal of the American Chemical Society & Physics & 18 & 8 & $-10$ \\
\rowcolor{gray!25}
Journal of the American Chemical Society & Engineering & 8 & 5 & $-3$ \\
\midrule
Physical Review Letters & Physics & 4 & 12 & $+8$ \\
\rowcolor{gray!25}
Physical Review Letters & Chemistry & 9 & 14 & $+5$ \\
\rowcolor{gray!25}
Physical Review Letters & Engineering & 10 & 10 & 0 \\
\rowcolor{gray!25}
Physical Review Letters & Mathematics & 23 & 4 & $-19$ \\
\midrule
Annals of Mathematics & Mathematics & 1 & 6 & $+5$ \\
\midrule
The American Historical Review & History & 1 & 6 & $+5$ \\
\midrule
The Philosophical Review & Philosophy & 1 & 8 & $+7$ \\
\bottomrule
\end{tabular}

\caption{\textbf{Field-Level Preference Rankings vs. Journal Impact Factor Rankings for Field Flagship Venues.} {\normalfont White rows indicate fields where the venue is a flagship, while gray-shaded rows indicate fields where the venue is in at least 10\% of respondents' consideration sets, but not a flagship. Diff. = Rank (JIF) $-$ Rank (Pref.); positive values indicate that JIF undervalues the venue relative to the field's preferences, while negative values indicate that JIF overvalues the venue. Only venues chosen by at least 10\% of a field's respondents are included in both the table and the underlying rankings.
}}
\label{stab:venue_rankings_flagships}
\addcontentsline{toc}{subsubsection}{Table \ref{stab:venue_rankings_flagships}: Field-Level Preference Rankings vs. Journal Impact Factor Rankings for Field Flagship Venues}
\end{table}

\FloatBarrier

\section{Survey Data}
\label{sec:SI-survey}

\subsection{Sample of Academics}
\label{sample_frame}

Our sample of faculty comes from the Academic Analytics Research Center (AARC), specifically their census of tenured or tenure-track assistant, associate, and full professors at United States PhD-granting institutions circa 2020~\citeS{wapman2022quantifying}. While there are almost $250,000$ faculty in this dataset our sample frame consists of \NSurveyed~faculty for whom we have email addresses~\citeS{morgan2022socioeconomic,spoon2023gender}. \NSurveyedStarted~individuals from this sample frame started the survey (\PercSurveyedStart\%) and \NSurveyedCompleted~completed the main portion (\PercSurveyedCompleted\%). In addition, of the \NStudied~participants studied \NStudiedNotSurveyed~were not on the AARC list (\PercStudiedNotSurveyed\%), but invited by participants who had completed the survey. 
Survey completion rate by field is shown in Table~\ref{tab:si_field_rate}.

Table~\ref{tab:field_breakdown} compares participant demographics with those reported by AARC, including gender, academic rank, and institutional prestige percentile. Even though the sample frame for the survey was based on the 2020 version of AARC data, for this analysis of representativeness we use the 2023 AARC data. The AARC sample is skewed toward senior academics. Because our dataset predates recent hires and recent promotions to Associate Professor, the resulting sample underrepresents Assistant Professors. This relative seniority likely contributes to a higher proportion of men in our sample.

In Table~\ref{tab:field_breakdown}, prestige percentiles are reported as means, with standard deviations in parentheses. They reflect the average institutional prestige (on a percentile scale where 0 indicates highest prestige and 100 the lowest) of all institutions employing participants within a given field. See Materials and Methods section for the construction of prestige percentiles.


\begin{table}[h]
    \centering
    \begin{tabular}{lccc}
        \toprule
        Assigned Field & Completed & Invited & \% Completion Rate \\
        \midrule
        Business & \BusinessResponseCount & \BusinessEmailedCount & \BusinessResponseRate \\ 
        Psychology & \PsychologyResponseCount & \PsychologyEmailedCount & \PsychologyResponseRate \\ 
        Biology & \BiologyResponseCount & \BiologyEmailedCount & \BiologyResponseRate \\ 
        Physics and Astronomy & \PhysicsAstronomyResponseCount & \PhysicsAstronomyEmailedCount & \PhysicsAstronomyResponseRate \\ 
        Sociology & \SociologyResponseCount & \SociologyEmailedCount & \SociologyResponseRate \\ 
        Mathematics & \MathematicsResponseCount & \MathematicsEmailedCount & \MathematicsResponseRate \\
        Computer science & \ComputerScienceResponseCount & \ComputerScienceEmailedCount & \ComputerScienceResponseRate \\ 
        History & \HistoryResponseCount & \HistoryEmailedCount & \HistoryResponseRate \\ 
        Chemistry & \ChemistryResponseCount & \ChemistryEmailedCount & \ChemistryResponseRate \\ 
        Economics & \EconomicsResponseCount & \EconomicsEmailedCount & \EconomicsResponseRate \\ 
        Biochemistry & \BiochemistryResponseCount & \BiochemistryEmailedCount & \BiochemistryResponseRate \\ 
        Anthropology & \AnthropologyResponseCount & \AnthropologyEmailedCount & \AnthropologyResponseRate \\ 
        Mechanical Engineering & \MechanicalEngineeringResponseCount & \MechanicalEngineeringEmailedCount & \MechanicalEngineeringResponseRate \\ 
        Civil Engineering & \CivilEngineeringResponseCount & \CivilEngineeringEmailedCount & \CivilEngineeringResponseRate \\ 
        Philosophy & \PhilosophyResponseCount & \PhilosophyEmailedCount & \PhilosophyResponseRate \\ 
        \bottomrule
        \end{tabular}
    \caption{\textbf{Survey completion statistics by field.} \normalfont{The fields used here reflect an AARC taxonomy and thus differ slightly from those selected by participants during the survey which reflect an OpenAlex taxonomy.}}
    \label{tab:si_field_rate}
    \addcontentsline{toc}{subsubsection}{Table \ref{tab:si_field_rate}: Survey completion statistics by field}
\end{table}

\begin{table}[h!]
    \tiny
    \centering
    \begin{minipage}{0.45\textwidth}
    \begin{tabular}{lcc}
    \toprule
    \textbf{Academia} & \% AARC ($N=\AcademiaCountAARC$) & \% Sample ($N=\AcademiaCountPorC$) \\
    \midrule
         Men & \AcademiaMenPercentAARC & \AcademiaMenPercentPorC \\
         Women & \AcademiaWomenPercentAARC & \AcademiaWomenPercentPorC \\
         Unknown & \AcademiaUnknownPercentAARC & \AcademiaUnknownPercentPorC \\
    \midrule
         Assistant Professor & \AcademiaAssistantProfessorPercentAARC & \AcademiaAssistantProfessorPercentPorC \\
         Associate Professor & \AcademiaAssociateProfessorPercentAARC & \AcademiaAssociateProfessorPercentPorC \\
         Full Professor & \AcademiaFullProfessorPercentAARC & \AcademiaFullProfessorPercentPorC \\
    \midrule
        Prestige Percentile & \AcademiaMeanPrestigeAARC~(\AcademiaSdPrestigeAARC) & \AcademiaMeanPrestigePorC~(\AcademiaSdPrestigePorC) \\
    \midrule
    \midrule

    \textbf{Biology} & \% AARC ($N=\BiologyCountAARC$) & \% Sample ($N=\BiologyCountPorC$) \\
    \midrule
         Men & \BiologyMenPercentAARC & \BiologyMenPercentPorC \\
         Women & \BiologyWomenPercentAARC & \BiologyWomenPercentPorC \\
         Unknown & \BiologyUnknownPercentAARC & \BiologyUnknownPercentPorC \\
    \midrule
         Assistant Professor & \BiologyAssistantProfessorPercentAARC & \BiologyAssistantProfessorPercentPorC \\
         Associate Professor & \BiologyAssociateProfessorPercentAARC & \BiologyAssociateProfessorPercentPorC \\
         Full Professor & \BiologyFullProfessorPercentAARC & \BiologyFullProfessorPercentPorC \\
    \midrule
        Prestige Percentile & \BiologyMeanPrestigeAARC~(\BiologySdPrestigeAARC) & \BiologyMeanPrestigePorC~(\BiologySdPrestigePorC) \\
    \midrule
    \midrule

    \textbf{Business} & \% AARC ($N=\BusinessCountAARC$) & \% Sample ($N=\BusinessCountPorC$) \\
    \midrule
         Men & \BusinessMenPercentAARC & \BusinessMenPercentPorC \\
         Women & \BusinessWomenPercentAARC & \BusinessWomenPercentPorC \\
         Unknown & \BusinessUnknownPercentAARC & \BusinessUnknownPercentPorC \\
    \midrule
         Assistant Professor & \BusinessAssistantProfessorPercentAARC & \BusinessAssistantProfessorPercentPorC \\
         Associate Professor & \BusinessAssociateProfessorPercentAARC & \BusinessAssociateProfessorPercentPorC \\
         Full Professor & \BusinessFullProfessorPercentAARC & \BusinessFullProfessorPercentPorC \\
    \midrule
        Prestige Percentile & \BusinessMeanPrestigeAARC~(\BusinessSdPrestigeAARC) & \BusinessMeanPrestigePorC~(\BusinessSdPrestigePorC) \\
    \midrule
    \midrule

    \textbf{Chemistry} & \% AARC ($N=\ChemistryCountAARC$) & \% Sample ($N=\ChemistryCountPorC$) \\
    \midrule
         Men & \ChemistryMenPercentAARC & \ChemistryMenPercentPorC \\
         Women & \ChemistryWomenPercentAARC & \ChemistryWomenPercentPorC \\
         Unknown & \ChemistryUnknownPercentAARC & \ChemistryUnknownPercentPorC \\
    \midrule
         Assistant Professor & \ChemistryAssistantProfessorPercentAARC & \ChemistryAssistantProfessorPercentPorC \\
         Associate Professor & \ChemistryAssociateProfessorPercentAARC & \ChemistryAssociateProfessorPercentPorC \\
         Full Professor & \ChemistryFullProfessorPercentAARC & \ChemistryFullProfessorPercentPorC \\
    \midrule
        Prestige Percentile & \ChemistryMeanPrestigeAARC~(\ChemistrySdPrestigeAARC) & \ChemistryMeanPrestigePorC~(\ChemistrySdPrestigePorC) \\
    \midrule
    \midrule

    \textbf{Computer Science} & \% AARC ($N=\ComputerScienceCountAARC$) & \% Sample ($N=\ComputerScienceCountPorC$) \\
    \midrule
         Men & \ComputerScienceMenPercentAARC & \ComputerScienceMenPercentPorC \\
         Women & \ComputerScienceWomenPercentAARC & \ComputerScienceWomenPercentPorC \\
         Unknown & \ComputerScienceUnknownPercentAARC & \ComputerScienceUnknownPercentPorC \\
    \midrule
         Assistant Professor & \ComputerScienceAssistantProfessorPercentAARC & \ComputerScienceAssistantProfessorPercentPorC \\
         Associate Professor & \ComputerScienceAssociateProfessorPercentAARC & \ComputerScienceAssociateProfessorPercentPorC \\
         Full Professor & \ComputerScienceFullProfessorPercentAARC & \ComputerScienceFullProfessorPercentPorC \\
    \midrule
        Prestige Percentile & \ComputerScienceMeanPrestigeAARC~(\ComputerScienceSdPrestigeAARC) & \ComputerScienceMeanPrestigePorC~(\ComputerScienceSdPrestigePorC) \\
    \midrule
    \midrule

    \textbf{Economics} & \% AARC ($N=\EconomicsCountAARC$) & \% Sample ($N=\EconomicsCountPorC$) \\
    \midrule
         Men & \EconomicsMenPercentAARC & \EconomicsMenPercentPorC \\
         Women & \EconomicsWomenPercentAARC & \EconomicsWomenPercentPorC \\
         Unknown & \EconomicsUnknownPercentAARC & \EconomicsUnknownPercentPorC \\
    \midrule
         Assistant Professor & \EconomicsAssistantProfessorPercentAARC & \EconomicsAssistantProfessorPercentPorC \\
         Associate Professor & \EconomicsAssociateProfessorPercentAARC & \EconomicsAssociateProfessorPercentPorC \\
         Full Professor & \EconomicsFullProfessorPercentAARC & \EconomicsFullProfessorPercentPorC \\
    \midrule
        Prestige Percentile & \EconomicsMeanPrestigeAARC~(\EconomicsSdPrestigeAARC) & \EconomicsMeanPrestigePorC~(\EconomicsSdPrestigePorC) \\
    \midrule
    \midrule
    
    \textbf{Engineering} & \% AARC ($N=\EngineeringCountAARC$) & \% Sample ($N=\EngineeringCountPorC$) \\
    \midrule
         Men & \EngineeringMenPercentAARC & \EngineeringMenPercentPorC \\
         Women & \EngineeringWomenPercentAARC & \EngineeringWomenPercentPorC \\
         Unknown & \EngineeringUnknownPercentAARC & \EngineeringUnknownPercentPorC \\
    \midrule
         Assistant Professor & \EngineeringAssistantProfessorPercentAARC & \EngineeringAssistantProfessorPercentPorC \\
         Associate Professor & \EngineeringAssociateProfessorPercentAARC & \EngineeringAssociateProfessorPercentPorC \\
         Full Professor & \EngineeringFullProfessorPercentAARC & \EngineeringFullProfessorPercentPorC \\
    \midrule
        Prestige Percentile & \EngineeringMeanPrestigeAARC~(\EngineeringSdPrestigeAARC) & \EngineeringMeanPrestigePorC~(\EngineeringSdPrestigePorC) \\

    \bottomrule
    \end{tabular}
    \end{minipage}%
    \hfill
    \begin{minipage}{0.45\textwidth}
    \begin{tabular}{lcc}
    \toprule
    \textbf{History} & \% AARC ($N=\HistoryCountAARC$) & \% Sample ($N=\HistoryCountPorC$) \\
    \midrule
         Men & \HistoryMenPercentAARC & \HistoryMenPercentPorC \\
         Women & \HistoryWomenPercentAARC & \HistoryWomenPercentPorC \\
         Unknown & \HistoryUnknownPercentAARC & \HistoryUnknownPercentPorC \\
    \midrule
         Assistant Professor & \HistoryAssistantProfessorPercentAARC & \HistoryAssistantProfessorPercentPorC \\
         Associate Professor & \HistoryAssociateProfessorPercentAARC & \HistoryAssociateProfessorPercentPorC \\
         Full Professor & \HistoryFullProfessorPercentAARC & \HistoryFullProfessorPercentPorC \\
    \midrule
        Prestige Percentile & \HistoryMeanPrestigeAARC~(\HistorySdPrestigeAARC) & \HistoryMeanPrestigePorC~(\HistorySdPrestigePorC) \\
    \midrule
    \midrule

    \textbf{Mathematics} & \% AARC ($N=\MathematicsCountAARC$) & \% Sample ($N=\MathematicsCountPorC$) \\
    \midrule
         Men & \MathematicsMenPercentAARC & \MathematicsMenPercentPorC \\
         Women & \MathematicsWomenPercentAARC & \MathematicsWomenPercentPorC \\
         Unknown & \MathematicsUnknownPercentAARC & \MathematicsUnknownPercentPorC \\
    \midrule
         Assistant Professor & \MathematicsAssistantProfessorPercentAARC & \MathematicsAssistantProfessorPercentPorC \\
         Associate Professor & \MathematicsAssociateProfessorPercentAARC & \MathematicsAssociateProfessorPercentPorC \\
         Full Professor & \MathematicsFullProfessorPercentAARC & \MathematicsFullProfessorPercentPorC \\
    \midrule
        Prestige Percentile & \MathematicsMeanPrestigeAARC~(\MathematicsSdPrestigeAARC) & \MathematicsMeanPrestigePorC~(\MathematicsSdPrestigePorC) \\
    \midrule
    \midrule

    \textbf{Medicine} & \% AARC ($N=\MedicineCountAARC$) & \% Sample ($N=\MedicineCountPorC$) \\
    \midrule
         Men & \MedicineMenPercentAARC & \MedicineMenPercentPorC \\
         Women & \MedicineWomenPercentAARC & \MedicineWomenPercentPorC \\
         Unknown & \MedicineUnknownPercentAARC & \MedicineUnknownPercentPorC \\
    \midrule
         Assistant Professor & \MedicineAssistantProfessorPercentAARC & \MedicineAssistantProfessorPercentPorC \\
         Associate Professor & \MedicineAssociateProfessorPercentAARC & \MedicineAssociateProfessorPercentPorC \\
         Full Professor & \MedicineFullProfessorPercentAARC & \MedicineFullProfessorPercentPorC \\
    \midrule
        Prestige Percentile & \MedicineMeanPrestigeAARC~(\MedicineSdPrestigeAARC) & \MedicineMeanPrestigePorC~(\MedicineSdPrestigePorC) \\
    \midrule
    \midrule

    \textbf{Philosophy} & \% AARC ($N=\PhilosophyCountAARC$) & \% Sample ($N=\PhilosophyCountPorC$) \\
    \midrule
         Men & \PhilosophyMenPercentAARC & \PhilosophyMenPercentPorC \\
         Women & \PhilosophyWomenPercentAARC & \PhilosophyWomenPercentPorC \\
         Unknown & \PhilosophyUnknownPercentAARC & \PhilosophyUnknownPercentPorC \\
    \midrule
         Assistant Professor & \PhilosophyAssistantProfessorPercentAARC & \PhilosophyAssistantProfessorPercentPorC \\
         Associate Professor & \PhilosophyAssociateProfessorPercentAARC & \PhilosophyAssociateProfessorPercentPorC \\
         Full Professor & \PhilosophyFullProfessorPercentAARC & \PhilosophyFullProfessorPercentPorC \\
    \midrule
        Prestige Percentile & \PhilosophyMeanPrestigeAARC~(\PhilosophySdPrestigeAARC) & \PhilosophyMeanPrestigePorC~(\PhilosophySdPrestigePorC) \\
    \midrule
    \midrule

    \textbf{Physics} & \% AARC ($N=\PhysicsCountAARC$) & \% Sample ($N=\PhysicsCountPorC$) \\
    \midrule
         Men & \PhysicsMenPercentAARC & \PhysicsMenPercentPorC \\
         Women & \PhysicsWomenPercentAARC & \PhysicsWomenPercentPorC \\
         Unknown & \PhysicsUnknownPercentAARC & \PhysicsUnknownPercentPorC \\
    \midrule
         Assistant Professor & \PhysicsAssistantProfessorPercentAARC & \PhysicsAssistantProfessorPercentPorC \\
         Associate Professor & \PhysicsAssociateProfessorPercentAARC & \PhysicsAssociateProfessorPercentPorC \\
         Full Professor & \PhysicsFullProfessorPercentAARC & \PhysicsFullProfessorPercentPorC \\
    \midrule
        Prestige Percentile & \PhysicsMeanPrestigeAARC~(\PhysicsSdPrestigeAARC) & \PhysicsMeanPrestigePorC~(\PhysicsSdPrestigePorC) \\
    \midrule
    \midrule

    \textbf{Psychology} & \% AARC ($N=\PsychologyCountAARC$) & \% Sample ($N=\PsychologyCountPorC$) \\
    \midrule
         Men & \PsychologyMenPercentAARC & \PsychologyMenPercentPorC \\
         Women & \PsychologyWomenPercentAARC & \PsychologyWomenPercentPorC \\
         Unknown & \PsychologyUnknownPercentAARC & \PsychologyUnknownPercentPorC \\
    \midrule
         Assistant Professor & \PsychologyAssistantProfessorPercentAARC & \PsychologyAssistantProfessorPercentPorC \\
         Associate Professor & \PsychologyAssociateProfessorPercentAARC & \PsychologyAssociateProfessorPercentPorC \\
         Full Professor & \PsychologyFullProfessorPercentAARC & \PsychologyFullProfessorPercentPorC \\
    \midrule
        Prestige Percentile & \PsychologyMeanPrestigeAARC~(\PsychologySdPrestigeAARC) & \PsychologyMeanPrestigePorC~(\PsychologySdPrestigePorC) \\
    \midrule
    \midrule

    \textbf{Sociology} & \% AARC ($N=\SociologyCountAARC$) & \% Sample ($N=\SociologyCountPorC$) \\
    \midrule
         Men & \SociologyMenPercentAARC & \SociologyMenPercentPorC \\
         Women & \SociologyWomenPercentAARC & \SociologyWomenPercentPorC \\
         Unknown & \SociologyUnknownPercentAARC & \SociologyUnknownPercentPorC \\
    \midrule
         Assistant Professor & \SociologyAssistantProfessorPercentAARC & \SociologyAssistantProfessorPercentPorC \\
         Associate Professor & \SociologyAssociateProfessorPercentAARC & \SociologyAssociateProfessorPercentPorC \\
         Full Professor & \SociologyFullProfessorPercentAARC & \SociologyFullProfessorPercentPorC \\
    \midrule
        Prestige Percentile & \SociologyMeanPrestigeAARC~(\SociologySdPrestigeAARC) & \SociologyMeanPrestigePorC~(\SociologySdPrestigePorC) \\
    \bottomrule
    \end{tabular}
    \end{minipage}
    \caption{AARC vs PorC demographics: percentages for gender and rank, and means (SD in parentheses) for institutional prestige.}
    \label{tab:field_breakdown}
    \addcontentsline{toc}{subsubsection}{Table \ref{tab:field_breakdown}: AARC vs PorC demographics}
\end{table}

\subsection{Dataset of Publication Venues}

We collected the majority of our data on publication venues from OpenAlex, ``A fully-open index of scholarly works, authors, venues, institutions, and concepts'' that is an expansion of the no longer maintained Microsoft Academic Graph \citeS{priem2022openalex}. Downloading the OpenAlex venues snapshot at the end of 2022 yielded approximately $225,000$ venues. 
We then worked to select a subset of these venues deemed suitable for inclusion in our project. 
Removing venues with fewer than $200$ cataloged works (publications), those that had no OpenAlex concept (field/topic) attached to them, those that were not labeled with an OpenAlex type of either ``journal'' or ``conference'', and a small number of venues deemed irrelevant by inspection (e.g. {\it Scientific American}) produced a list of approximately $100,000$ venues. 
Next, many venues were examined manually and either renamed or dropped based on being, for example, an old version of a current or discontinued venue. Finding more duplicated or irrelevant venues, removing venues that lacked an ISSN, and requiring that venues have at least $500$ works reduced the list to approximately $45,000$ venues.

Finally, approximately $20,000$ venues were added (back) to the above list based on two procedures.
First, we analyzed the OpenAlex publication histories of the academics in our sample frame (described above). We found publication histories for approximately $40,000$ academics and used these to retrieve venues we had removed from our list. Second, we used DBLP.org and csrankings.org to find conferences and journals in the field of computer science, where OpenAlex appeared to be limited in its coverage. This required a substantial amount of manual data collection and venue renaming to integrate these other data sources with OpenAlex. These processing steps resulted in a final, curated list of approximately $65,000$ venues.

\subsection{Relevant Survey Questions}
For a full description of the survey, see SI Section~\ref{sec:SI-porc-platform}. For the description of survey administration, see SI Section~\ref{sec:SI-conducting-the-survey}.

\begin{itemize}

\item \textit{Current academic rank:} Dropdown menu with the following options: Undergraduate, Masters Student, PhD Student, Postdoctoral Researcher, Assistant Professor, Associate Professor, Full Professor, Other. 
\item \textit{Field (best approximation):}  Nineteen options are given for ``Field'' and participants must select one of these. (The final set of fields was selected by constraining the sample only to fields with over 100 participants.)

\item \textit{Aspirations:} Participants specify three venues by searching for the venue name and selecting it from our database. Participants do not have the ability to add venues not in our database.
    \begin{itemize}
        \item \textit{In your field, what's a top venue you'd like to publish in?} 
        \item \textit{If you weren't aiming for a top venue, where might you publish?}
        \item \textit{If you didn't aim as high as either of these venues, where might you publish?}
    \end{itemize}

\item \textit{Would you personally like to, hope to, or want to publish in...} Participants indicate whether or not they would like to publish in each of 20 venues the survey recommendation algorithm presents to them. The participants can press the following buttons: ``NO,'' ``YES,'' and ``UNDO.'' At the end of this stage, participants can directly add venues they would like to publish in.

\item \textit{Where would you prefer to publish?} Participants make pairwise comparisons between venues they indicated they would like to publish in. The participants have the option to select one venue over the other, press ``INDIFFERENT,'' or ``UNDO.''

\item \textit{What is your gender? Please select one.} ``Woman,'' ``Man,'' ``Non-binary,'' ``Prefer to self describe'' (free text response), ``Prefer not to say.''

\item \textit{Current institution or employer (if not listed, select ``Other'')} Participants select an institution from our database. The list of institutions is combined from AARC data~\citeS{wapman2022quantifying} and College Scorecard data~\citeS{collegescorecard2023}.

\end{itemize}

\clearpage

\section{Linear Regression: Prestige, Gender, and Career Stage}

To analyze how aspirations and preferences vary with researchers' characteristics, we estimate the following linear regression model:
\begin{equation}
    Y_i = \beta_0 + \beta_1 \text{AssociateProf}_i + \beta_2 \text{FullProf}_i + \beta_3 \text{Prestige}_i + \beta_4 \text{Gender}_i + \varepsilon_i    
\end{equation}
where $Y_i$ denotes the outcome for individual $i$, which refers to either the mean field-level rank of the individual's top aspiration (shown in blue in Fig.~\ref{fig:researcher_characteristics}) or their top preference (shown in yellow in Fig.~\ref{fig:researcher_characteristics}); $\text{AssociateProf}_i$ and $\text{FullProf}_i$ are dummy variables indicating an individual's academic title, with assistant professors as the reference category; $\text{Prestige}_i$ denotes the prestige decile associated with an individual's current institution (1 = highest prestige, 10 = lowest prestige); $\text{Gender}_i$ indicates the gender of the respondent (0 = man, 1 = woman); and $\varepsilon_i$ is the error term.

This analysis uses a smaller, more restricted sample due to three limitations. First, some respondents could not be linked to an AARC profile, preventing us from determining their institution's prestige rank. Second, we could not assign gender labels to all respondents. When gender labels were available, we included only respondents labeled as woman or man, excluding those identifying with other genders due to their limited representation in our sample. Third, a database error restricted the venue options available for the top-tier aspiration to a subset of respondents. Because the regression analyses use aspiration data, we only include respondents who were not affected by this error. Overall, we therefore only consider \ParticipantsWithAspirations~respondents here, while all other analyses except for those involving publication history rely on responses from the full survey sample of \ParticipantsInRelevantFields~academics.

\begin{figure}[h]
    \centering
    \includegraphics[width=\linewidth]{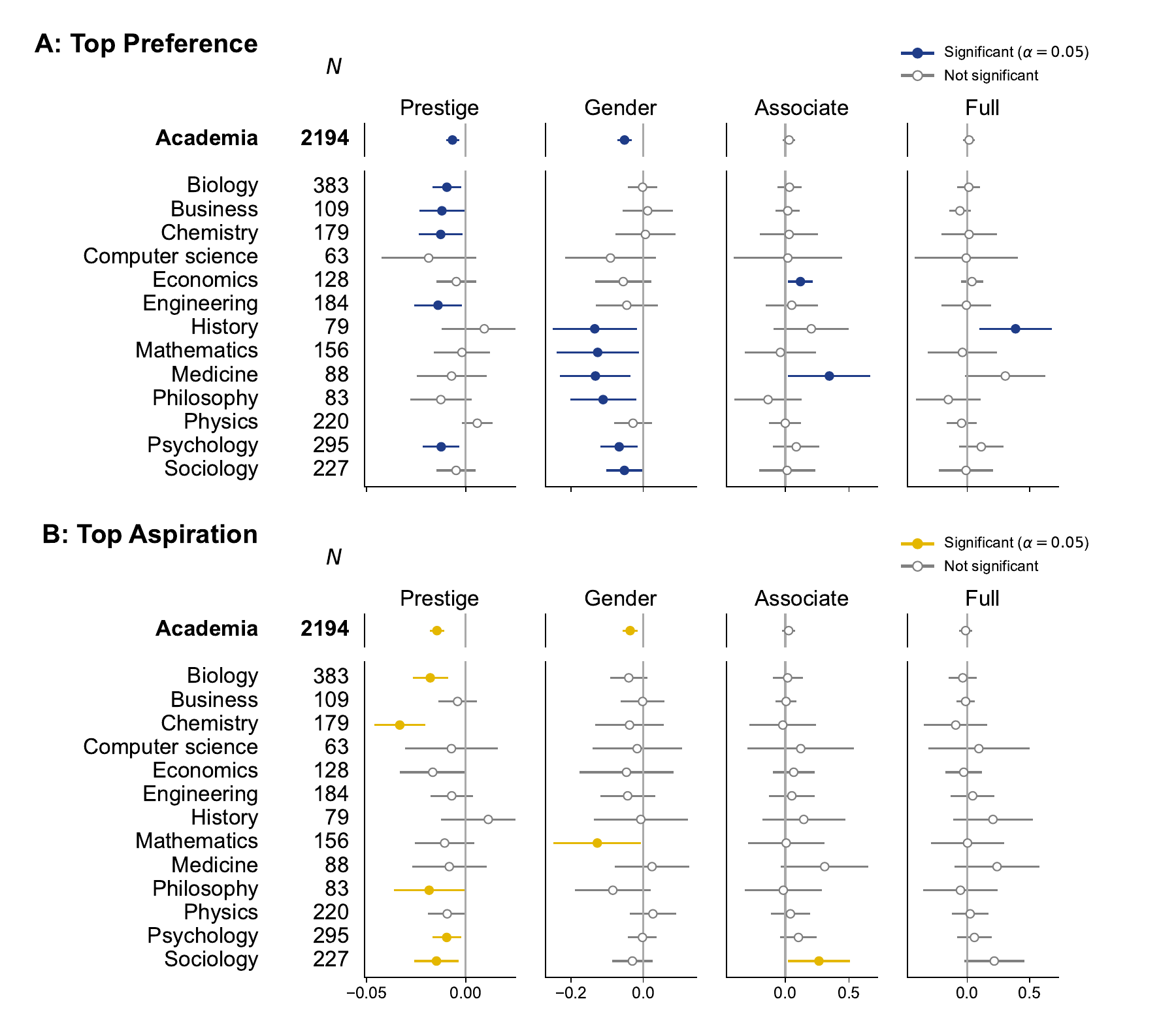}
    \caption{\textbf{Full OLS estimates for prestige, gender, and career stage.} (A) Estimates for top preference. (B) Estimates for top aspiration. Each row shows coefficients from the linear regression model in Eq.~S1, estimated for all respondents (``Academia'') and for each field individually. Prestige coefficients indicate the change in normalized venue rank per one-decile increase in institutional prestige. Gender coefficients indicate the difference for women relative to men. Associate and Full Professor coefficients are relative to Assistant Professors. Colored markers indicate statistical significance at $\alpha = 0.05$; gray markers indicate non-significance. Error bars are 95\% confidence intervals. N indicates the number of respondents in each regression. The main text (Fig.~4C--F) presents the prestige and gender columns; the career stage columns are shown here for completeness.}
    \label{sfig:full_prestige_gender_regression}
    \addcontentsline{toc}{subsubsection}{Figure \ref{sfig:full_prestige_gender_regression}: Full OLS estimates for prestige, gender, and career stage}
\end{figure}

The main text focuses on prestige and gender; Fig.~\ref{sfig:full_prestige_gender_regression} presents the full regression results, including estimates for associate and full professors relative to assistant professors. We do not find a significant effect of career stage on researchers' top preferences or aspirations at the aggregate level. While in a few individual fields associate or full professors are more likely to select higher-ranked venues as their top preference or aspiration, the career stage estimates are not significant across the majority of fields.

\clearpage

\section{Computing Continuous Ranks and Ordinal Ranks} 
\label{si:adjacency}

SpringRank infers a latent score for each venue from a directed weighted network in which
edge weight $A_{ij}$ records the number of times venue $i$ was preferred over venue $j$.
We construct this matrix from pairwise comparison responses as follows.
For each comparison in which a respondent expressed a strict preference, we increment
$A_{ij}$ by $1$, where $i$ is the preferred venue and $j$ is the other.
When a respondent indicated indifference between two venues, we increment both $A_{ij}$
and $A_{ji}$ by $0.5$, treating an indifferent response as a half-win for each venue.
Indifferent responses constituted $8.1\%$ of all pairwise comparisons.

We compute rankings at three levels of aggregation.
\begin{itemize}
    \item \textit{Individual rankings} ($\alpha = 0$) are fit independently for each respondent
using only that respondent's own comparisons.
    \item \textit{Field-level rankings} ($\alpha = 20$) are computed leave-one-out: for each
respondent, we fit SpringRank on all comparisons from respondents in the same field
\emph{excluding} that respondent's own, then record the resulting scores for the
venues that respondent compared.
This design ensures that field-level scores used to evaluate a respondent's preferences
are not contaminated by their own responses.
A non-leave-one-out field consensus ranking (also $\alpha = 20$, fit on all field
comparisons jointly) is used to position nodes in the network visualizations
(Figs.~\ref{fig:networks} and \ref{sfig:network-sketches}).
    \item Finally, \textit{global rankings} ($\alpha = 20$) apply the same leave-one-out
procedure but pool comparisons across all 13 fields, yielding an academia-wide
consensus score used as a cross-field baseline in Fig.~\ref{fig:jif_comparison_accuracy}.
\end{itemize}

For the comparison between the JIF-based venue ranks and the PorC-based venue ranks (Fig.~\ref{fig:jif_comparison_accuracy}), we consider ordinal ranks. Because the JIF and PorC scores use different scales, focusing on differences in ordinal rank (rather than absolute values) provides a more meaningful comparison between the two rankings. For the visual comparison of selected interdisciplinary and flagship venues (Fig. \ref{fig:jif-deviations-field}B), we restrict the analysis to venues selected by at least 10\% of respondents within a field and for which a JIF score is available. We select \textit{Nature,} \textit{Science,} and \textit{PNAS} as interdisciplinary venues because they are widely recognized across all fields (Fig.~\ref{fig:overlap}). To determine field-specific flagship venues, we consider the top aspiration that most respondents in a given field selected, excluding the interdisciplinary venues \textit{Nature,} \textit{Science,} and \textit{PNAS.}

\clearpage

\section{Field Composition Alone Cannot Account for the Observed Relationships Between Gender/Prestige and Top Preferences/Aspirations}

\begin{figure}[h]
    \centering
    \includegraphics[width=.7\linewidth]{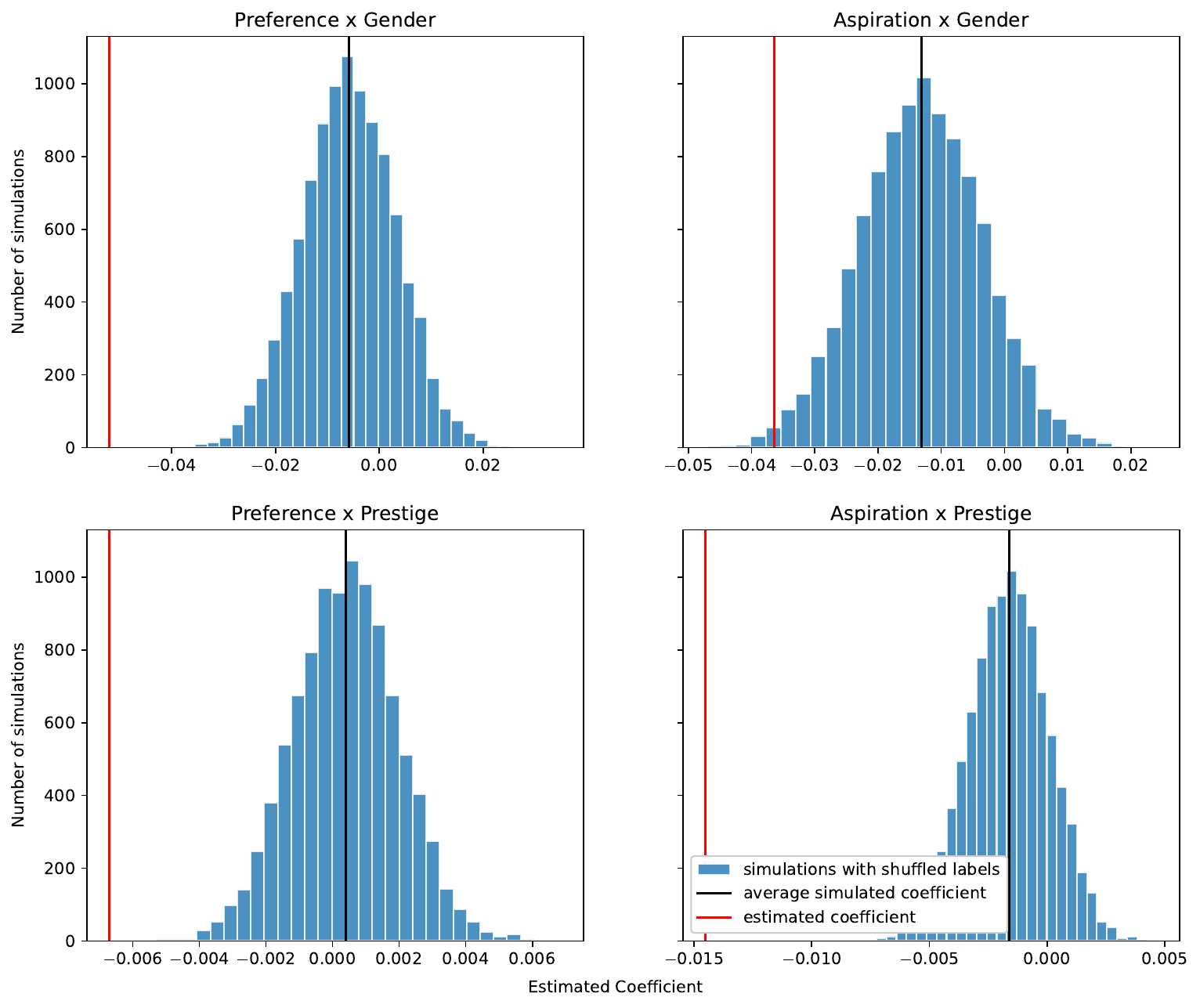}
    \caption{\textbf{Simulation analysis suggests that field composition alone cannot account for the observed relationships between gender/prestige and top preferences/aspirations.} Histograms show the distribution of regression coefficients from 10,000 simulated datasets in which gender or prestige values were permuted within fields, preserving each field's aggregate composition but removing within-field correlations. The dashed line indicates the mean simulated coefficient; the solid red line indicates the observed coefficient from the actual data (Eq.~S1). In all four panels, the observed coefficient falls outside the simulated distribution, confirming that the relationships reported in Fig.~4 of the main text are not attributable to differences in gender or prestige composition across fields.} 
    \label{fig:regression_robustness}
    \addcontentsline{toc}{subsubsection}{Figure \ref{fig:regression_robustness}: Simulation analysis suggests that field composition alone cannot account for the observed relationships between gender/prestige and top preferences/aspirations}
\end{figure}

To test if the observed relationships between gender/prestige and top preferences/aspirations could be attributed to mere differences in how gender/prestige vary with field we conducted 10,000 regressions on simulated datasets where either gender or prestige values were permuted within fields. This preserved the aggregate gender/prestige composition of each field but removes any within field correlations. The results indicate that the observed relationships between gender/prestige and top preferences/aspirations are not attributable to mere differences in how fields are composed (Fig.~\ref{fig:regression_robustness}).

\clearpage
\section{Journal Impact Factor Availability and Prediction Accuracy}

\begin{table}[h]
    \centering
    \begin{tabular}{ll}
        \toprule
        Field & \% JIF \\
        \midrule
        Medicine & \MedicinePercentJIF \\
        Biology & \BiologyPercentJIF \\
        Chemistry & \ChemistryPercentJIF \\
        Mathematics & \MathematicsPercentJIF \\
        Economics & \EconomicsPercentJIF \\
        Psychology & \PsychologyPercentJIF \\
        Sociology & \SociologyPercentJIF \\
        Business & \BusinessPercentJIF \\
        History & \HistoryPercentJIF \\
        Engineering & \EngineeringPercentJIF \\
        Philosophy & \PhilosophyPercentJIF \\
        Physics & \PhysicsPercentJIF \\
        Computer science & \ComputerSciencePercentJIF \\
        \bottomrule
        \end{tabular}
    \caption{\textbf{JIF Coverage by Field.} \normalfont{Percentage of field-specific venues (relative to the participant consideration sets, not relative to the database) that have a Journal Impact Factor documented by Clarivate~\protect\citeS{clarivate2024}. The low coverage for Computer Science reflects the field's reliance on conferences, which are not assigned Journal Impact Factors.}}
    \label{tab:si_jif}
    \addcontentsline{toc}{subsubsection}{Table \ref{tab:si_jif}: JIF Coverage by Field}
\end{table}

The Journal Impact Factor quantifies the average number of citations received by articles published in a given venue over the past two years. We collect data on each venue's current Journal Impact Factor from Clarivate's Journal Citation Reports~\citeS{clarivate2024}. However, JIF data is not available for all venues -- Table~\ref{tab:si_jif} shows the percentage of venues per field for which we could retrieve JIF data. For most fields, JIF data is available for approximately 80–90\% of field-specific venues. Computer science is a notable exception, with coverage at only 57\%. This lower percentage reflects the field's reliance on conferences---which do not have a journal impact factor associated with them---rather than journals as primary publication venues.

\begin{figure}[h!]
    \centering
    \includegraphics[width=.7\linewidth]{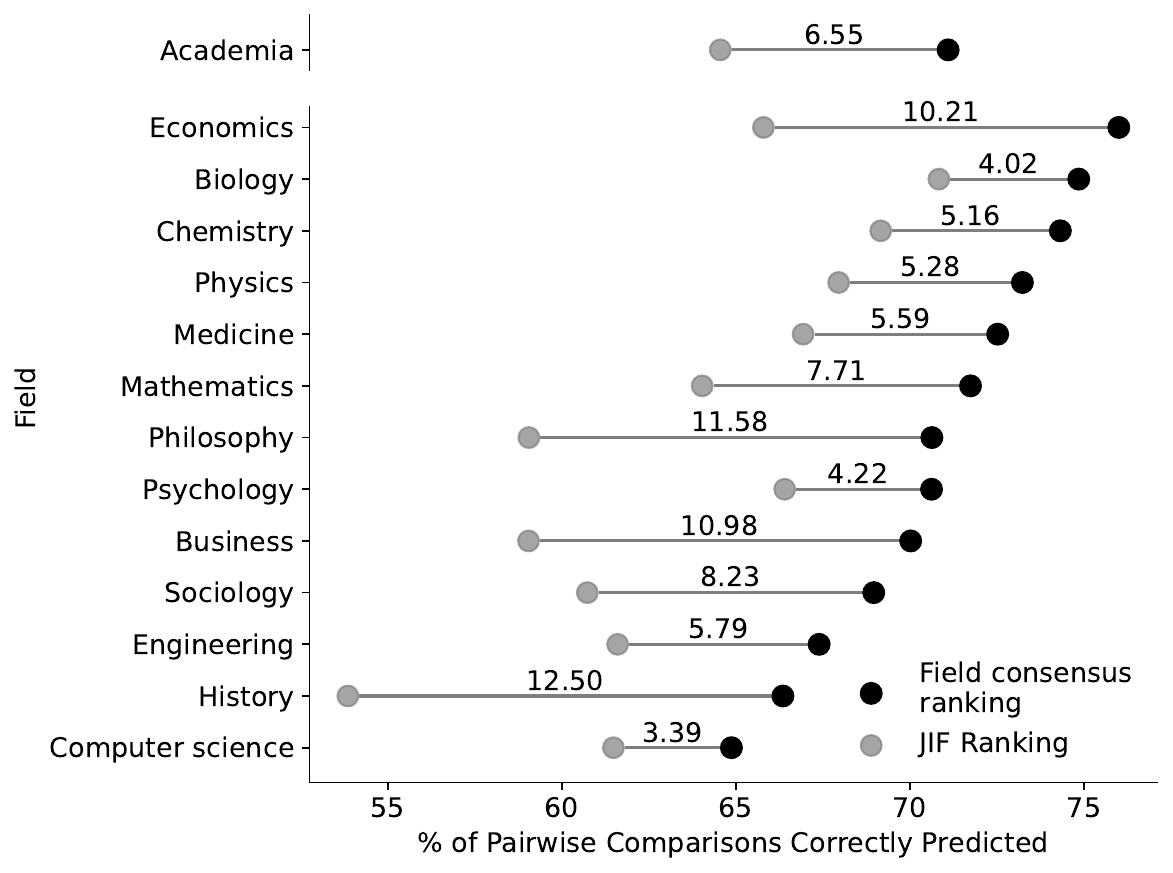}
    \caption{\textbf{JIF vs Consensus Prediction accuracy.} For all pairwise comparisons made by participants in a field for which both venues have a JIF, we assess whether choosing the venue with the higher JIF value or the PorC field consensus rank matches the participant's actual choice. For each participant, the field consensus ranking is computed with that participant's responses held out (leave-one-out). Numeric annotations indicate the difference in prediction accuracy (in percentage points) between the field consensus ranking and JIF.} 
    \label{fig:jif_comparison_accuracy}
    \addcontentsline{toc}{subsubsection}{Figure \ref{fig:jif_comparison_accuracy}: JIF vs Consensus Prediction accuracy}
\end{figure}

To what extent can publication preferences be explained by JIF? To address this, we calculate the percentage of pairwise venue comparisons in each field that can be correctly predicted by choosing the venue with the higher JIF (Fig.~\ref{fig:jif_comparison_accuracy}). We compare this to the predictive accuracy of our field consensus ranking, which consistently outperforms JIF. While it is unsurprising that aggregate rankings offer stronger predictive power, this result underscores that JIF alone does not fully capture academic publication preferences. A subtle point is that the survey algorithm prioritized comparisons that were most informative—typically between venues ranked closely together. As a result, if all participants had evaluated all possible pairs, the predictive accuracy of both JIF and the field consensus ranking would likely increase. 

\begin{figure}[!ht]
\centering
    \includegraphics[width=0.7\linewidth]{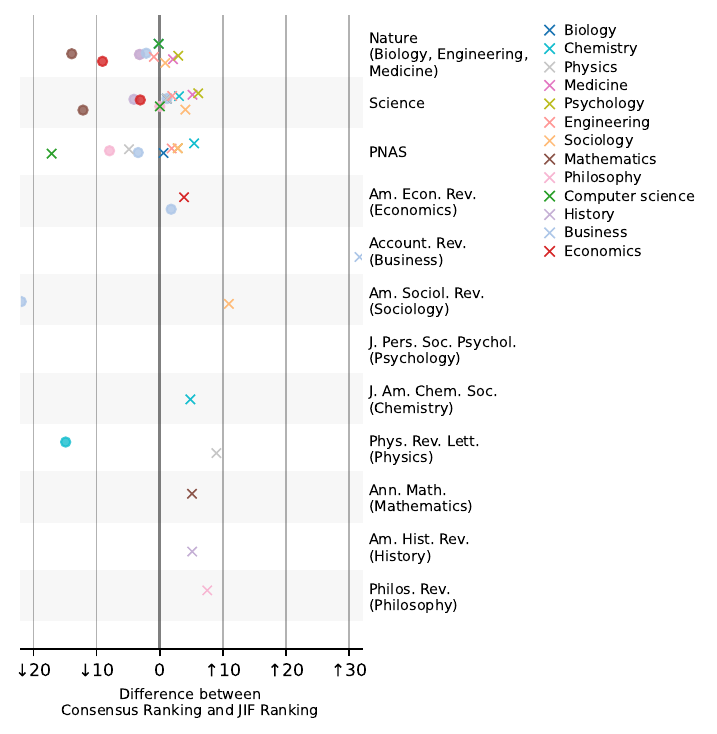}
    \caption{
    \textbf{Difference between the JIF-based ordinal ranking and the field consensus ranking.} We visualize both general-interest venues (\textit{Nature}, \textit{Science}, \textit{PNAS}) and field-specific flagship venues, defined as the most commonly selected top aspiration in each field. Only venues chosen by at least 10\% of a field's participants are included in both the visualization and the underlying rankings. For general-interest venues, colored $\times$'s indicate fields where 25\% or fewer of participants selected the venue (see Fig.~\ref{fig:overlap}). For flagship venues, colored $\times$'s mark the field for which the venue is the top aspiration. 
    }
    \label{fig:jif-deviations-field}
        \addcontentsline{toc}{subsubsection}{Figure \ref{fig:jif-deviations-field}: Difference between the JIF-based ordinal ranking and the field consensus ranking}
\end{figure}

Fig.~\ref{fig:jif-deviations-field} shows field-specific differences for top aspirations and general-interest venues, showcasing between-field relationships through the flagship venues. For example, participants in both Economics and Business aspire to publish in \textit{The American Economic Review}, but not a sufficient fraction of economists aspire to publish in \textit{The Accounting Review}.

\clearpage
\section{Most Common Top Aspirations}

\begin{table}[!ht]
    \centering
\begin{tabular}{lrrr}
\toprule
Field & Top Unique Venues & Mid Unique Venues & Low Unique Venues \\
\midrule
Economics & \EconomicsTopUnique & \EconomicsMidUnique & \EconomicsLowUnique \\
Philosophy & \PhilosophyTopUnique & \PhilosophyMidUnique & \PhilosophyLowUnique \\
Chemistry & \ChemistryTopUnique & \ChemistryMidUnique & \ChemistryLowUnique \\
Physics & \PhysicsTopUnique & \PhysicsMidUnique & \PhysicsLowUnique \\
Business & \BusinessTopUnique & \BusinessMidUnique & \BusinessLowUnique \\
Medicine & \MedicineTopUnique & \MedicineMidUnique & \MedicineLowUnique \\
History & \HistoryTopUnique & \HistoryMidUnique & \HistoryLowUnique \\
Mathematics & \MathematicsTopUnique & \MathematicsMidUnique & \MathematicsLowUnique \\
Sociology & \SociologyTopUnique & \SociologyMidUnique & \SociologyLowUnique \\
Biology & \BiologyTopUnique & \BiologyMidUnique & \BiologyLowUnique \\
Computer science & \ComputerTopUnique & \ComputerMidUnique & \ComputerLowUnique \\
Engineering & \EngineeringTopUnique & \EngineeringMidUnique & \EngineeringLowUnique \\
Psychology & \PsychologyTopUnique & \PsychologyMidUnique & \PsychologyLowUnique \\
\bottomrule
\end{tabular}
\caption{\textbf{Number of unique aspirations by field.} \normalfont{For each of the three aspiration tiers elicited by the survey — top (a top venue you'd like to publish in''), mid (if you weren't aiming for a top venue''), and low (``if you didn't aim as high as either of these'') — the number of distinct venues named by respondents. Fields are ordered by the number of unique top-tier aspirations, from most concentrated (Economics) to most dispersed (Psychology). See Table~\ref{tab:top_venues} for the most common top aspirations in each field.}}
\label{tab:unique_aspirations}
\addcontentsline{toc}{subsubsection}{Table \ref{tab:unique_aspirations}:Number of unique aspirations by field}
\end{table}

Table~\ref{tab:unique_aspirations} shows the number of unique aspirations for each field and each of the three aspiration levels. The fields are ordered by the number of unique venues among the top aspiration. While there are some permutations in the order, the overall trend is consistent with Table~\ref{tab:top_venues} and the high-consensus to low-consensus spectrum we show in the main Results section.

Table~\ref{tab:top_venues} shows most selected top venues by field and the percentage of participants in each field that selected each venue, ordered by the percentage of participants that selected the most aspired to venue in that field. Again, we see a distinction between high-consensus fields like Economics, where 41.9\% of participants selected \textit{The American Economic Review}, whereas in computer science, the most aspired to venue was selected only by 3.6\% of participants.

\begin{table}[!ht]
    \centering
    \small
\begin{adjustbox}{max width=\textwidth}
    \begin{tabular}{@{}llr|llr@{}}
    \toprule
    Field & Most Selected Top Venues & \% Sel. & Field & Most Selected Top Venues & \% Sel. \\
    \midrule
    \multirow{5}{*}{Economics}
        & \EconomicsTopNameA & \EconomicsTopPercentA 
        & \multirow{5}{*}{Chemistry}
        & \ChemistryTopNameA & \ChemistryTopPercentA \\
        & \EconomicsTopNameB & \EconomicsTopPercentB 
        && \ChemistryTopNameB & \ChemistryTopPercentB \\
        & \EconomicsTopNameC & \EconomicsTopPercentC 
        && \ChemistryTopNameC & \ChemistryTopPercentC \\
        & \EconomicsTopNameD & \EconomicsTopPercentD 
        && \ChemistryTopNameD & \ChemistryTopPercentD \\
        & \EconomicsTopNameE & \EconomicsTopPercentE 
        && \ChemistryTopNameE & \ChemistryTopPercentE \\
    \midrule
    \multirow{5}{*}{Sociology}
        & \SociologyTopNameA & \SociologyTopPercentA 
        & \multirow{5}{*}{Biology}
        & \BiologyTopNameA & \BiologyTopPercentA \\
        & \SociologyTopNameB & \SociologyTopPercentB 
        && \BiologyTopNameB & \BiologyTopPercentB \\
        & \SociologyTopNameC & \SociologyTopPercentC 
        && \BiologyTopNameC & \BiologyTopPercentC \\
        & \SociologyTopNameD & \SociologyTopPercentD 
        && \BiologyTopNameD & \BiologyTopPercentD \\
        & \SociologyTopNameE & \SociologyTopPercentE 
        && \BiologyTopNameE & \BiologyTopPercentE \\
    \midrule
    \multirow{5}{*}{History}
        & \HistoryTopNameA & \HistoryTopPercentA 
        & \multirow{5}{*}{Business}
        & \BusinessTopNameA & \BusinessTopPercentA \\
        & \HistoryTopNameB & \HistoryTopPercentB 
        && \BusinessTopNameB & \BusinessTopPercentB \\
        & \HistoryTopNameC & \HistoryTopPercentC 
        && \BusinessTopNameC & \BusinessTopPercentC \\
        & \HistoryTopNameD & \HistoryTopPercentD 
        && \BusinessTopNameD & \BusinessTopPercentD \\
        & \HistoryTopNameE & \HistoryTopPercentE 
        && \BusinessTopNameE & \BusinessTopPercentE \\
    \midrule
    \multirow{5}{*}{Physics}
        & \PhysicsTopNameA & \PhysicsTopPercentA 
        & \multirow{5}{*}{Medicine}
        & \MedicineTopNameA & \MedicineTopPercentA \\
        & \PhysicsTopNameB & \PhysicsTopPercentB 
        && \MedicineTopNameB & \MedicineTopPercentB \\
        & \PhysicsTopNameC & \PhysicsTopPercentC 
        && \MedicineTopNameC & \MedicineTopPercentC \\
        & \PhysicsTopNameD & \PhysicsTopPercentD 
        && \MedicineTopNameD & \MedicineTopPercentD \\
        & \PhysicsTopNameE & \PhysicsTopPercentE 
        && \MedicineTopNameE & \MedicineTopPercentE \\
    \midrule
    \multirow{5}{*}{Mathematics}
        & \MathematicsTopNameA & \MathematicsTopPercentA 
        & \multirow{5}{*}{Engineering}
        & \EngineeringTopNameA & \EngineeringTopPercentA \\
        & \MathematicsTopNameB & \MathematicsTopPercentB 
        && \EngineeringTopNameB & \EngineeringTopPercentB \\
        & \MathematicsTopNameC & \MathematicsTopPercentC 
        && \EngineeringTopNameC & \EngineeringTopPercentC \\
        & \MathematicsTopNameD & \MathematicsTopPercentD 
        && \EngineeringTopNameD & \EngineeringTopPercentD \\
        & \MathematicsTopNameE & \MathematicsTopPercentE 
        && \EngineeringTopNameE & \EngineeringTopPercentE \\
    \midrule
    \multirow{5}{*}{Philosophy}
        & \PhilosophyTopNameA & \PhilosophyTopPercentA 
        & \multirow{5}{*}{Psychology}
        & \PsychologyTopNameA & \PsychologyTopPercentA \\
        & \PhilosophyTopNameB & \PhilosophyTopPercentB 
        && \PsychologyTopNameB & \PsychologyTopPercentB \\
        & \PhilosophyTopNameC & \PhilosophyTopPercentC 
        && \PsychologyTopNameC & \PsychologyTopPercentC \\
        & \PhilosophyTopNameD & \PhilosophyTopPercentD 
        && \PsychologyTopNameD & \PsychologyTopPercentD \\
        & \PhilosophyTopNameE & \PhilosophyTopPercentE 
        && \PsychologyTopNameE & \PsychologyTopPercentE \\
    \midrule
    & & 
        & \multirow{5}{*}{Comp.\ Sci.}
        & \ComputerTopNameA & \ComputerTopPercentA \\
        & & 
        && \ComputerTopNameB & \ComputerTopPercentB \\
        & & 
        && \ComputerTopNameC & \ComputerTopPercentC \\
        & & 
        && \ComputerTopNameD & \ComputerTopPercentD \\
        & & 
        && \ComputerTopNameE & \ComputerTopPercentE \\
    \bottomrule
    \end{tabular}
\end{adjustbox}
\caption{\textbf{Most common top aspirations by field.} {\normalfont For each field, the five most frequently selected top-tier aspiration venues and the percentage of respondents selecting each. Fields are ordered by the selection percentage of the most common top aspiration, from Economics (41.9\%) to Computer Science (3.6\%), reading down the left column then the right. The number of respondents and unique top aspirations per field are reported in Table~\ref{tab:unique_aspirations}.}}
\label{tab:top_venues}
        \addcontentsline{toc}{subsubsection}{Table \ref{tab:top_venues}:Most common top aspirations by field}
\end{table}

\clearpage

\section{Publish or Comparish: A Platform for Pairwise Comparisons}
\label{sec:SI-porc-platform}

This section presents a detailed account of the design of the online platform for pairwise comparisons of publication venues. The core functionalities of this platform have been extracted and made available as an open-source software package for others to build on and use~\citeS{vanbuskirk2024wiserank}.

\subsection{Consent Form}

\begin{figure}[h]
    \centering
        \includegraphics[width=.49\textwidth]{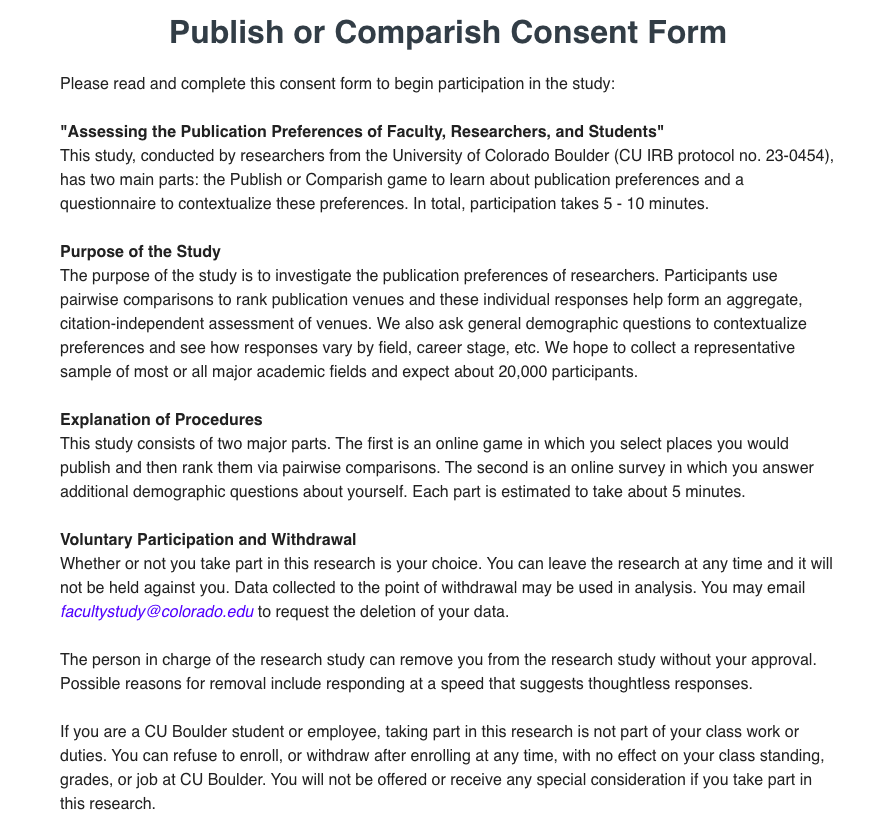}
        \includegraphics[width=.49\textwidth]{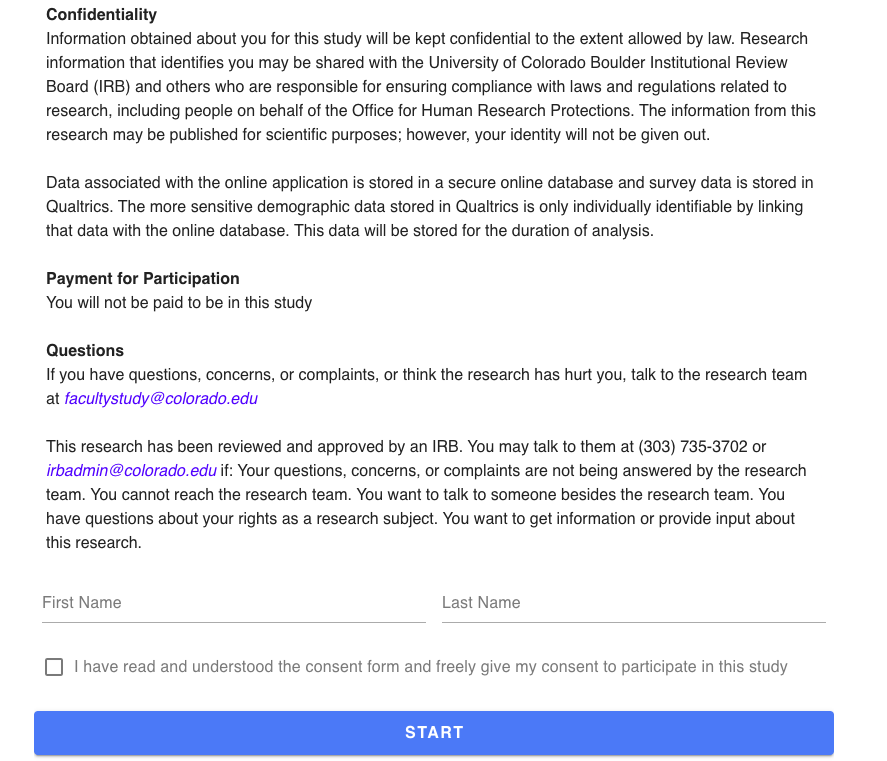}
        \caption[Consent Form]{IRB required consent form with first and last name fields. The form is broken up here for presentation purposes but online is displayed as a single, long page.}
        \label{fig:consent}
\end{figure}

The first page that is common to all participants regardless of how they come to participate in our survey is the consent form (Fig.~\ref{fig:consent}). This form follows the format required by CU Boulder's IRB office. We also collect each participant's first and last name here. This allows us to query OpenAlex behind the scenes and, if possible, collect the participant's publication history. This publication history is used in a later stage of the survey.

\clearpage

\subsection{Academic Rank, Field, and Three Significant Venues}

On the second webpage, participants select their ``Current Academic Rank'' from the following options: Undergraduate, Masters Student, PhD Student, Postdoctoral Researcher, Assistant Professor, Associate Professor, Full Professor, Other. Nineteen options are given for ``Field" and participants must select one of these. The options for ``Subfield" update based on the selected field and an ``Other'' option is provided at the subfield level. The fields and subfields come from the (now deprecated) \href{https://docs.openalex.org/api-entities/concepts}{OpenAlex Concepts}~\citeS{priem2022openalex}.

Participants are also asked three questions about where they would like publish. These questions are meant to elicit from each participant a venue that is, in their mind, a first-tier venue, a second-tier venue, and a third-tier venue. Participants specify a venue by searching for a venue's name and selecting it from our database. Participants do not have the ability to add venues not in our database here or at any other place in the survey. We are careful to remove already selected venues from subsequent participant searches.


\begin{figure}[h]
    \centering
        \includegraphics[width=.7\textwidth]{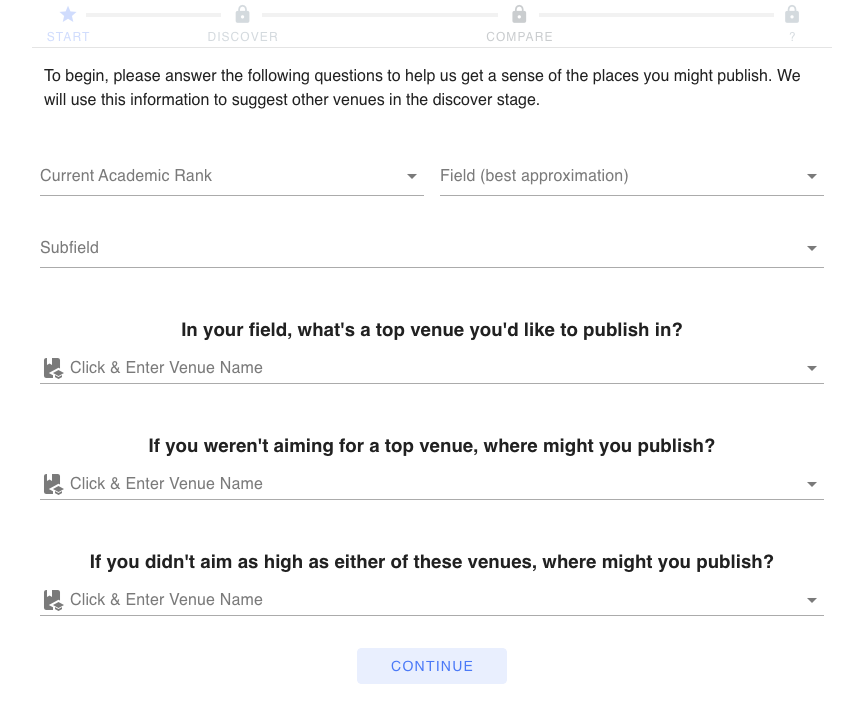}
        \caption[Start Page]{``Start page" where participants provide their current academic rank, field, subfield, and an initial set of $3$ venues where they would like to publish.}
        \label{fig:rank_field}
\end{figure}

\clearpage

\subsection{Venue Discovery: Where Would You Like to Publish?}
\label{venue_discovery}

\begin{figure}[h]
    \centering
        \includegraphics[width=.75\textwidth]{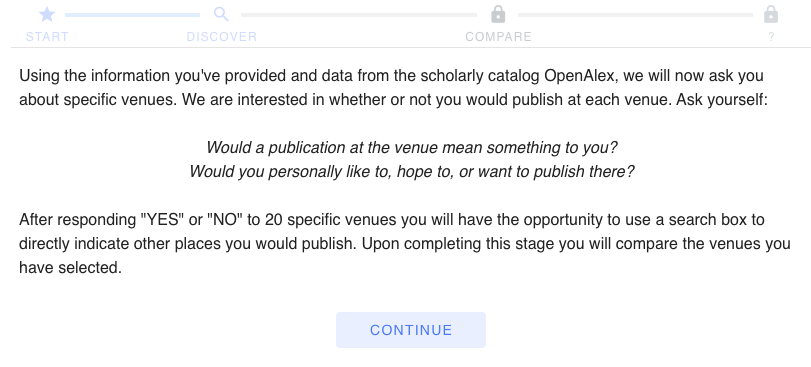}
        \caption[Discovery Stage Instructions]{Instructions for venue discovery stage.}
        \label{fig:disc_dir}
\end{figure}

Before beginning the venue discovery stage participants are shown instructions (Fig.~\ref{fig:disc_dir}). We give participants multiple questions that frame how they should think about the venues they encounter: Would a publication at the venue mean something to them? Would they personally like to, hope to, or want to publish there? The rationale is that by setting the instructions apart on this page participants get the necessary context before they are overwhelmed or simply start clicking options.

\begin{figure}[h]
    \centering
        \includegraphics[width=.75\textwidth]{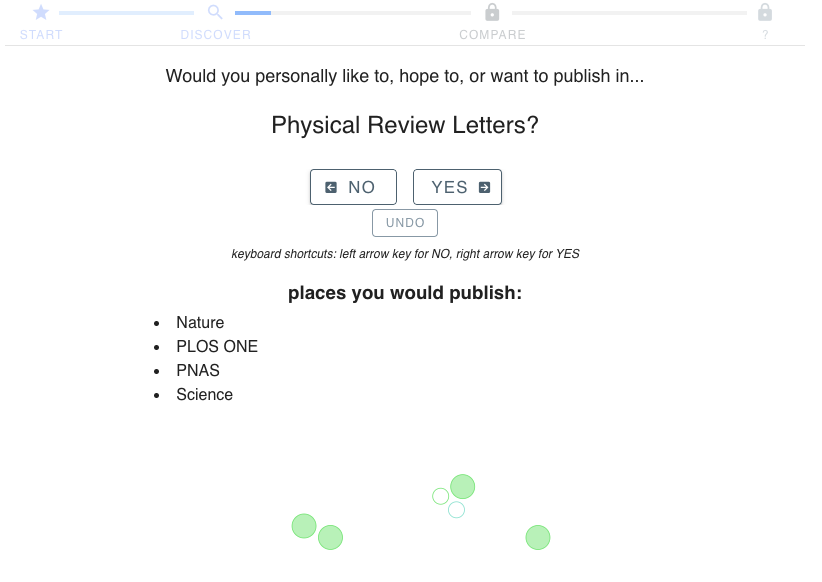}
        \caption[Discovery Stage]{The main portion of the discovery stage where participants specify which venues they would like to publish in. At the bottom of the page, participants can see an embedding of venues they have been asked about, with filled in circles representing those they indicated they would like to publish in.}
        \label{fig:disc_main}
\end{figure}

In the main portion of the discovery stage participants simply indicate whether or not they would like to publish in each of $20$ venues we select and present to them. The interface is fairly straightforward (Fig.~\ref{fig:disc_main}) but it is worth calling out a few aspects. First, the progress bar across the top is present throughout the app and captures the participant's current state within each stage as well as the survey overall. Second, the ``No'' and ``Yes'' responses can be triggered with keyboard shortcuts, as indicated. Third, the undo button was added as a result of beta tester feedback.

Venues the participant has selected are shown as a list and visualized in an embedding alongside those places the participant rejected (Fig.~\ref{fig:disc_main}). This embedding was constructed using a citation network between venues (built using OpenAlex data) and a methodology that treats random walks on a network like sentences in a corpus, allowing one to use basic Word2Vec like embedding techniques to get vector representations of venues \citeS{peng2021neural}. The dimensionality of the resulting vector representation is further reduced using t-SNE~\citeS{van2008visualizing,wattenberg2016how}. The embedding of venues is not only useful for visualization but is likely of direct scientific interest: it covers substantially more venues than the previous embedding of scientific venues that used this method and can be linked to metadata from OpenAlex and our survey~\citeS{peng2021neural}.

\subsection{Recommendation Algorithm}

We also attempted to use the embedding as the basis for our journal recommendation algorithm in this discovery stage. However, tests showed that directly making use of citation counts better predicted what venues would be in the publication histories of academics in our sample frame. As a result, we simply add up normalized counts of how often each possible venue is cited by the venues a participant has said they would like to publish in and the most cited venue is chosen for recommendation. However, we do not always return this venue to the participant. After the three initial questions of the discovery stage, if we have found past publications of the participant on OpenAlex through our direct query of their name, we serve a maximum of five of these found venues in descending order of the number of works each has published. After those five, five from our citation-based recommendation algorithm that do not show up in the found venues are served. Then, for the remainder of this stage we ask about the venue suggested by our recommendation algorithm only if it has more published works than any venue in the remaining set of found venues not yet asked about. Otherwise, we ask about the found venue with the most published works. These steps are in place to ensure that we ask everyone about some venues that do not appear in their publication history on OpenAlex.

An additional technical note: if a participant at any point rejects three of the venues found in their publication history on OpenAlex, we cease to use that pool of venues. The reason for this caution is that it is possible we pair a participant with the wrong OpenAlex profile and want to correct course rather than continue to recommend irrelevant venues.

\begin{figure}[h]
    \centering
        \includegraphics[width=.75\textwidth]{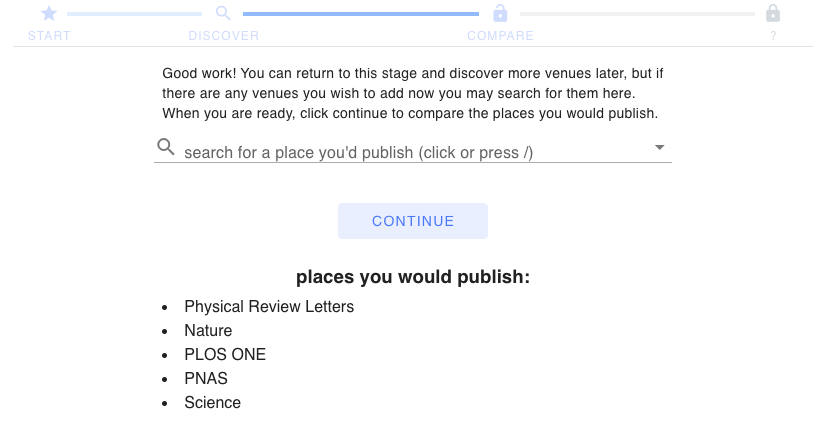}
        \caption[Opportunity to Directly Add Venues]{At the end of the discovery stage participants have the chance to directly add other venues they would like to publish in. The end of this stage and others is marked by the filling of the respective stage's progress bar and the visual ``unlocking'' of the next stage.}
        \label{fig:disc_end}
\end{figure}

At the end of the discovery stage, participants are given the opportunity to directly search for and add other venues in which they would like to publish. This ensures that we do not miss any venues from our database a participant feels are important to consider (Fig.~\ref{fig:disc_end}).

\clearpage

\subsection{Pairwise Comparisons}
\label{s:comps}

\begin{figure}[h]
    \centering
    \includegraphics[width=.75\textwidth]{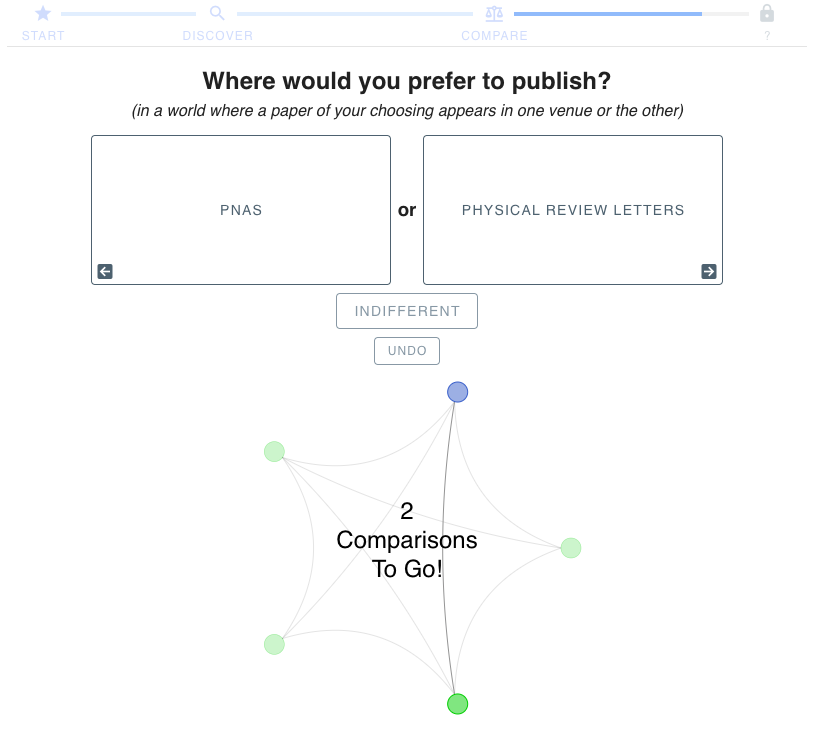}
    \caption[Pairwise Comparison Stage]{The interface for making pairwise comparisons. In addition to the ability to indicate a preference for one venue or the other, affordances are provided that allow the participant to express indifference or ``undo'' their most recently entered response. A visual is also provided that further helps participants mark their progress and motivates the making of more comparisons.}
    \label{fig:comps}
\end{figure}

With selected venues in hand, participants are prepared to make pairwise comparisons between venues. This is arguably the most central part of the data collection process and as such the interface for pairwise comparisons~(shown in Fig.~\ref{fig:comps}) was designed with great care. We aimed to facilitate unbiased but efficient data collection with a simple but powerful interface. Participants may once again make use of keyboard shortcuts, an undo button, and have a progress bar as well as dynamic visualizations to mark their progress. Of most significance here is how venues are paired for comparison. For some time during development the pairing of venues was done randomly. However, feedback from beta testers once again inspired us to develop a more sophisticated approach. The resulting schedule for comparisons is meant to reduce the number of comparisons between very high and very low ranking venues and focus on comparing venues similar in rank. 

Comparisons are organized into three distinct rounds, each of which provides pairs of venues through a different intentional process. In the first round of comparisons, venues are paired at random. No prior information is used to facilitate pairing. While using prior information could lead us to more quickly resolve a ranking and avoid collecting easily inferred data it could also introduce unwanted bias. In the second round, all ``winners'' from round one are paired with other ``winners'' until only one undefeated venue remains. The same is done for the ``losers'' of round one: ``losers'' are paired until only one venue has not been selected over another. In this process, preference is always towards selecting the venue that has been compared the fewest times. If, after these two rounds, some venues have still not been compared $3$ times they are paired with whatever venue is closest in terms of a ranking computed quickly on the backend using SpringRank~\citeS{de2018physical}.

Each venue must be compared three times (or all possible comparisons must be made) for a participant to complete this comparison stage. We estimated that this would give sufficient data to resolve individual preferences to an adequate degree without keeping participants from engaging with the remaining stages of the survey. 
A nudge for the participant to continue appears once participants have compared all venues three times. However, they are allowed to continue making comparisons until they exhaust all $\binom{N}{2}$ pairings. Then, participants are forced to complete the survey as shown in Fig.~\ref{fig:comps_end}.

\subsection{The Reliability of the Pairwise Comparison Pairing Algorithm}
We make use of the approximately 300 participants in the PorC study that completed all $\binom{N}{2}$ comparisons of their selected items to investigate the reliability of our custom pairing algorithm. We can compare a SpringRank model fit with various fractions of the $\binom{N}{2}$ comparisons actually made to a fit on fractions of the $\binom{N}{2}$ comparisons after shuffling. Shuffling the order in which comparisons were made and fitting a model with subsets of these shuffled comparisons approximates the convergence to the final ranking if venues were paired randomly without repetition. 

\begin{figure}[H]
    \centering
    \includegraphics[width=0.48\textwidth]{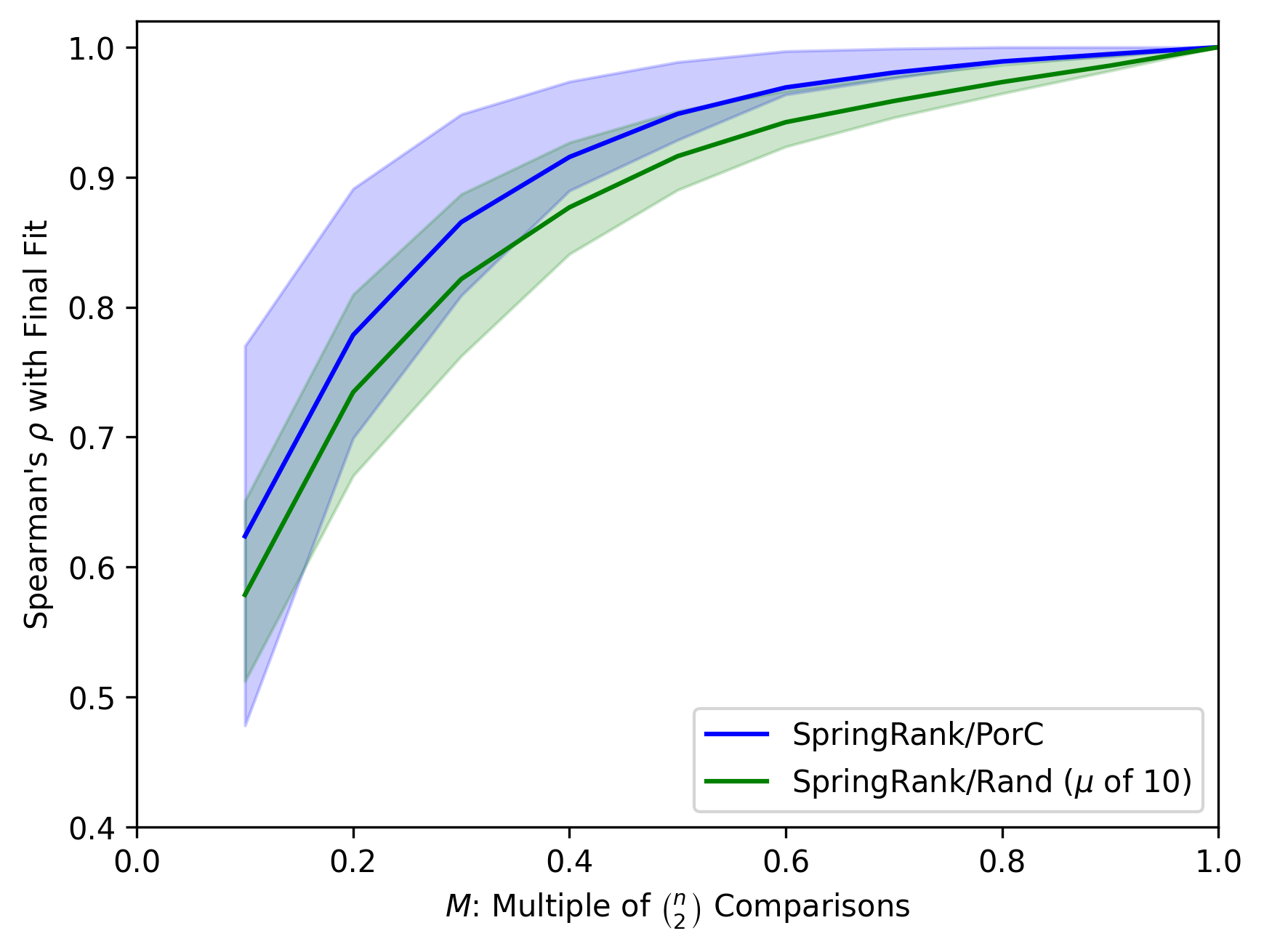}  
    \includegraphics[width=0.48\textwidth]{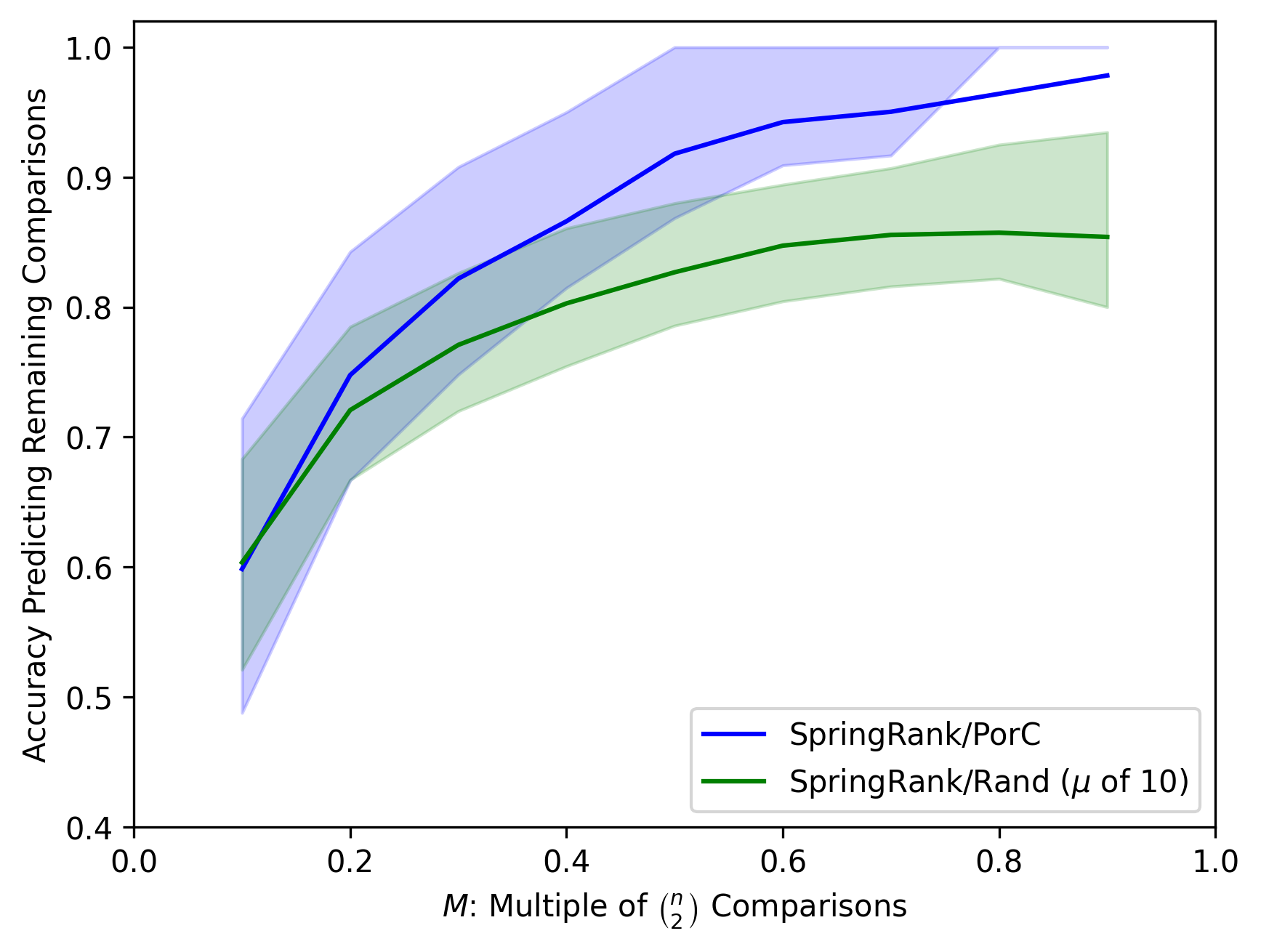}
    \caption[Empirical rank correlations support the Publish or Comparish pairing algorithm.]{The average empirical convergence (left) and predictive accuracy (right) using different subsets of the data from participants that completed all $\binom{N}{2}$ comparisons compared to a counterfactual meant to represent random pairing. Intervals span the middle 60\% of individual results. Each participant is compared to the mean of 10 random shufflings which reduces the spread of the counterfactual data. 
    }
    \label{fig:shuffle}
\end{figure}

\begin{figure}[h]
    \centering
    \includegraphics[width=.75\textwidth]{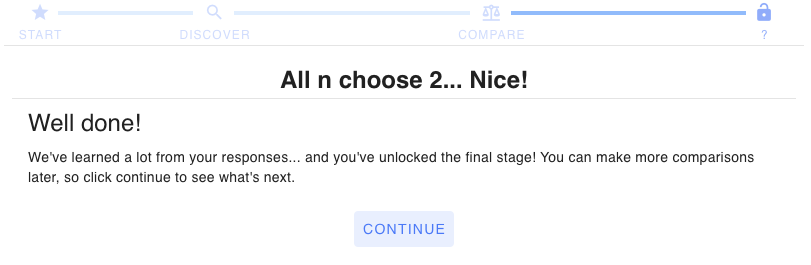}
    \caption[Completing the Survey]{When a participant has exhausted all possible comparisons they are forced to continue to the end of the survey.}
    \label{fig:comps_end}
\end{figure}

Fig.~\ref{fig:shuffle} shows the quality of fit for each scenario measured two ways. First, the Spearman's $\rho$ between each counterfactual ranking and the ranking fit to all $\binom{N}{2}$ comparisons is shown. This reflects how quickly the ranking settles down to a steady state. Second, we document the ability of each counterfactual ranking to predict the remaining unused comparisons. The order in which the custom pairing algorithm asks participants pairwise comparison questions allows for quicker convergence and more accurate prediction of unseen comparisons than random pairing.

.

\subsection{Simulation of Pairwise Comparisons}
\label{si:sims}

Due to the nonrandom nature of the pairing algorithm that determined which pairs of venues participants would compare, we compared the observed pairwise comparison data to data that would be produced were participants to respond at random. In creating these simulated datasets we used the same pairing algorithm as discussed in section~\ref{s:comps} and also endowed each simulated participant with the same proclivity towards indicating indifference between two venues as its observed counterpart. The result of each simulation is a dataset with the exact same number of comparisons made by each participant over the exact same venues, but with preferences randomized but for the bias of our survey's pairing algorithm. 

These simulations provide a model against which we compared our calculations of individuals' self-inconsistencies, in which a pairwise choice contradicted that individual's ranking, also called a rank violation. On average, the higher ranked venues in inconsistent pairs were ranked \ObsAvgUpsetRank~within rescaled ranks, where 1 represents the most preferred venue and 0 the least. Contrasting the observed data, the higher-ranked venue involved in a ranking violation in the simulated data, as described above, had an average rank of \SimAvgUpsetRank~in the simulated data. This confirms that, when individuals make inconsistent choices, those inconsistencies involve less preferred venues.

\subsection{Self-Consistency of Individual Responses}
\label{si:self-consistency}

We define a self-inconsistency as any pairwise comparison in which a respondent's  choice disagrees with their inferred SpringRank ranking. Such disagreements can arise  from two sources: truly intransitive survey responses (e.g., preferring A over B, B  over C, yet C over A) or inferred inconsistencies in which the ranking algorithm  assigns a higher score to a venue that the respondent chose against in a specific  comparison (e.g., estimating a higher utility for B than A despite a response  preferring A over B). We do not distinguish between these two sources in our analysis.

Combined inconsistencies of both types were observed in only  \PercentOfCompsUpset\% of comparisons (range: 1.6\%\textendash3.2\% across  fields), and \PercentWithoutUpset\% of respondents were entirely self-consistent.  Inconsistencies that do occur tend to involve lower-ranked venues: the higher-ranked  venue in inconsistent pairs had an average rescaled rank of \ObsAvgUpsetRank, compared  with \SimAvgUpsetRank~under a null model in which responses were randomized while  preserving the adaptive pairing algorithm (Section~\ref{si:sims}). This confirms that  preferences for lower-ranked venues are less clearly resolved and thus more likely to  produce inconsistent pairwise choices. Importantly, the narrow range of inconsistency  rates across fields (1.6\%\textendash3.2\%) indicates that variation in field-level  consensus is driven by genuinely different preferences among self-consistent  individuals, not by differences in response quality across fields.

\clearpage

\subsection{Completing the Survey}
\label{survey_completion}

\begin{figure}[h]
    \centering
    \includegraphics[height=.24\textwidth]{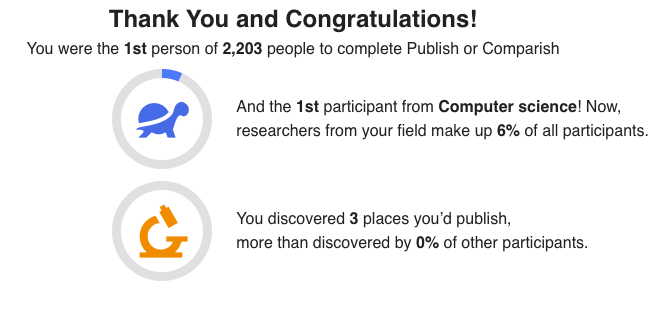}
    \\
    \includegraphics[height=.29\textwidth]{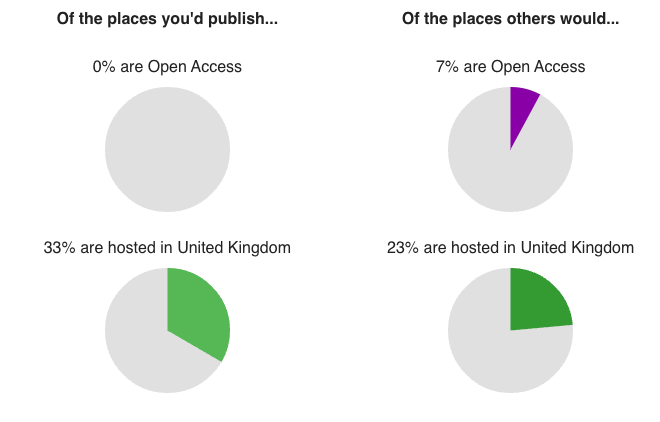}
    \includegraphics[height=.29\textwidth]{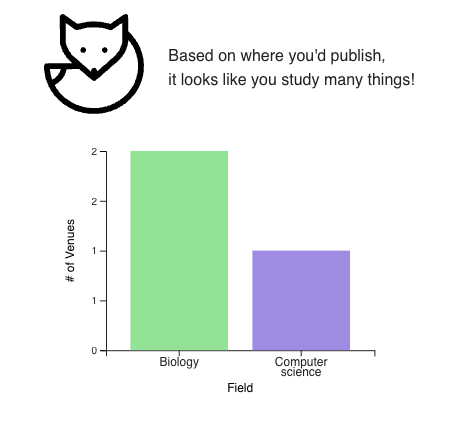}
    \includegraphics[height=.19\textwidth]{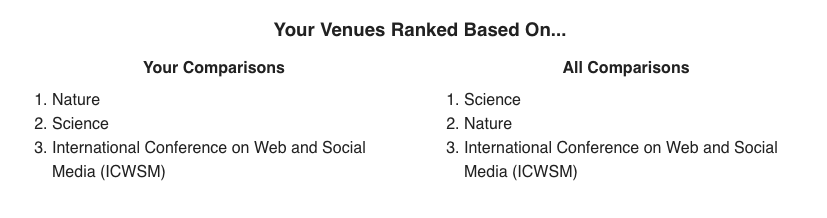}
    \caption[Visualization of Personal Statistics]{Visualizations of personalized statistics that form a slide show when a participant completes the survey.}
    \label{fig:viz_slide}
\end{figure}

When a participant completes the survey they are shown a slide show of statistics and visualizations (Fig.~\ref{fig:viz_slide}) as a way of thanking them and sparking further interest in the study. Sustaining their engagement is important as the following page of the survey includes two additional requests. We first ask them to take the additional questionnaire of demographic questions (discussed in section~\ref{survey}) and then to share the survey with colleagues. As of October 2024, almost 100 participants were invited by another participant and almost half~(1,700) have completed the additional questionnaire.

\begin{figure}[h]
    \centering
    \includegraphics[width=.75\textwidth]{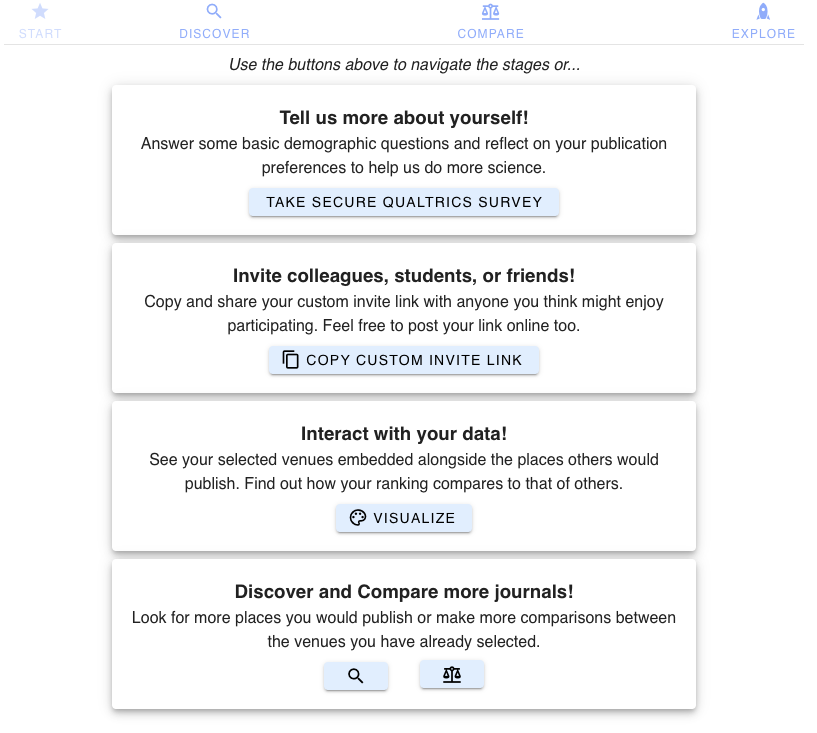}
    \caption[Navigation Page]{The navigation page that a participant sees when they complete the survey, after the slide show of visualizations (Fig.~\ref{fig:viz_slide}).}
    \label{fig:nav}
\end{figure}

\begin{figure}[h]
    \centering
    \includegraphics[width=.75\textwidth]{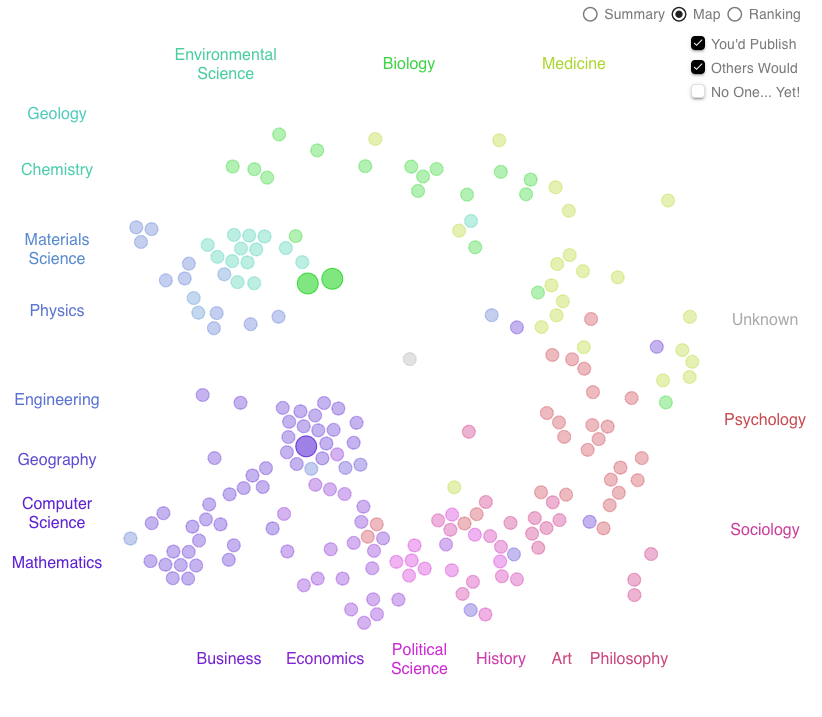}
    \caption[Map of Selected Venues]{A map of venues selected by the participant and others. The venues selected by the participant are larger and have a slightly greater opacity. The participant can toggle on or off different sets of venues to explore the embedding and how they compare to others.}
    \label{fig:map}
\end{figure}

\begin{figure}[h]
    \centering
    \includegraphics[width=.45\textwidth]{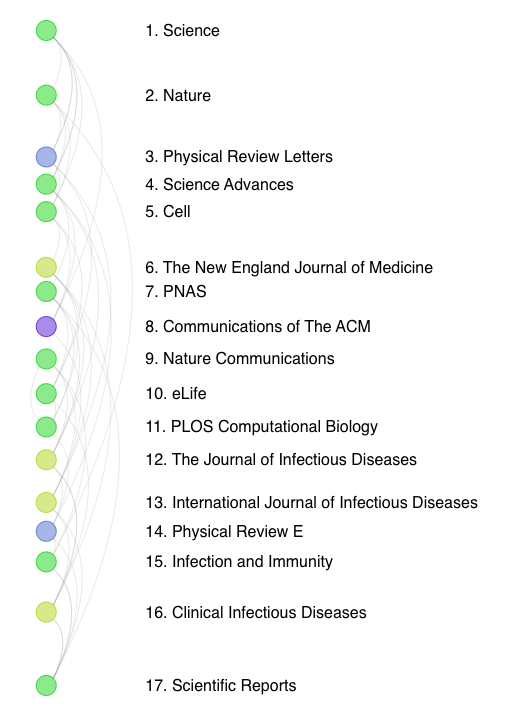}
    \includegraphics[width=.47\textwidth]{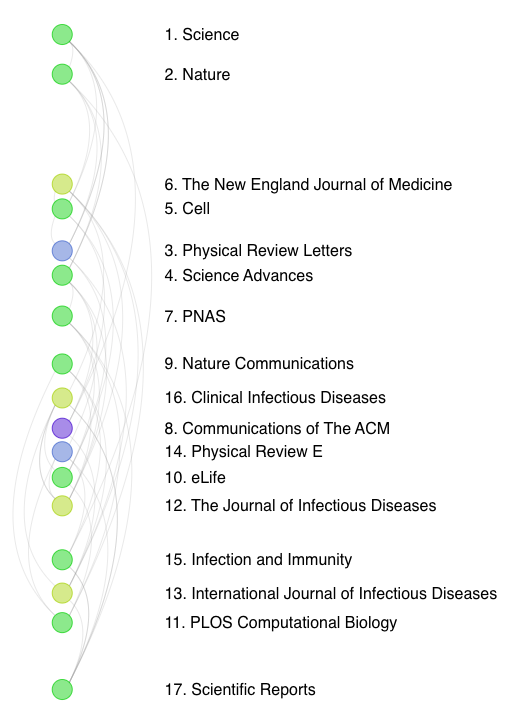}
    \caption[Rankings]{Two rankings, one based only on comparisons made by the participant (left) and the other based on all comparisons between the venues selected by the participant (right).}
    \label{fig:ranking}
\end{figure}

Finally, a few more visualizations are available to participants if they click the ``Visualize'' button. We show a more developed and interactive version of their venue embedding with field labels and venues selected by other participants~\ref{fig:map}. We also show two rankings of the participant's selected venues~\ref{fig:ranking}. Both rankings are formed using SpringRank: one uses only the participant's pairwise comparisons while the other uses all comparisons between the selected venues~\citeS{de2018physical}. This gives the participant an interesting way to compare their perspective to that of others without compromising the privacy of any individuals.

\clearpage

\section{Conducting the Survey}
\label{sec:SI-conducting-the-survey}
\label{survey}

This section details some of the practical aspects of conducting our survey including recruitment and collecting additional demographic information via Qualtrics.

\subsection{Recruitment}

\begin{figure}[h]
    \centering
    \includegraphics[width=.75\textwidth]{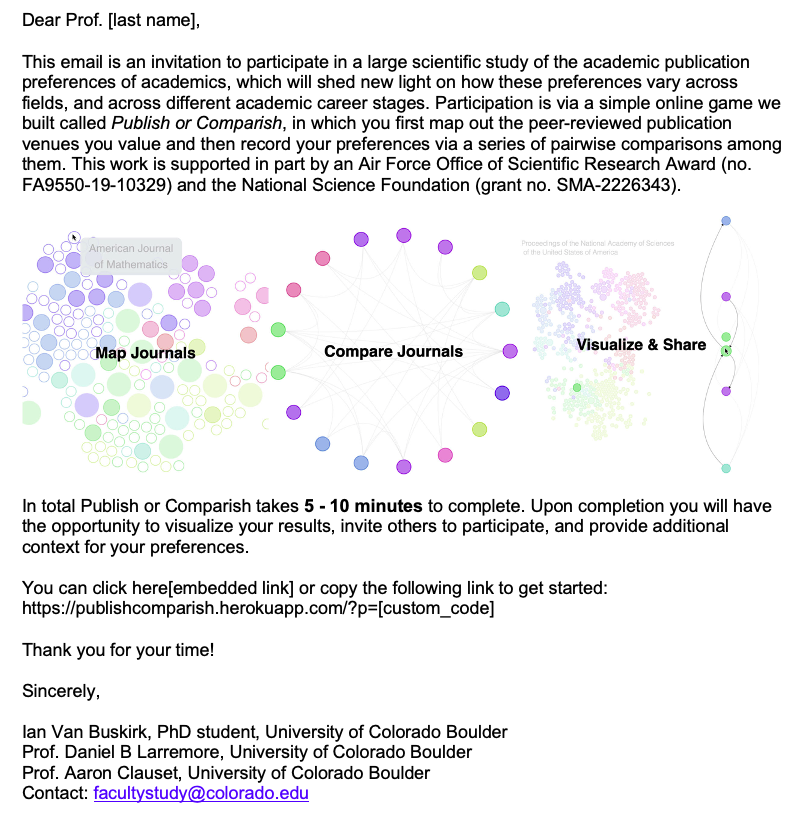}
    \caption[Email Template]{IRB approved email template used to invite faculty to participate in our survey.}
    \label{fig:email}
\end{figure}

The main recruitment method for the study has been emailing the faculty in our sample frame (see section~\ref{sample_frame}). We sent emails based on the template in Fig.~\ref{fig:email} to approximately 70,000 faculty. Emails were sent via Qualtrics in batches of about 3,000 in January and February of $2024$ and follow-up emails were sent in September and October of $2024$. 
The participants were given the option to invite other participants. However, participants recruited by snowball sampling constitute a small fraction of our sample: in total, $158$ individuals completed the survey via snowball sampling, $140$ of whom were of appropriate academic rank to appear in the final analysis. 


\subsection{Qualtrics Questionnaire}
\label{qualtrics}

\begin{figure}[h]
    \centering
    \begin{minipage}[T]{\textwidth}
        \centering
        \includegraphics[width=.45\textwidth]{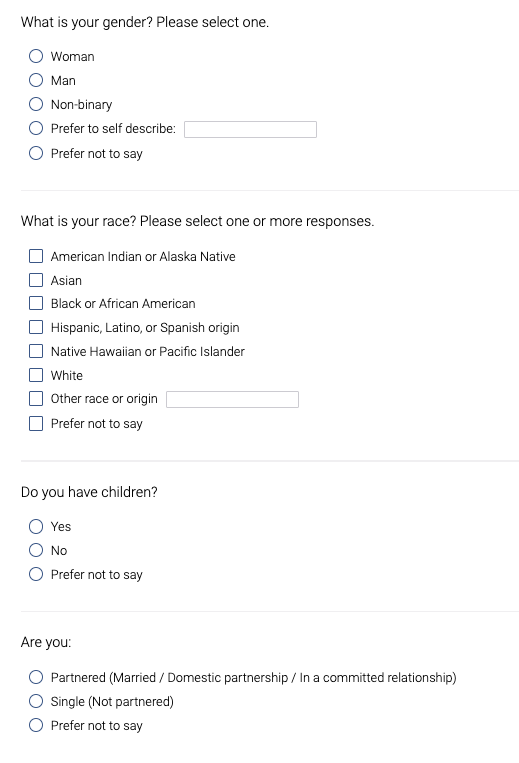}
        \includegraphics[width=.45\textwidth]{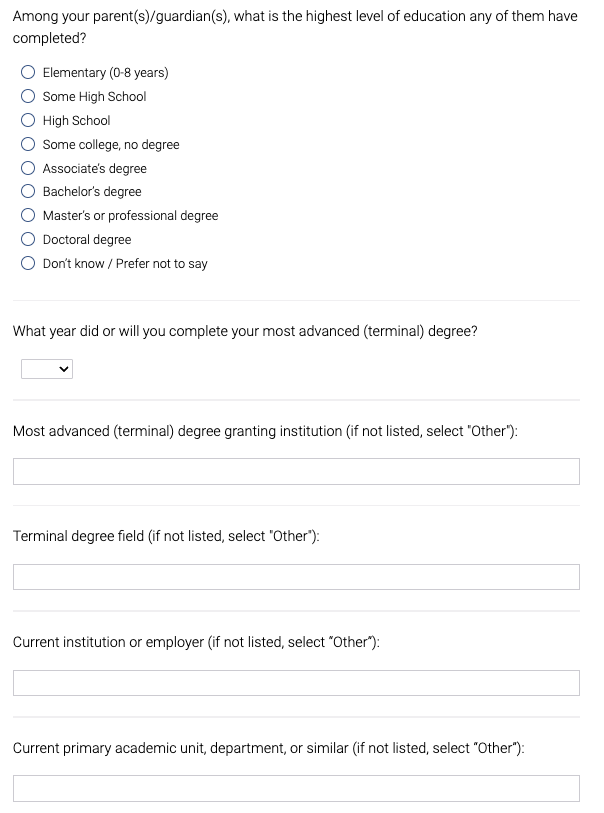}
        \caption[Qualtrics Questionnaire]{The first two pages of the questionnaire hosted on Qualtrics to collect more demographic and qualitative data on publication preferences.}
    \label{fig:quest1}
    \end{minipage}
\end{figure}

\begin{figure}[h]
    \centering
    \begin{minipage}[T]{\textwidth}
        \centering
        \includegraphics[width=.45\textwidth]{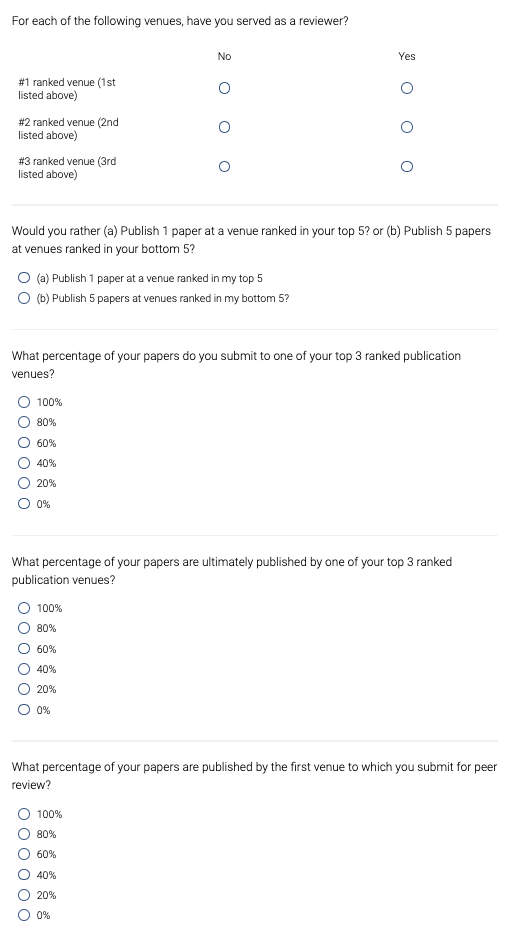}
        \includegraphics[width=.45\textwidth]{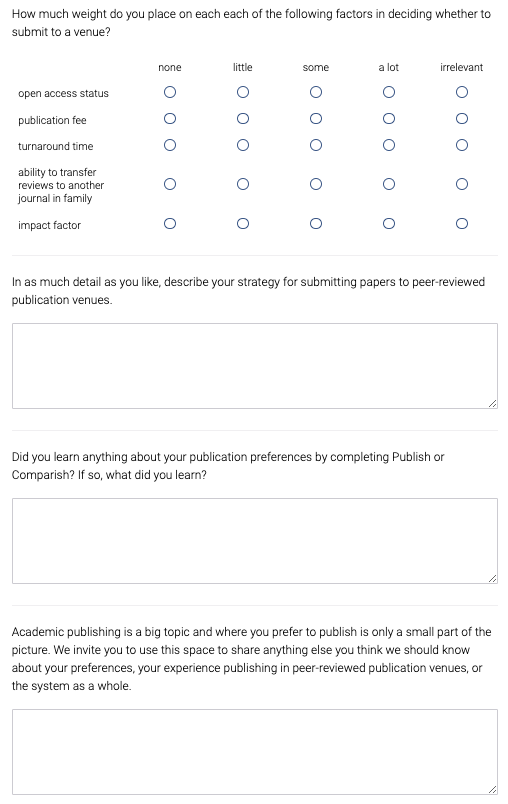}
    \end{minipage}
    \caption[Qualtrics Questionnaire (Cont.)]{The last two pages of the Qualtrics questionnaire. Although not visible here, the Qualtrics API is used to show the venues the participant selected in the Publish or Comparish web application alongside these questions.}
    \label{fig:quest2}
\end{figure}

Because the data captured via our online platform is of primary interest and sufficient on its own for many analyses, we elected to make the demographics portion separate and present it only after participants have completed the main survey (see section~\ref{survey_completion}). Nonetheless the additional information gives meaningful context to the responses of those that elect to complete it. The questionnaire captures both general demographic data and asks publication specific questions. The full questionnaire is shown in Fig.~\ref{fig:quest1} and Fig.~\ref{fig:quest2}.


\clearpage

\section{Supporting Information References}

\bibliographystyleS{unsrt} 
\bibliographyS{prefs}

\end{document}